\documentclass[useAMS,usenatbib,fleqn]{mnras}

\usepackage{graphicx,color}
\usepackage{mathtools}	
\usepackage{amsmath}
\usepackage{amssymb}
\usepackage{bm}
\usepackage{subfig}             
\usepackage{ulem}
\usepackage{fixltx2e} 
\usepackage{pstricks} 
   \usepackage[british]{babel}             
   \usepackage{newtxtext}                  
   \usepackage[slantedGreek]{newtxmath}    
   \usepackage[T1]{fontenc}                

\def\apj{ApJ}

\def\mnras{MNRAS}
\def\aap{A\&A}
\def\apjl{ApJ}

\def\physrep{PhR}
\def\pre{PhRvE}
\def\prl{PhRvL}

\def\nat{Nature}

\def\ssr{SSRv}
\def\araa{ARA\&A}

\def\gapfd{GApFD}
\def\an{AN}

\renewcommand{\vec}[1]{{{\mbox{\boldmath $#1$}}}}

\newcommand{\Exp}[1]{{\rm e}^{#1}}

\newcommand{\del}{\partial}

\newcommand{\bfDel}{\bm{\nabla}}
\newcommand{\bmDel}{\bm{\nabla}}

\newcommand{\bfOmega}{\vec{\Omega}}

\newcommand{\eps}{\epsilon}

\newcommand{\Emf}{\bm{\mathcal{E}}}

\newcommand{\Flux}{\bm{\mathcal{F}}}
\newcommand{\flux}{\mathcal{F}}

\newcommand{\bfJ}{\bm{J}}
\newcommand{\bfu}{\bm{u}}
\newcommand{\bfb}{\bm{b}}

\newcommand{\bfz}{\bm{z}}
\newcommand{\Emfr}{\mathcal{E}_r}
\newcommand{\Emfp}{\mathcal{E}_\phi}
\newcommand{\Emfz}{\mathcal{E}_z}
\newcommand{\Emfi}{\mathcal{E}_i}

\newcommand{\mean}[1]{\overline{#1}}
\newcommand{\meanv}[1]{\bm{#1}}

\newcommand{\eq}{_\mathrm{eq}}					
\newcommand{\f}{_\mathrm{0}}					
\newcommand{\kin}{_\mathrm{k}}			   		
\newcommand{\magn}{_\mathrm{m}}			   		
\newcommand{\turb}{_\mathrm{t}}			   		
\newcommand{\crit}{_\mathrm{c}}			   		
\newcommand{\const}{\mathrm{const}}			   	
\newcommand{\udiff}{^\mathrm{d}}			   	

\newcommand{\VC}{_\mathrm{VC}}			   		
\newcommand{\uVC}{^\mathrm{VC}}			   		

\newcommand{\cro}{\times}
\newcommand{\Rm}{\mathcal{R}_\mathrm{m}}

\newcommand{\mbr}{B_r}
\newcommand{\mbp}{B_\phi}
\newcommand{\mbz}{B_z}
\newcommand{\mbfr}{B_{\mathrm{0},r}}
\newcommand{\mbfp}{B_{\mathrm{0},\phi}}

\newcommand{\alphatilde}{\widetilde{\alpha}}

\newcommand{\Dtilde}{\widetilde{D}}

\newcommand{\Coriolis}{\mathrm{Co}}
\newcommand{\Strouhal}{\mathrm{St}}

\newcommand{\cyc}{_\mathrm{cyc}}

  \newcommand{\kms}{\,{\rm km\,s^{-1}}}
  \newcommand{\kmskpc}{\,{\rm km\,s^{-1}\,kpc^{-1}}}

  \newcommand{\kpc}{\,{\rm kpc}}

  \newcommand{\Myr}{\,{\rm Myr}}
  
  \newcommand{\Gyr}{\,{\rm Gyr}}

  \newcommand{\yr}{\,{\rm yr}}     
  
\title[Galactic Dynamos and the Magnetic R\"{a}dler Effect]
{Non-linear Galactic Dynamos and the Magnetic R\"{a}dler Effect}
\author[L.\ Chamandy, N.\ Singh]{Luke Chamandy,$^{1}$\thanks{lchamandy@pas.rochester.edu}
\& Nishant K. Singh$^{2}$\thanks{singh@mps.mpg.de}\\
$^{1}$Department of Physics and Astronomy, University of Rochester, Rochester NY, 14618, USA\\
$^{2}$Max Planck Institute for Solar System Research,
Justus-von-Liebig-Weg 3, D-37077 G\"ottingen, Germany
}

\begin{document}

\maketitle

\begin{abstract}
We show that the magnetic analogue of the R\"{a}dler effect of mean-field dynamo theory leads to a non-linear backreaction
that quenches a large-scale galactic dynamo, 
and can result in saturation of the large-scale magnetic field at near-equipartition with turbulent kinetic energy density.
In a rotating fluid containing small-scale magnetic fluctuations, 
anisotropic terms in the mean electromotive force are induced via the Coriolis effect
and these terms lead to a reduction of the growth rate in a predominantly $\alpha\Omega$-type galactic dynamo \citep{Chamandy+Singh17}.
By including the generation of small-scale magnetic fluctuations by turbulent tangling of the large-scale
magnetic field, one obtains a negative feedback effect that quenches the dynamo and leads to the saturation of the large-scale field.
This saturation mechanism is found to be competitive with the dynamical $\alpha$-quenching mechanism for realistic galactic parameter values.
Furthermore, in the context of the dynamical $\alpha$-quenching model, 
a separate non-linear term is obtained which has the same form as the 
helicity flux term of \citet{Vishniac+Cho01}, but which depends on the strength of small-scale magnetic fluctuations.
We briefly discuss the observational implications of the magnetic R\"{a}dler effect for galaxies.
\end{abstract}

\label{firstpage}
\defcitealias{Chamandy+Singh17}{Paper~I}
\defcitealias{Brandenburg+Subramanian05a}{BS05}
\defcitealias{Radler+03}{RKR}
\defcitealias{Chamandy+14b}{CSSS}
\defcitealias{Chamandy+Taylor15}{CT}
\defcitealias{Squire+Bhattacharjee15c}{SB15}
\defcitealias{Rogachevskii+Kleeorin07}{RK07}
\section{Introduction}
\label{sec:introduction}
Mean-field dynamo theory has had success in explaining the origin and
properties of large-scale magnetic fields of galaxies
\citep{Ruzmaikin+88,Beck+96,Shukurov07,Chamandy+16}.
In this theory, a key quantity is the mean electromotive
force $\Emf= \mean{\bfu\cro\bfb}$, 
where $\bfu$ and $\bfb$ are the small-scale turbulent velocity and
magnetic fields, respectively, and overbar denotes mean.
Here small-scale refers to scales smaller than the correlation length
$l$ of the turbulent velocity field,
while large scales are based on averages over scales much larger than $l$ but much smaller than the system size.%
\footnote{See \citet{Zhou+17} for a detailed discussion about averaging in mean-field dynamos.}
The quantity $\Emf$ can be written as a series in spatial derivatives of the large-scale or mean magnetic field $\meanv{B}$
with `turbulent transport' coefficients depending on correlations of small-scale fluctuating quantities \citep{Moffatt78, Krause+Radler80}.
In early works, these coefficients were found to depend on the statistical properties of the small-scale velocity field,
but contributions arising from the small-scale magnetic field were usually neglected in their derivations.

The mean emf $\Emf$ contains a term $\alpha\meanv{B}$, for instance, 
which is primarily responsible for generating poloidal mean magnetic field from toroidal,
and $\alpha$ is found to be proportional to the mean small-scale kinetic helicity density.%
\footnote{In general, the relevant term in the expression for $\Emfi$ is $\alpha_{ij}B_j$.
Adopting cylindrical coordinates $(r,\phi,z)$, 
with $\bfz$ along the galactic rotation axis, 
$\alpha_{\phi\phi}$ is responsible for generating poloidal from toroidal field.}
Such helical turbulence can be generated by vertical stratification and large-scale rotation, 
ubiquitous properties of disc galaxies.
However, oppositely signed mean small-scale magnetic helicity density builds up as a byproduct
of large-scale dynamo action, and a term involving the associated mean small-scale current helicity
acts to `catastrophically' quench the $\alpha$ effect 
\citep{Pouquet+76,Kleeorin+Ruzmaikin82,Gruzinov+Diamond94,Bhattacharjee+Yuan95}.
This scenario is averted in nature likely because there is a flux 
of the mean small-scale magnetic helicity density away from the dynamo-active region, 
allowing the mean field to saturate at near-equipartition with turbulent kinetic energy density 
\citep{Kleeorin+Rogachevskii99,Blackman+Field00,Kleeorin+00,Vishniac+Cho01,Field+Blackman02,Blackman+Field02}.
The simplest such flux terms are the advective \citep{Subramanian+Brandenburg06} and diffusive \citep{Mitra+10} fluxes,
which have been shown in models to lead to the expected saturation 
\citep[e.g.][hereafter \citetalias{Chamandy+14b}]{Kleeorin+02,Shukurov+06,Sur+07b,Chamandy+14b}.

Large-scale plasma motions such as the galactic rotation and the shear associated with the gradient of this rotation along galactocentric radius
have important roles in the mean-field dynamo.
The radial shear is primarily responsible for generating toroidal from poloidal field 
through the so-called $\Omega$-effect,
but shear also generates anisotropy of the turbulence,
which leads to additional terms in $\Emf$.
One such effect, the so-called `shear-current' effect, 
has been proposed as a driver of dynamo action
that, unlike the $\alpha$ effect, operates even without the presence of mean small-scale helicity.
Likewise, aside from being responsible, along with stratification,
for the generation of an $\alpha$ effect, 
the Coriolis force from large-scale rotation also generates anisotropy, 
resulting in additional terms in $\Emf$.
The R\"{a}dler or $\bfOmega\times\bfJ$ effect is one such effect that,
like the shear-current effect, has been proposed as yet another driver of dynamo action.
However, it has not been shown that the shear-current or R\"{a}dler effects
can lead to sustained dynamo action in a realistic
setting \citep{Rogachevskii+Kleeorin03,Rogachevskii+Kleeorin04,
Brandenburg+08a,Sridhar+Subramanian09a,Sridhar+Subramanian09b,
Sridhar+Singh10,Singh+Sridhar11}.
Yet another mechanism involving the Moffatt drift, which is expected to
exist in presence of statistically anisotropic $\alpha$ fluctuations,
was recently proposed that could enable a large-scale dynamo with
or without the shear \citep{Sridhar+Singh14,Singh16,Jingade+18}.

Essentially independently of the large-scale dynamo,
a small-scale or fluctuation dynamo is thought to be present in galaxies 
(\citealt{Biermann+Schluter51,Kraichnan+Nagarajan67,Kazantsev68,Meneguzzi+81,Kulsrud+Anderson92}; for reviews see \citealt{Brandenburg+Subramanian05a}, 
hereafter \citetalias{Brandenburg+Subramanian05a}; \citealt{Brandenburg+12a}).
This dynamo operates on the small-scale field, causing it to amplify 
exponentially with an $\mathrm{e}$-folding time of the order of the shortest eddy turnover time of the turbulence,
since we are dealing here with high magnetic Prandtl number flows \citepalias{Brandenburg+Subramanian05a}.
The small-scale magnetic field will then saturate near equipartition.
A high degree of compressibility as might be expected in some cases would likely lead to growth rates and saturation values 
a few times smaller than in the non-compressive or midly compressive case, assuming transonic turbulence \citep{Federrath+11}.
However, the $\mathrm{e}$-folding time would in any case be expected to be much smaller than the turnover time scale of energy-carrying eddies,
which has been estimated to be of the order $10^7\yr$ \citep{Shukurov07}.
The $\mathrm{e}$-folding time of the large scale field is expected to be greater than the galactic rotation period,
of order $10^8\yr$.
When the small-scale field saturates, the large-scale field is thus expected to still be very weak, 
and in the kinematic regime of mean-field dynamo action.%
\footnote{Although the mean of the small-scale field formally vanishes, 
it can be shown using a more careful approach taking into account the typical scale separation in galaxies,
that its mean strength will be a few orders of magnitude below the rms small-scale field strength.
This residual field provides a seed for the large-scale dynamo \citep{Ruzmaikin+88,Beck+94,Subramanian+Brandenburg14,Zhou+17}.}
Thus, magnetic fluctuations are likely present at near-equipartition levels during mean-field dynamo action,
and their effects should be taken into account in galactic dynamo models.

What has been considered only relatively recently 
(\citealt{Radler+03}, hereafter \citetalias{Radler+03}; \citetalias{Brandenburg+Subramanian05a}; 
\citealt{Squire+Bhattacharjee15b}; \citealt{Squire+Bhattacharjee15c}; \citealt{Chamandy+Singh17}, hereafter \citetalias{Chamandy+Singh17})
is the inclusion of the dependence of small-scale magnetic fluctuations
on terms stemming from the inclusion of effects that produce anisotropy.
While the R\"{a}dler effect, for instance, 
turns out to depend on the mean small-scale kinetic energy density,
including magnetic fluctuations in its derivation leads to analogous terms
proportional to the mean small-scale magnetic energy density. 

In \citetalias{Chamandy+Singh17} we explored the relevance of the magnetic analogue 
of the R\"{a}dler effect for galactic dynamos in the kinematic regime of the mean field.
It was found that for realistic values of the ratio of mean small-scale magnetic and kinetic energy densities $\xi$,
the magnetic R\"{a}dler effect leads to an effective partial suppression of the $\alpha$ effect.
This can result in a small to dramatic reduction in the dynamo growth rate,
or in some cases even negate dynamo action, depending on the values of certain parameters, especially $\xi$.

Likewise, earlier works have proposed that a near-equipartition small-scale magnetic field that developed on time scales 
much smaller than the time scale for large-scale dynamo growth (the latter is of the order of the galactic rotation period) 
could hamper large-scale dynamo action.
\citet{Kulsrud+Anderson92} speculated
that strong magnetic fluctuations would lead to non-linear feedback that would arrest mean-field dynamo action in galaxies. 
In particular, they suggested that once equipartition was reached between small-scale magnetic field and turbulence, 
turbulent energy would be converted to small-scale magnetic field and heat, leaving ``little left'' for the mean-field dynamo. 
They thus concluded that observed large-scale fields must be of primordial origin.
However, rather convincing arguments against a primordial origin of the observed large-scale fields of galaxies, 
and in favour of a dynamo origin, have since become well-established
(\citealt{Beck+96}; \citetalias{Brandenburg+Subramanian05a}; \citealt{Shukurov07,Gressel+08a,Gressel+08b,Gent+13b}).

Despite arguments to the contrary \citep{Tobias+Cattaneo13}, 
there is little reason to doubt that the fluctuation dynamo operates in galaxies,
and very quickly builds up small-scale fluctuations to near-equipartition levels \citep{Kolokolov+11,Singh+17}.
At the same time, \citet{Sur+08,Subramanian+Brandenburg14} were able to show 
that a dynamo resembling the expected mean-field dynamo was indeed operating 
in their direct numerical simulations (DNS) even in the presence of a fluctuation dynamo.

In light of these results and others, 
we take the view that large-scale galactic dynamo action must occur in the presence of strong magnetic fluctuations.
This still leaves ample room to explore the effects of near-equipartition small-scale magnetic fluctuations on the mean-field dynamo mechanism,
and here we focus on one such possible effect.

In this paper, we continue our exploration of the magnetic R\"{a}dler effect
with an extension of our model into the non-linear regime in the mean-field $\meanv{B}$.
In Section~\ref{sec:model}, we present the basic galactic dynamo model.
Then in Section~\ref{sec:xiVC} we elaborate on the nature of a new term 
that arises from the magnetic R\"{a}dler effect in the dynamical $\alpha$-quenching non-linearity;
we provide a derivation of this term under more general considerations in Appendix~\ref{sec:xiVC_general}.
Model parameters are then discussed in Section~\ref{sec:parameters}.
Following this, we present the results of our basic dynamo model that includes the magnetic R\"{a}dler effect in Section~\ref{sec:results},
and briefly compare them with approximate analytic solutions, derived in Appendix~\ref{sec:analytic}.
We then go on to consider the more realistic case that includes turbulent tangling of the magnetic field to produce small-scale fluctuations 
in Section~\ref{sec:dynamical_xi},
and numerical dynamo solutions that incorporate this tangling are presented in Section~\ref{sec:tangling_results}.
Finally, we discuss our results in Section~\ref{sec:discussion} and
present a summary and conclusions in Section~\ref{sec:conclusions}.
Appendix~\ref{sec:driving} explores the limiting behaviour of solutions when small-scale magnetic fluctuations are made to be large.

\section{Basic model}
\label{sec:model}

The basic kinematic model for the large-scale dynamo is presented in \citetalias{Chamandy+Singh17}, 
and we refer the reader to that work for details. 
Here we summarize the basic model but focus on the new non-linear terms in the mean-field $\meanv{B}$.
We begin by writing down the mean induction equation
\begin{equation}
  \label{dynamo}
  \frac{\del\meanv{B}}{\del t}= \bmDel\cro\left( \meanv{U}\cro\meanv{B} +\Emf\right),
\end{equation}
where $\Emf=\mean{\bfu\cro\bfb}$.
Microscopic diffusion has been neglected as the magnetic Reynolds number $\Rm$ 
is many orders of magnitude larger than unity in galaxies.
In any case, under the first-order smoothing or quasilinear approximation 
adopted below, turbulent diffusion and Ohmic diffusion combine linearly,
with the diffusivity being equal to the sum of turbulent and microscopic diffusivities.
We adopt the first-order smoothing approximation (FOSA) in our model for simplicity.
Adopting instead the minimal $\tau$ approximation (MTA) as in \citetalias{Brandenburg+Subramanian05a}
would mean that $\Emf$ would be solved for using an additional equation involving $\del\Emf/\del t$.
It has been shown \citep{Chamandy+13a,Chamandy+13b} that this memory effect can be important
for galactic dynamos that include non-axisymmetric spiral forcing, 
but that it can generally be neglected for the axisymmetric case as long as the dynamo growth or decay time $1/\lambda$
is much larger than the relaxation time $\tau_\mathrm{MTA}$ \citepalias[see also][]{Chamandy+Singh17}.
This is satisfied to a reasonable degree in the models explored in this work, and non-axisymmetry is neglected.
To maintain consistency between results from MTA and FOSA,
$\tau_\mathrm{MTA}$ is interpreted to be equal to the correlation time $\tau$ of turbulence \citep[c.f.][]{Brandenburg+Subramanian05b}.

\subsection{Mean electromotive force}
\label{sec:emf}
For the mean electromotive force, 
we adopt the expression computed in Section~10.3 of \citetalias{Brandenburg+Subramanian05a} (see also \citetalias{Radler+03}),
for the case where mean kinetic helicity density is induced by slow rotation
(formally, $\Omega\tau\ll1$) and weak stratification of the turbulence 
(e.g.~$\del/\del z\ll1/l$),
while the density $\rho$ is assumed to be constant and the turbulence incompressible.
Large-scale shear was also omitted from the derivation of $\Emf$ to make their calculation tractable,
but differential rotation plays a key role in our dynamo model by providing the $\Omega$ effect
mentioned in Section~\ref{sec:introduction}.
Keeping with the notation of \citetalias{Brandenburg+Subramanian05a} and \citetalias{Chamandy+Singh17},
we adopt units such that $\rho=1$ and $\mu\f=1$ so that $\bfb$ is the Alfven velocity.

In our galaxy model, 
we make use of cylindrical coordinates $(r,\phi,z)$ with the angular velocity equal to $\Omega\bm{\hat{z}}$.
We also apply the slab approximation, 
where $\del/\del r$ and $(1/r)\del/\del\phi$ are neglected,
except for the radial shear $\del\Omega/\del r$.
Thus, the problem is 1D in $z$, 
but \citet{Chamandy16} showed that by simply ``stitching together'' 
local \textit{saturated} galactic dynamo solutions which depend parametrically on $r$,
it is possible to reproduce remarkably well axisymmetric global solutions obtained
from the full set of axisymmetric equations that include $\del/\del r$ terms.
Of course, such a comparison has not yet been done when including the magnetic R\"{a}dler effect,
but there is no reason to expect that local saturated solutions would not rather accurately
approximate locally the global solutions.
Therefore, we restrict ourselves to 1D solutions both for simplicity and numerical expediency,
but global models would be useful in the future, e.g. to explore non-axisymmetry.

Under the slab approximation, transport coefficients reduce to scalars or pseudo-scalars.
Defining $u\equiv \left(\mean{\bfu^2}\right)^{1/2}$ and $b\equiv \left(\mean{\bfb^2}\right)^{1/2}$, we obtain \citepalias[c.f.][]{Brandenburg+Subramanian05a}
\begin{align}
  \label{Emfr_slab}
  \Emfr&= \alpha\mbr +\eta\frac{\del\mbp}{\del z} -\gamma\mbp -\delta'\frac{\del\mbr}{\del z},\\
  \label{Emfp_slab}
  \Emfp&= \alpha\mbp -\eta\frac{\del\mbr}{\del z} +\gamma\mbr -\delta'\frac{\del\mbp}{\del z},
\end{align}
with
\begin{align}
  \label{alpha_slab}
  &\alpha= \tfrac{1}{3}\tau\mean{(\bfDel\cro\bfb)\cdot\bfb} 
     -\tfrac{4}{5}\Omega\tau^2\frac{\del}{\del z}\left(u^2 -\tfrac{1}{3}b^2\right),\\
  \label{eta_slab}
  &\eta= \tfrac{1}{3}\tau u^2,\\
  \label{gamma_slab}
  &\gamma= -\tfrac{1}{6}\tau\frac{\del}{\del z}\left(u^2 -b^2\right),\\
  \label{deltakappa_slab}
  &\delta'= -\tfrac{2}{5}\Omega\tau^2b^2.
\end{align}
Strictly speaking $\bfu$ is the turbulent velocity of a presumed initial turbulent state, 
while there is no such restriction on $\bfb$ \citepalias{Brandenburg+Subramanian05a}.
Following \citet{Brandenburg+08a} we have used the notation $\delta'$ 
to represent the combined effects of the $\delta$ and $\kappa$ coefficients 
\citepalias[for details see][]{Chamandy+Singh17}.
Note that $\delta'$ is proportional to $\Omega$ and to $b^2$ 
and is not explicitly dependent on $\bfu$;
hence it is formally akin to the R\"{a}dler effect but with $u^2$ replaced by $b^2$.
It is convenient to define the ratio of the mean small-scale energies as
\begin{equation}
  \label{xi}
  \xi\equiv \frac{b^2}{u^2}.
\end{equation}

\subsection{Further simplification of the model}
\label{sec:simplification}
In order to isolate the magnetic R\"{a}dler effect we keep the dynamo model as simple as possible.
We take most parameters, including the correlation time scale $\tau$ and rms turbulent speed $u$, to be independent of $z$.  
We assume $\gamma=0$ and adopt a heuristic prescription to replace the second term on the right-hand-side of equation~\eqref{alpha_slab} 
that does not depend explicitly on the stratification of $u^2$ or $b^2$.
The first term on the right-hand-side of equation~\eqref{alpha_slab},
the mean current helicity term, is referred to as the magnetic $\alpha$ effect $\alpha\magn$,
discussed in Section~\ref{sec:xiVC},
while the second term is the kinetic $\alpha$ effect $\alpha\kin$.%
\footnote{This is standard nomenclature although $\alpha\kin$
now contains a term that depends on $b^2$, 
making the qualifier `kinetic' somewhat inappropriate, but we retain it for convenience.}
Therefore we can write
\begin{equation}
  \alpha= \alpha\kin +\alpha\magn.
\end{equation}
As is rather common in the literature, we adopt the functional form 
\begin{equation}
  \label{alpha_kin}
  \alpha\kin= \alpha\f \sin\left(\frac{\pi z}{h}\right),
\end{equation}
which satisfies the constraint that the mean kinetic helicity density must be an odd function of $z$.
The magnetic part $\alpha\magn$ is assumed to be generated by the large-scale dynamo,
and we solve for it self-consistently as described below.
Solutions are generally only weakly dependent on the precise functional form chosen for $\alpha\kin$.
In the same spirit, we also assume the simple form for the mean
velocity field $\meanv{U}= (0,r\Omega,0)$,
with $r$ and $\Omega$ parameters, so we neglect outflows for example.
All of these assumptions were also made in \citetalias{Chamandy+Singh17}, 
but there we considered the kinematic stage of large-scale dynamo action, so $\alpha\magn$ was negligible.

Non-linear effects become important as the mean field nears equipartition strength.
In our units with $\mu\f=\rho=1$, the equipartition field strength 
\begin{equation}
  \label{Beq}
  B\eq= u,
\end{equation}
and is constant with $z$ and $t$.
Making $B\eq$ $z-$dependent would have only a very small effect on dynamo solutions.

With these simplifications equation~\eqref{dynamo} reduces to
\begin{equation}
  \label{Br}
  \frac{\del B_r   }{\del t}=              -\frac{\del}{\del z}(\alpha B_\phi) 
                             +\frac{\del}{\del z}\left(\delta'\frac{\del B_\phi}{\del z}\right) +\eta\frac{\del^2B_r   }{\del z^2},
\end{equation}
\begin{equation}
  \label{Bp}
  \frac{\del B_\phi}{\del t}= -q\Omega B_r +\frac{\del}{\del z}(\alpha B_r   ) 
                             -\frac{\del}{\del z}\left(\delta'\frac{\del B_r}{\del z}\right) +\eta\frac{\del^2B_\phi}{\del z^2},
\end{equation}
where $q= -\del\ln\Omega/\del\ln r$ is the shear parameter,
and $q=1$ for a flat rotation curve.
Note that since $\bfDel\cdot\meanv{B}=0$ and we assume $h\ll r$, we have $\mbz\ll\mbr$, $\mbp$.
Not only will $\mbz$ be small, but it can be reconstructed from solutions using $\bfDel\cdot\meanv{B}$ \citep[e.g.][]{Chamandy16},
so we need not explicitly include $\mbz$ in our analysis.
We employ vacuum boundary conditions which for a thin galactic disc 
can be approximated as $\mbr=\mbp=0$ \citep{Ruzmaikin+88} at the disc surface $z=\pm h$.
For most runs, we employ 51 grid points in the region $-h\le z\le h$, which is more than needed for convergence,
but to present profiles with $z$ for certain runs, we increase this to 201 grid points.
Seed fields are of order $10^{-4}$ of the equipartition value, 
small enough that saturated solutions are completely insensitive to the seed field.

\section{A new non-linear effect within the dynamical $\alpha$-quenching framework}
\label{sec:xiVC}
As alluded to above,
an important constraint on the dynamo comes from the fact that the total magnetic helicity is approximately conserved in a system with $\Rm\gg1$.
This leads to a dynamical equation for $\alpha\magn$ \citep{Shukurov+06,Sur+07b},
\begin{equation}
  \label{dynamical_quenching}
  \frac{\del\alpha\magn}{\del t}= 
    -\frac{2\eta}{l^2}\left(\frac{\Emf\cdot\meanv{B}}{B\eq^2} +\frac{\alpha\magn}{\Rm}\right) -\bfDel\cdot\Flux,
\end{equation}
where $\Flux$ is a flux density of $\alpha\magn$ 
and henceforth the term involving $\Rm$ is neglected since it is expected to be negligible compared to the other terms.
The simplest flux term is the advective flux \citep{Subramanian+Brandenburg06,Shukurov+06}.
Since we are dealing with axisymmetric mean flows with only an azimuthal component
so that any systematic outflow is neglected, this flux vanishes.
A diffusive flux, given by
\begin{equation}
  \Flux\udiff= -\kappa\turb \bfDel\alpha\magn,
\end{equation}
with $\kappa\turb$ a turbulent diffusivity of the same order as 
the turbulent diffusivity $\eta$ of the mean magnetic field, 
might be expected on physical grounds \citep{Kleeorin+00,Kleeorin+02} and has been found to exist in numerical simulations \citep{Mitra+10}.

Intriguingly, we shall see that including $\delta'$-dependent terms in $\Emf$ leads to new contributions in the $\Emf\cdot\meanv{B}$ term
in equation~\eqref{dynamical_quenching} that have the same form as a flux term 
known as the Vishniac-Cho (VC) flux \citep{Vishniac+Cho01}, and thus the two effects can effectively be combined.

\subsection{Notes on the generalized Vishniac-Cho flux}
\label{sec:VC}
Let us consider a version of the VC flux that is generalized from that of \citet{Vishniac+Cho01}.
This generalized version is derived in \citet{Subramanian+Brandenburg04,Brandenburg+Subramanian05c,Subramanian+Brandenburg06}.
When $\bfDel\cdot\meanv{U}=0$, which is satisfied in our model (where outflows are absent),
the generalized VC flux of $\alpha\magn$ has the form \citep{Brandenburg+Subramanian05c}
\begin{equation}
  \label{generalized_VC}
   \flux\uVC_i= \frac{1}{3}\tau C\VC\eps_{ijl}\Strouhal^2\mathsf{S}_{lk}B_jB_k,
\end{equation}
where we define the Strouhal number as $\Strouhal\equiv \tau u/l$,%
\footnote{We have chosen to redefine $C\VC$ by factoring out $\Strouhal^2$ 
from the definition \citep[c.f.~Appendix~A of][]{Brandenburg+Subramanian05c,Sur+07b}.}
and where we have multiplied the expression
for the flux of mean small-scale current helicity in \citet{Brandenburg+Subramanian05c} 
by $\tau/3$ since $\alpha\magn= (1/3)\tau\mean{(\bfDel\cro\bfb)\cdot\bfb}$ in our units.
Here $C\VC$ is a dimensionless coefficient of order unity, and
\begin{equation}
  \mathsf{S}_{lk}= \tfrac{1}{2}(U_{l,k}+U_{k,l})
\end{equation}
is the mean rate of strain tensor, with comma denoting the spatial derivatives.
For the velocity field we have assumed, the components are
\begin{align}
  \flux\uVC_r   &= -\tfrac{1}{6}\tau C\VC \Strouhal^2(1-q)\Omega\mbr\mbz,                  \\
  \flux\uVC_\phi&=  \tfrac{1}{6}\tau C\VC \Strouhal^2(1-q)\Omega\mbp\mbz,                  \\
  \flux\uVC_z   &=  \tfrac{1}{6}\tau C\VC \Strouhal^2(1-q)\Omega\left(\mbr^2-\mbp^2\right).
\end{align}
Note that the dominant component is $\flux\uVC_z$, since $\mbz$ is small,
but that $\Flux\uVC=0$ for a strictly flat rotation curve.
The expression for $\flux\uVC_z$ is consistent with the expression used in \citet{Sur+07b},
if only the part due to shear is included (so that $1-q\rightarrow -q$) and $C\VC$ is set equal to unity.%
\footnote{Their choice $C\VC=1$ is based on results of \citet{Subramanian+Brandenburg06}.
However, we find that there is a typographical error in equation~(11a) of the published version of that paper, 
not present in the arXiv version (astro-ph/0509392v2), and that correcting this error leads to the estimate $C\VC=2$.}
Under the slab approximation, and neglecting terms involving $\mbz$ (Appendix~\ref{sec:xiVC_general}),
the flux enters the right-hand-side of equation~\eqref{dynamical_quenching} as
\begin{equation}
  \label{VCterm}
  -\bfDel\cdot\Flux\uVC= \frac{C\VC\Strouhal^2\eta}{B\eq^2}(q-1)\Omega\left(\mbr\frac{\del\mbr}{\del z} -\mbp\frac{\del\mbp}{\del z}\right),
\end{equation}
where we have made use of equations~\eqref{eta_slab} and \eqref{Beq}.

\subsection{Non-linear magnetic R\"{a}dler term}
\label{mr_nonlinear}
As for the term from the magnetic R\"{a}dler effect that stems 
from the $\Emf\cdot\meanv{B}$ term in equation~\eqref{dynamical_quenching}, 
this can be written down with the help of equations~\eqref{Emfr_slab} and \eqref{Emfp_slab}.
The relevant contributions involving $\delta'$ are
\begin{equation}
  \label{EdotB_mr}
  \frac{\del\alpha\magn}{\del t} 
    = \ldots +\frac{2\eta\delta'}{l^2B\eq^2}\left(\mbr\frac{\del\mbr}{\del z} +\mbp\frac{\del\mbp}{\del z}\right),
\end{equation}
where dots represent other terms. 
Using equations~\eqref{deltakappa_slab} and \eqref{xi}, equation~\eqref{EdotB_mr} can be written as
\begin{equation}
  \label{deltakappaterm}
  \frac{\del\alpha\magn}{\del t} 
    = \ldots -\frac{4}{5}\frac{\Strouhal^2\eta\Omega\xi}{B\eq^2}\left(\mbr\frac{\del\mbr}{\del z} +\mbp\frac{\del\mbp}{\del z}\right).
\end{equation}

\subsection{Combining the VC and magnetic R\"{a}dler terms}
The terms in equations~\eqref{VCterm} and \eqref{deltakappaterm} have the same form.
They can be combined to give 
\begin{equation}
  \label{combined_slab}
  \begin{split}
  \frac{\del\alpha\magn}{\del t} 
    = &\ldots +\frac{\Strouhal^2\eta\Omega}{B\eq^2}
    \left[\left(C\VC(q -1) -\frac{4}{5}\xi\right)\mbr\frac{\del\mbr}{\del z}\right.\\
      & \left.-\left(C\VC(q -1) +\frac{4}{5}\xi\right)\mbp\frac{\del\mbp}{\del z}\right].
  \end{split}
\end{equation}
In Appendix~\ref{sec:xiVC_general} we show that even in the general case where $r$- and $\phi$- spatial derivatives are retained, 
this equivalence in form persists, but equation~\eqref{combined_slab} contains the dominant terms for a thin disc geometry.
The correspondence in form between the generalized VC flux term and the $\delta'$-dependent part of the $\Emf\cdot\meanv{B}$ term 
is remarkable and probably hints at a deeper relationship between the two effects that is not yet understood.

Of the two terms in equation~\eqref{combined_slab}, the term proportional to $\mbp\del\mbp/\del z$ will typically dominate,
because $|\mbp|$ is usually a few times larger than $|\mbr|$, and both have comparable scale heights $\sim h$.
In that case, the new magnetic R\"{a}dler term acts like a shear-induced VC effect since it has the same sign as $q$
(whereas the term proportional to $\mbr\del\mbr/\del r$ acts like a rotation-induced VC effect, since it has opposite sign to $q$).
Thus, we expect that the shear-induced VC flux will effectively be enhanced by the magnetic R\"{a}dler effect.
The fact that VC terms are proportional to $q-1$, which can be small in galactic discs, 
means that $\xi$-dependent terms can be important even for $\xi$ as low as $\sim0.1$.
On the other hand, it also hints at the possibility that had shear been included
in the derivation of the magnetic R\"{a}dler terms, a term of the same form but proportional to $q$ (and perhaps of opposite sign)
might have been obtained.
A derivation like that in \citetalias{Radler+03} or \citetalias{Brandenburg+Subramanian05a}, 
but with \textit{differential} rotation included needs to be carried out to elucidate the general and most realistic case,
but this is beyond the scope of this work.

Putting it all together, the dynamical quenching equation~\eqref{dynamical_quenching} becomes
\begin{equation}
  \label{dynamical_quenching_modified}
  \begin{split}
  \frac{\del\alpha\magn}{\del t} 
    = & -\frac{\Strouhal^2\eta}{B\eq^2}\Bigg\{ \frac{2}{3\tau}\left[\frac{\alpha}{\eta}(\mbr^2+\mbp^2) 
        -\mbp\frac{\del\mbr}{\del z} +\mbr\frac{\del\mbp}{\del z}\right]\\
      & +\Omega\left[\left(C\VC(q-1) +\frac{4}{5}\xi\right)\mbp\frac{\del\mbp}{\del z}\right.\\
      & \left. -\left(C\VC(q-1) -\frac{4}{5}\xi\right)\mbr\frac{\del\mbr}{\del z}\right]\Bigg\}
                     +R_\kappa\eta\frac{\del^2\alpha\magn}{\del z^2},
  \end{split}
\end{equation}
where $R_\kappa\equiv \kappa\turb/\eta$, and we have assumed $\kappa\turb$ to be independent of $z$.

\subsection{Algebraic $\alpha$-quenching provides a useful comparison}
It was shown in \citetalias{Chamandy+14b} that solutions that instead employ the oft-used heuristic non-linearity
\begin{equation}
  \label{algebraic}
  \alpha= \frac{ \alpha\kin}{1+a\f B^2/B\eq^2},
\end{equation}
known as the algebraic $\alpha$ quenching formalism,
approximate well solutions with dynamical quenching with diffusive and/or advective fluxes of $\alpha\magn$
(but the comparison did not consider the VC flux or other effects such as the magnetic R\"{a}dler effect).
Moreover, it was found that the coefficient $a\f$ can be calibrated to make the correspondence between
the two prescriptions even closer, but below we adopt $a\f=1$ for simplicity.
Below we study solutions that invoke equation~\eqref{algebraic} for
comparison with the solutions that
are based on dynamical quenching, to help to isolate the effects of the new terms.


\begin{table*}
  \begin{center}
  \caption{Key parameters for numerical examples.
           For all models, $q=1$, $u=10\kms$, and $\Strouhal=1$.
           For models with dynamical $\alpha$-quenching, the parameter $R_\kappa=1$ unless indicated in the text.
           \label{tab:models}
          }
  \begin{tabular}{@{}lcccccccc@{}}
    \hline
    Model  &$\tau\;[\!\Myr]$ &$h\;[\!\kpc]$ &$\Omega\;[\!\kmskpc]$ &$R_\Omega$ &$R_\alpha$ &$q^{1/2}h/(\tau u)$ &$\Omega\tau$ &$D$     \\
    \hline                   
    A      &$10$             &$0.4$         &$30$                  &$-14.1$    &$0.92$     &$3.91$              &$0.31$       &$-13.0$ \\
    B      &$5$              &$0.2$         &$90$                  &$-21.1$    &$1.38$     &$3.91$              &$0.46$       &$-29.2$ \\
    C      &$10$             &$0.2$         &$90$                  &$-10.6$    &$2.76$     &$1.96$              &$0.92$       &$-29.2$ \\
    \hline
  \end{tabular}
  \end{center}
\end{table*}

\section{Model parameters}
\label{sec:parameters}
As parameter values can be rather different between and within galaxies,
it is important to study the parameter space of solutions.
Therefore, we solve the equations with different sets of parameter values,
which we refer to as models, listed in Table~\ref{tab:models}.
The dimensionless parameters in the table are turbulent Reynolds numbers:
$R_\alpha\equiv \alpha\f h/\eta$ characterizes the strength of the $\alpha$ effect
and $R_\Omega\equiv -q\Omega h^3/\eta$ the $\Omega$ effect (differential rotation).
The parameter $R_\kappa = \kappa\turb/\eta$ is the ratio of the turbulent diffusivity of $\alpha\magn$ to that of $\meanv{B}$,
and $R_\kappa=1$ in most cases, but we also explore $R_\kappa=0$ and $R_\kappa=0.3$ for some runs.
Also, the Strouhal number is $\Strouhal=1$, but we keep $R_\kappa$ and $\Strouhal$ as parameters in the analytic expressions.
Under the $\alpha\Omega$ approximation, and for the case where vertical and radial mean velocities 
can be neglected, the dynamo number $D=R_\alpha R_\Omega$ and $\xi$ are the control parameters of the kinematic problem.
The parameter $\xi$ is not listed in Table~\ref{tab:models} because an aim of this paper is to explore the dependence
of the solutions on $\xi$.

Models~A and B have sets of underlying galaxy parameters that are the same
as the kinematic models in \citetalias{Chamandy+Singh17},
except now there is the additional parameter $R_\kappa$ relevant for the non-linear regime 
(except for cases where equation~\eqref{algebraic} is invoked in place
of equation~\eqref{dynamical_quenching_modified}).
We also briefly consider the case $R_\kappa=0$, i.e.~no diffusive flux of small-scale magnetic helicity density.
Models~A and B can loosely be thought of as having conditions similar to galactocentric radii 
of $8\kpc$ and $2\kpc$ in the Milky Way, respectively.
To obtain quantities such as time in physical units,
each set of dimensionless parameters is obtained by setting the rms turbulent speed $u=10\kms$,
the shear parameter $q=-d\ln\Omega/d\ln r$, 
the turbulence correlation time $\tau$, 
the disc scale height $h$,
and the angular rotation rate $\Omega$ to the values listed in Table~\ref{tab:models}.
In addition we introduce Model~C, which is similar to Model~B, but with $\tau=10\Myr$,
resulting in a twice larger Coriolis number $\Coriolis=\Omega\tau=0.96$.
This value only marginally satisfies the assumption $\Omega\tau\ll1$
made in the original derivation of the magnetic R\"{a}dler effect \citepalias{Radler+03,Brandenburg+Subramanian05a}
but it is still worth exploring as a limiting case.

\section{Results}
\label{sec:results}
In this section we present solutions to the system of coupled equations~\eqref{Br}, \eqref{Bp} and 
\eqref{dynamical_quenching_modified} or \eqref{algebraic}.

\subsection{Effect on the mean magnetic field strength}
\label{sec:saturation}

\begin{figure*}
  \includegraphics[width=58mm,clip=true,trim=50 55  35 0]{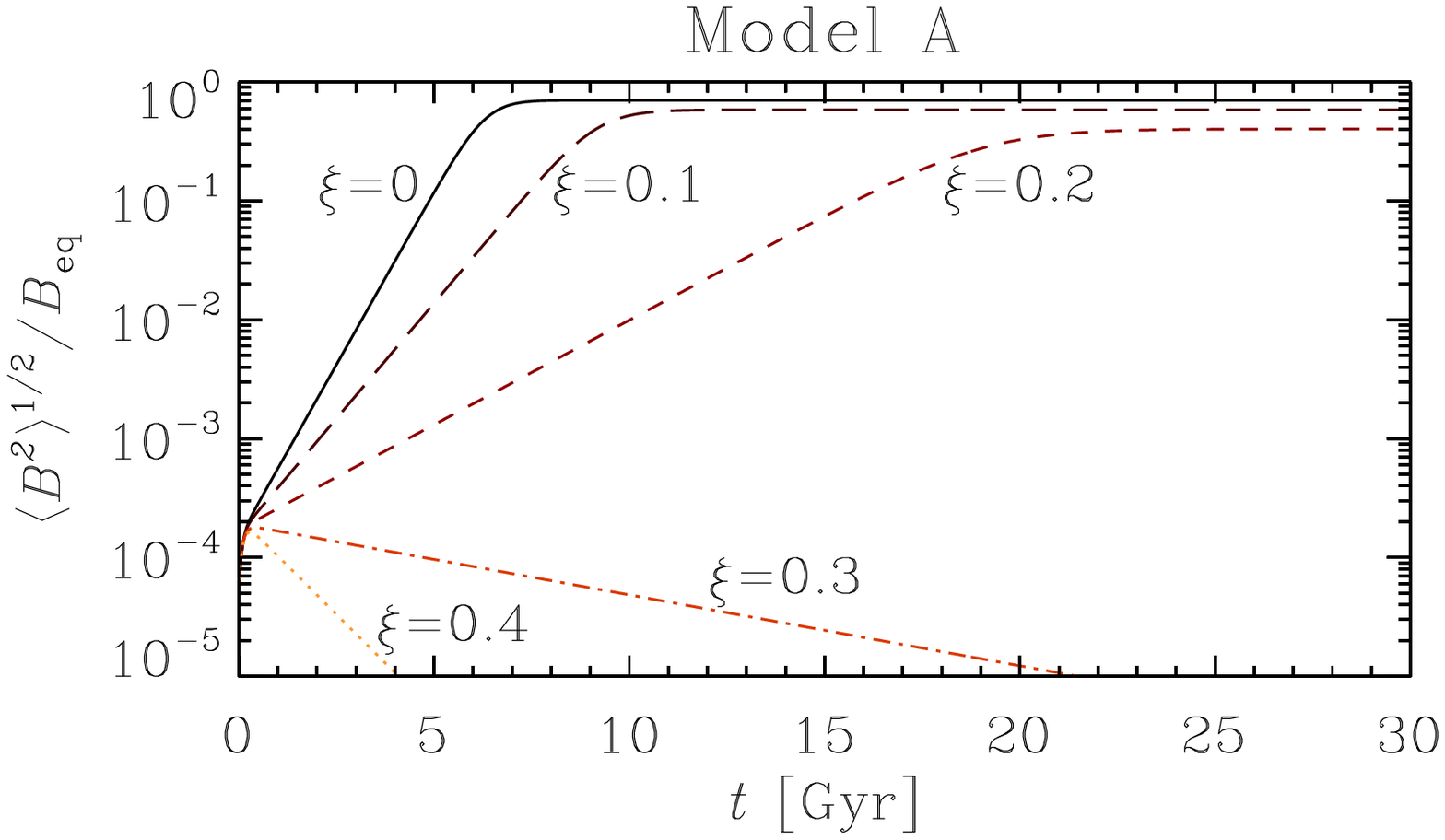}
  \includegraphics[width=58mm,clip=true,trim=50 55  35 0]{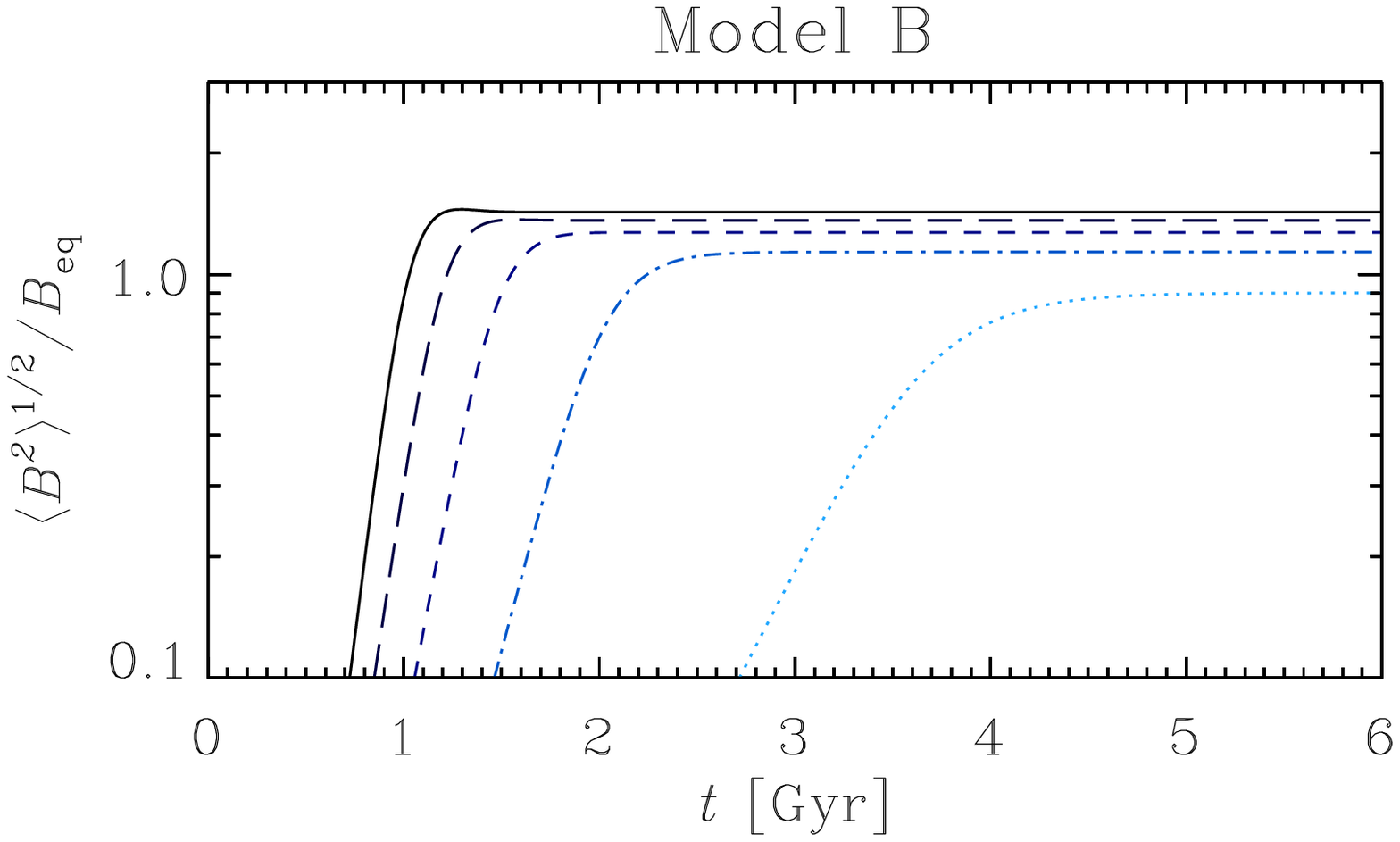}
  \includegraphics[width=58mm,clip=true,trim=50 55  35 0]{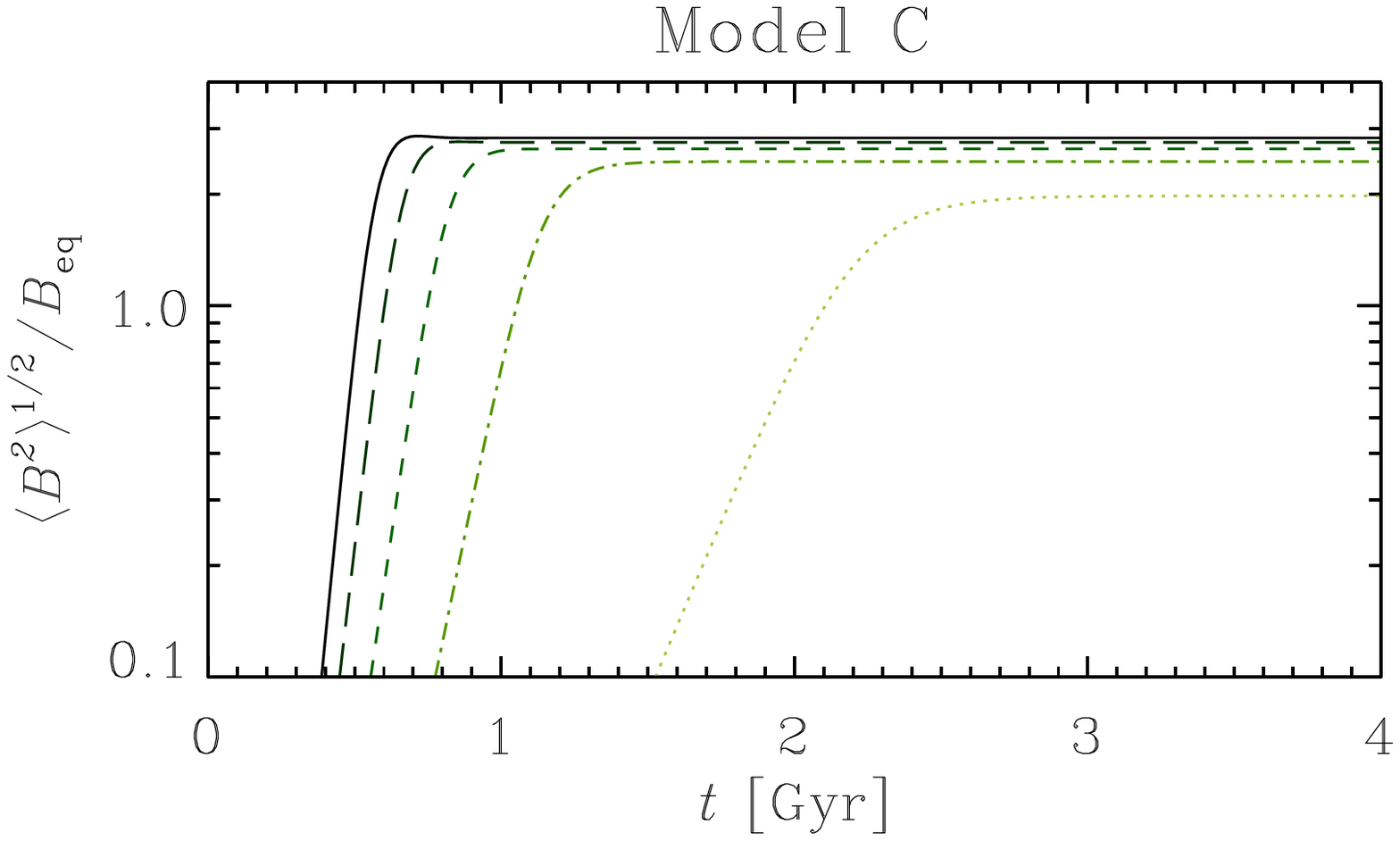}\\
  \includegraphics[width=58mm,clip=true,trim=50 55  35 0]{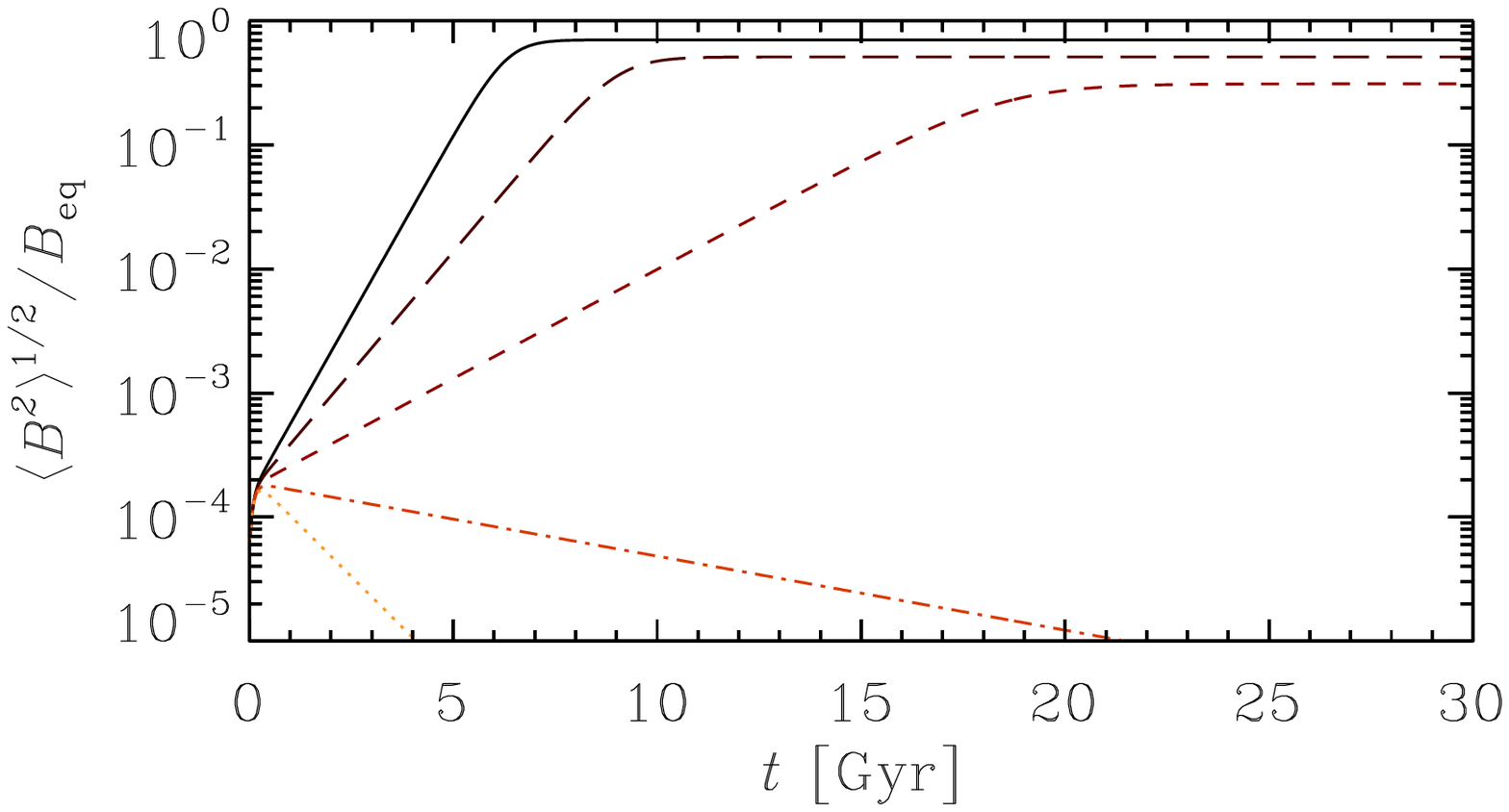}
  \includegraphics[width=58mm,clip=true,trim=50 55  35 0]{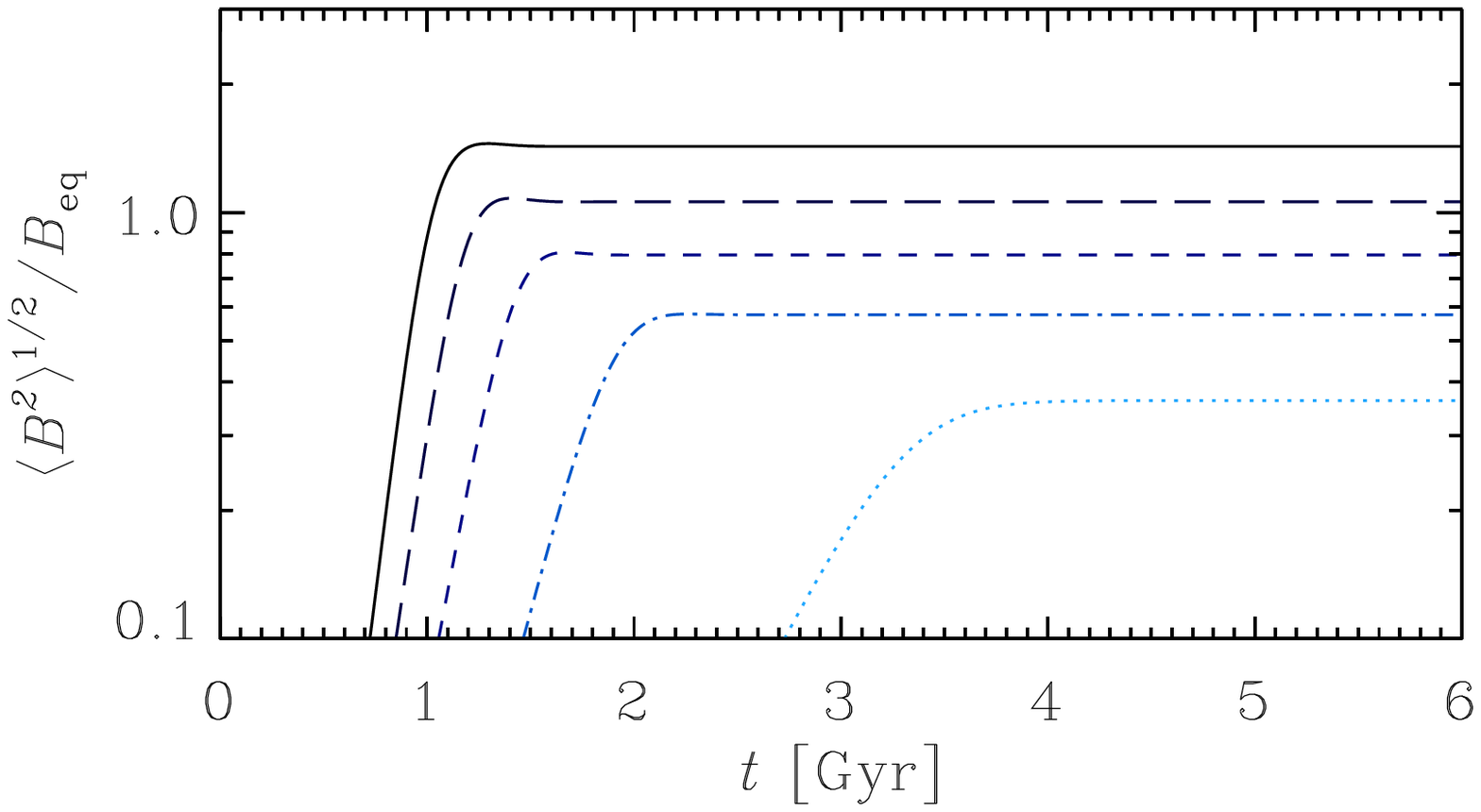}
  \includegraphics[width=58mm,clip=true,trim=50 55  35 0]{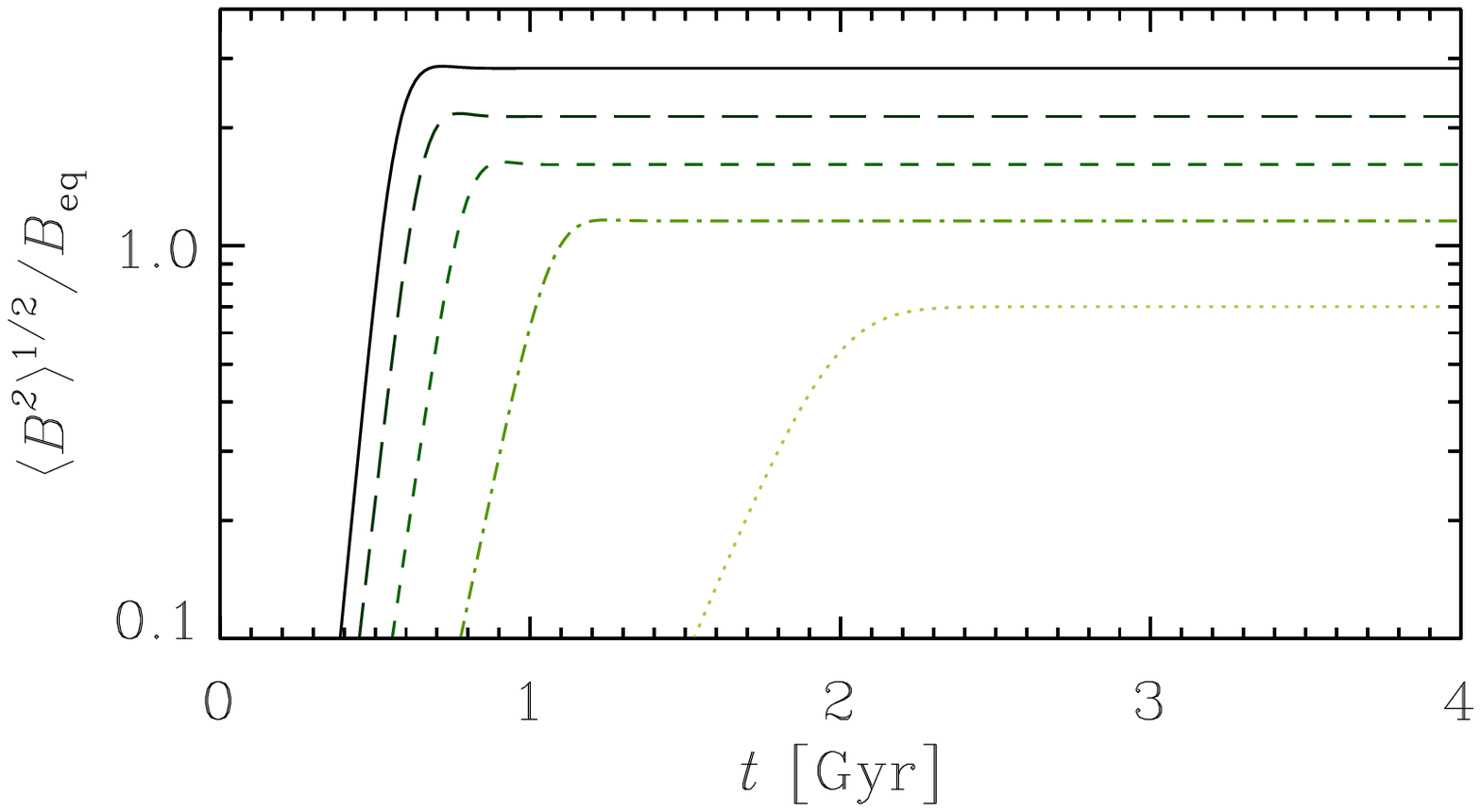}\\
  \includegraphics[width=58mm,clip=true,trim=50 0  35 0]{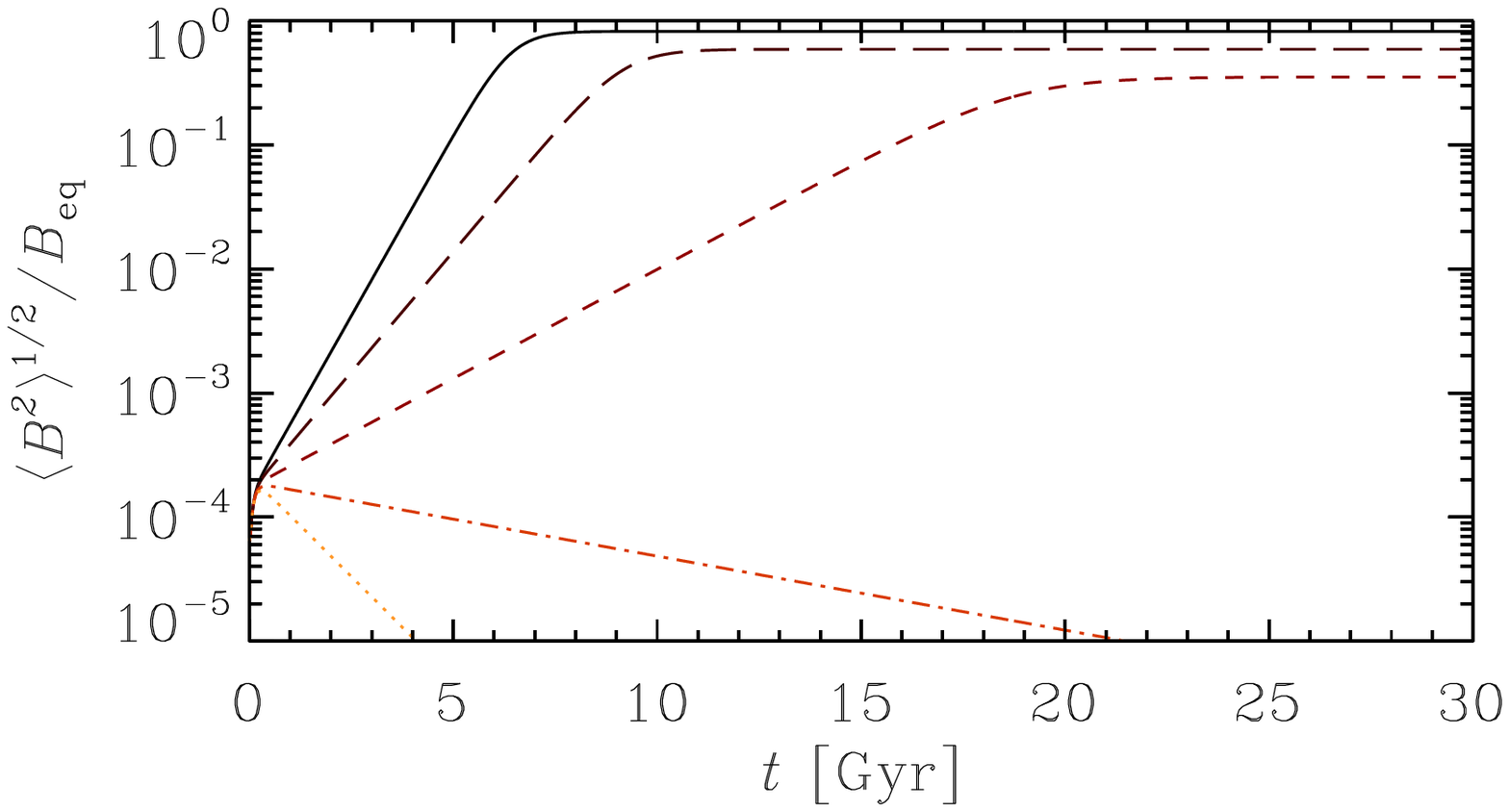}
  \includegraphics[width=58mm,clip=true,trim=50 0  35 0]{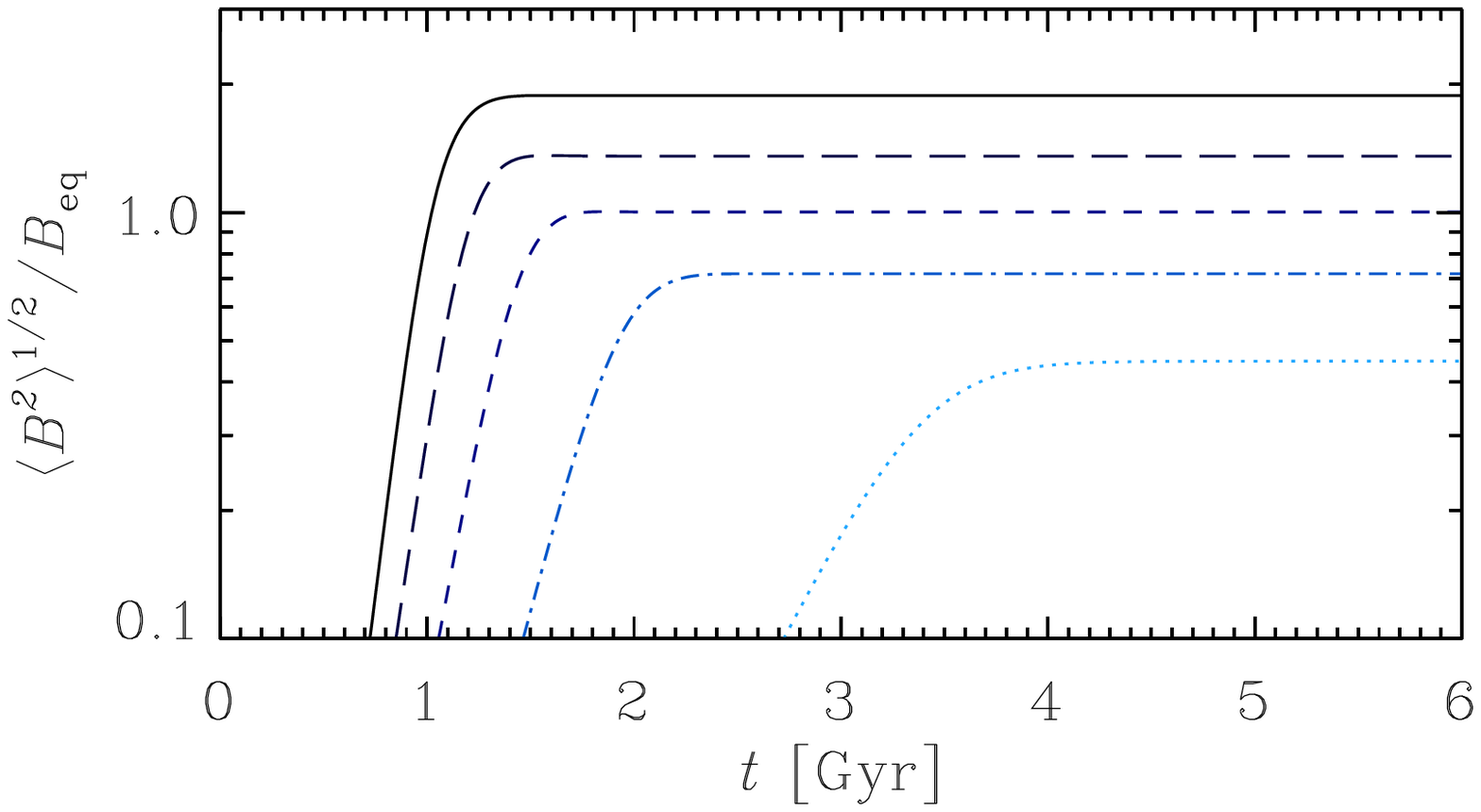}
  \includegraphics[width=58mm,clip=true,trim=50 0  35 0]{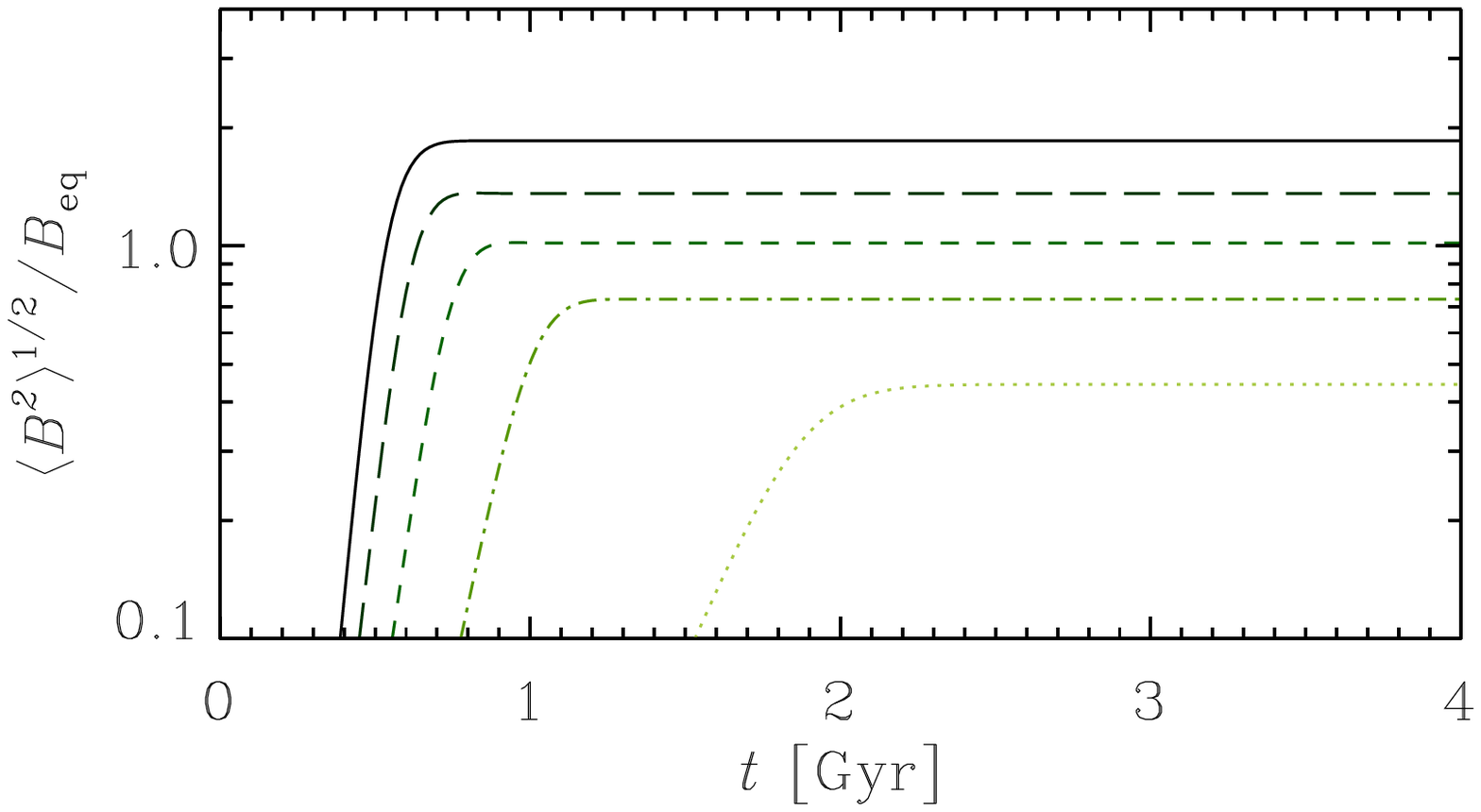}\\
  \caption{Mean magnetic field strength evolution 
           for Model~A (leftmost column),
           Model~B (middle column)
           and Model~C (rightmost column).
           For each model five different curves are shown corresponding to different values of $\xi$:
           $\xi=0$   (solid);
           $\xi=0.1$ (long dashed);
           $\xi=0.2$ (short dashed);
           $\xi=0.3$ (dashed-dotted);
           $\xi=0.4$ (dotted).
           Local growth times are not indicative of actual global growth times, 
           but local saturated solutions are known to approximate well global solutions locally.
           Note that in Model~A, the dynamo is subcritical for $\xi=0.3$ and $0.4$,
           which explains the exponential decay of the field in the kinematic regime of $\meanv{B}$.
           Top row: Full dynamical $\alpha$ quenching with $\xi$ terms in equation \eqref{deltakappaterm}.
           Middle row: Dynamical $\alpha$ quenching, now omitting $\xi$ terms.
           Bottom row: Algebraic $\alpha$ quenching.
           \label{fig:Btime_log_ss}
          }            
\end{figure*}

\begin{figure*}
  \includegraphics[width=58mm,clip=true,trim=50 0  35 0]{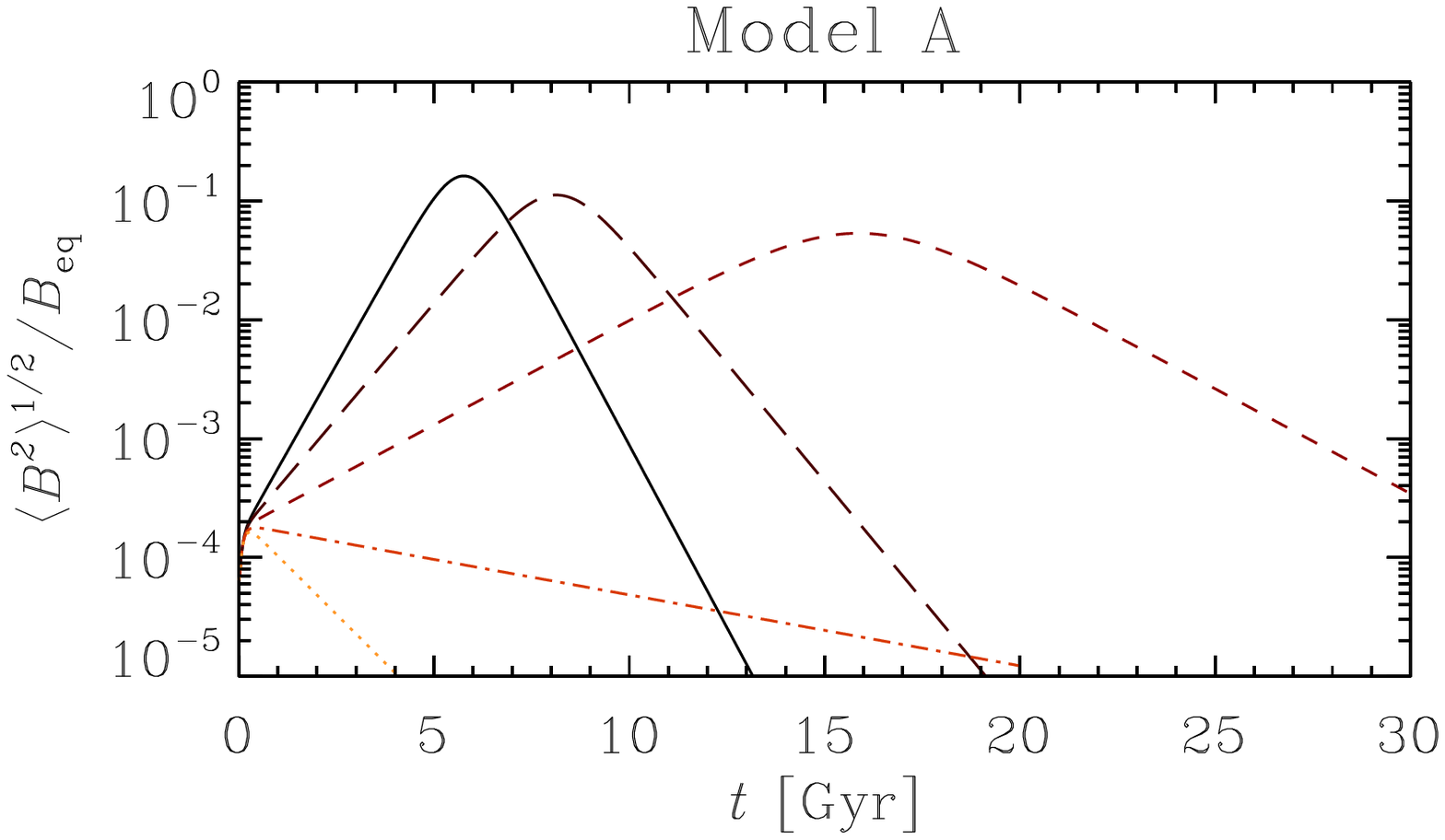}
  \includegraphics[width=58mm,clip=true,trim=50 0  35 0]{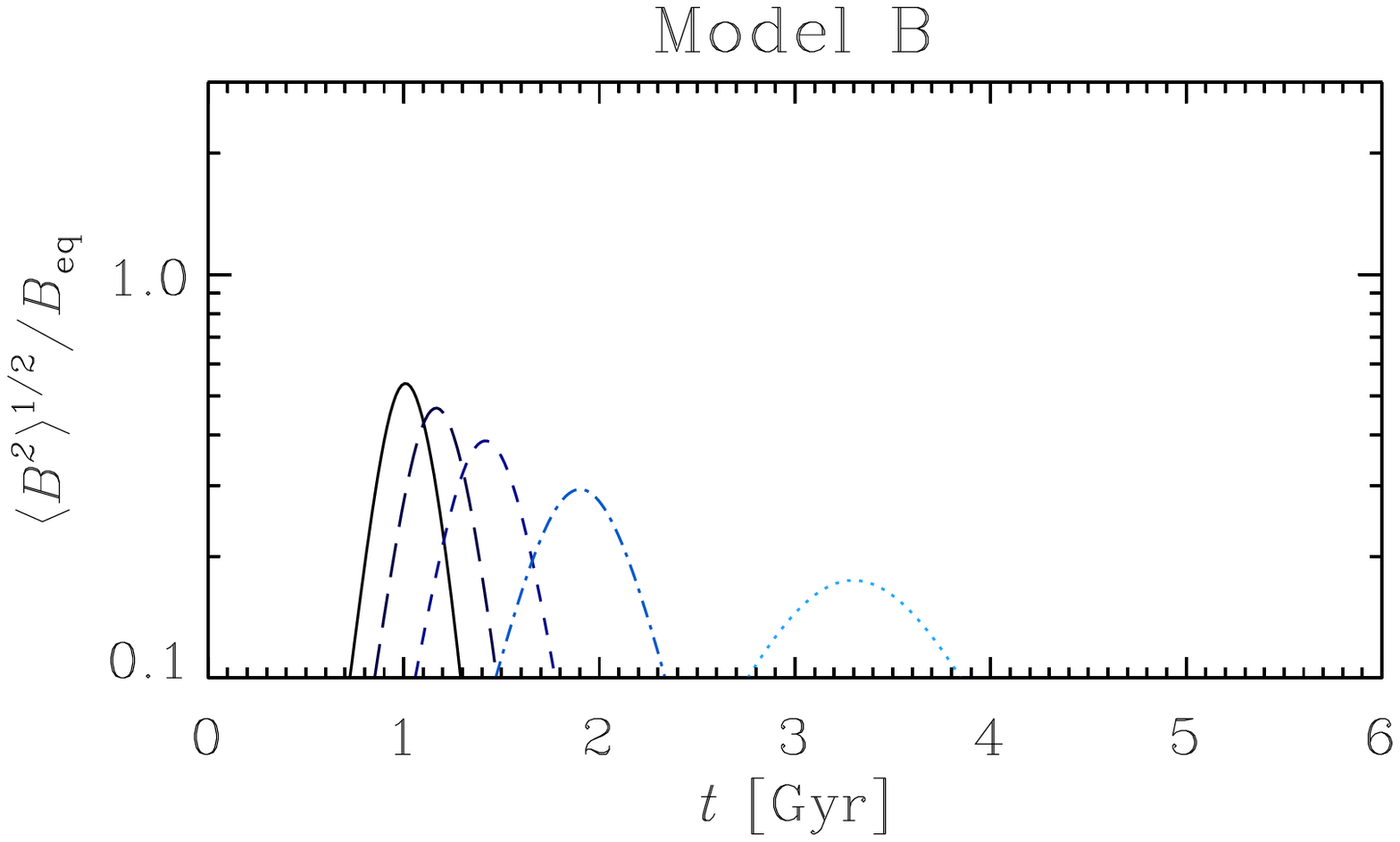}
  \includegraphics[width=58mm,clip=true,trim=50 0  35 0]{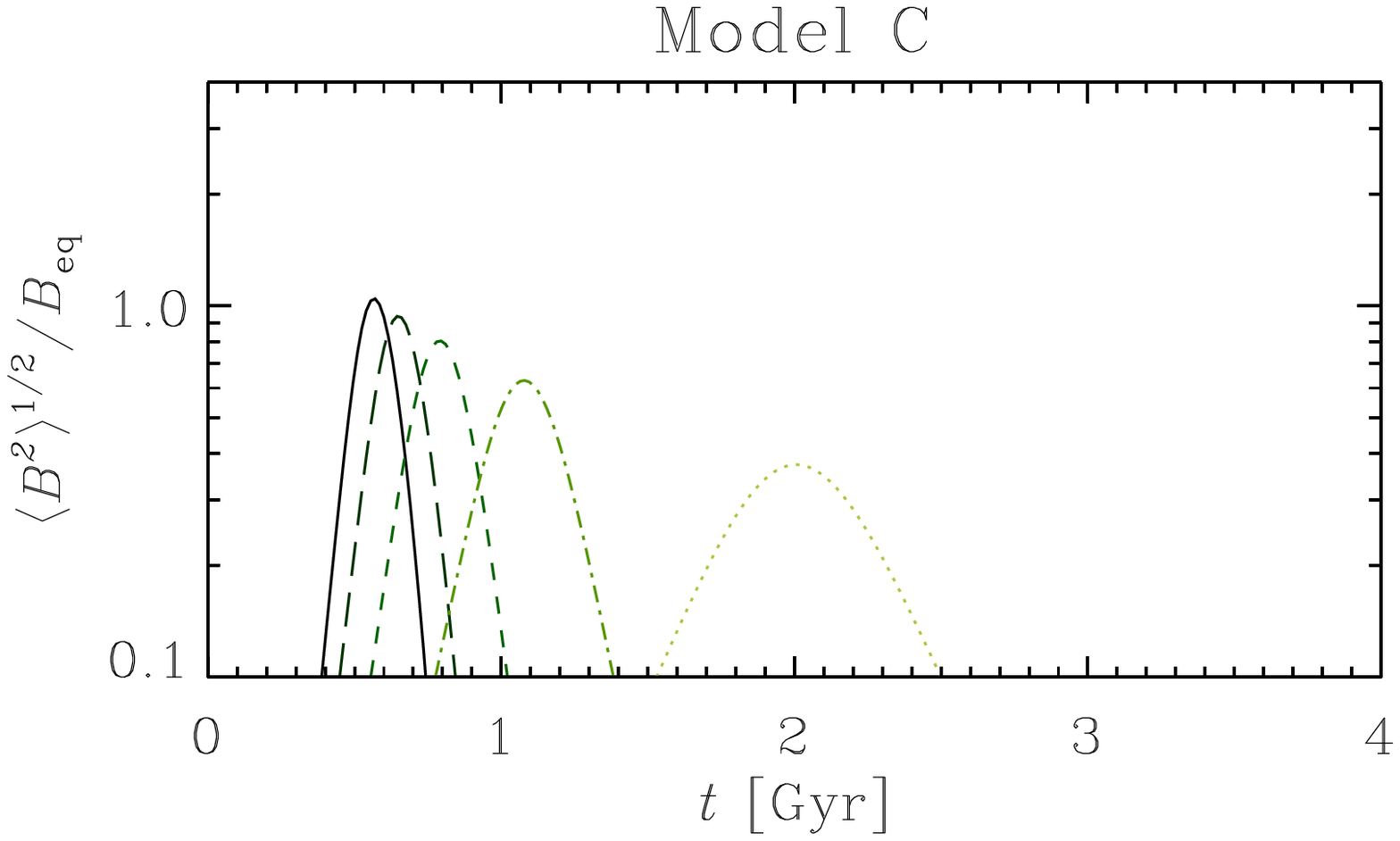}
  \caption{As top row of Figure~\ref{fig:Btime_log_ss} but with $R_\kappa=0$ instead of $1$,
           which results in `catastrophic quenching' of the field.
           \label{fig:Btime_Rkappa0_log_ss}
          }
\end{figure*}

In Figure~\ref{fig:Btime_log_ss} we show the evolution of the normalized
rms strength of the mean magnetic field, $\sqrt{\langle B^2\rangle}/B\eq$,
through to the saturated (steady) state, with $B\eq=\const$ being
the equipartition strength and the angular brackets denote an average
across the disc $-h\leq z\leq h$.
(Plots of the field strength at the midplane $B(0)$ are not shown but appear almost identical except for a different normalization.)
From left to right columns depict results from Models~A, B and C, while rows show three cases for each model.
From top to bottom, we have the case where $\xi$-terms are included in equation~\eqref{dynamical_quenching_modified},
where $\xi$-terms are not so included,
and where the equation for $\del\alpha\magn/\del t$ is replaced by the simpler algebraic $\alpha$-quenching non-linearity \eqref{algebraic}.

For all models, increasing $\xi$ leads to decreasing kinematic growth rates as discussed in \citetalias{Chamandy+Singh17},
and smaller saturated field strengths.
The latter is consistent with the former: the dynamo weakens with increasing $\xi$. 
The effect of a finite $\xi$ on the value of $B$ in the saturated state can be estimated analytically, 
as shown in Appendix~\ref{sec:analytic}.
As discussed below, these analytic estimates correctly predict the qualitative response of the saturated field strength to changes in $\xi$.

In Model~A, shown in the leftmost column, dynamical quenching leads to very similar results whether or not $\xi$ terms 
are kept (leftmost column, top row) or not kept (leftmost column, middle row) in the equation for $\del\alpha\magn/\del t$.
Algebraic quenching (leftmost column, bottom row) also gives very similar results for this model.
For the cases $\xi=0.3$ and $\xi=0.4$, the dynamo is \textit{subcritical}, 
and there is exponential \textit{decay} \citepalias[c.f.][]{Chamandy+Singh17},
whilst for the other cases there is growth and saturation.
The saturated field strengths are very similar between the runs with different quenching formalisms.
The saturation times are longer than a Hubble time but this is not really relevant.
The field at every radius will in reality grow together at the \textit{global} growth rate,
which is similar to (slightly smaller than) the largest \textit{local} growth rate in the galaxy, 
until the non-linear regime, at which time the local conditions determine the evolution
\citep{Ruzmaikin+88,Beck+94,Chamandy+13a,Chamandy16}.

Results of Model~B, shown in the middle column, are in some ways similar to those of Model~A, 
but there is a much more significant difference between the case where $\xi$ terms 
are included in the non-linearity (middle column, top row), and the two cases where they are not included
(middle column, middle and bottom rows).
Finally in Model~C, we see a scenario similar to what is seen for Model~B.
We also note that for Model~C, algebraic quenching (rightmost column, bottom row) gives smaller saturation strengths 
compared with the case of dynamical quenching that does not include $\xi$ terms in the equation for $\del\alpha\magn/\del t$ 
(rightmost column, middle row), 
but this is not surprising because the precise level of agreement between algebraic and dynamical quenching 
is known to depend sensitively on certain parameters \citepalias{Chamandy+14b}.

That the top and middle rows are very similar for Model~A but not for Models~B and C 
is qualitatively consistent with the analysis of Appendix~\ref{sec:analytic}.
There it is shown that the effect of the extra $\xi$-dependent term in $\Emf\cdot\meanv{B}$ is contained in the factor $1/\chi(p,\xi)$,
where $\chi(p,\xi)$ is given by equation~\eqref{chi}, and that the relevant term in $\chi(p,\xi)$ is $\propto D/D\crit$,
with $D\crit$ the critical dynamo number.
From Table~\ref{tab:models}, we see that $D$ is much larger for Models~B and C than for Model~A, 
while $D\crit$ is similar in all models.
Therefore, it is not surprising that this extra term (included in the top row only) makes a larger difference for Models~B and C.
As can be seen by comparing the field strength for Models~B and C in the top and middle rows of Figure~\ref{fig:Btime_log_ss},
this new VC flux-like term leads to a larger saturated field strength.
This behaviour is predicted qualitatively by the simple analysis of Appendix~\ref{sec:analytic}.
The panels of the top row suggest that the observed field strength should not depend very sensitively
on the ratio of small-scale magnetic to kinetic energy density $\xi$,
and varies by less than a factor of two between $\xi=0$ and $\xi=0.4$.
Since current observations of large-scale field strength in galaxies are uncertain to within a factor of a few,
dynamo models can be said to predict field strength to within the precision of observations 
even in cases where $\xi$ is not well-constrained.

Given that the new $\xi$-dependent part of the $\Emf\cdot\meanv{B}$ term is formally like a VC flux,
we might expect results to resemble those obtained when a VC flux is included with $\xi=0$.
\citet{Sur+07b} studied a model that included a strong VC flux (they effectively explored the case $C\VC(q-1)=1$).
They found that this term leads to a threshold effect, such that if the field becomes sufficiently strong,
the dynamo enters a new exponential growth phase. 
This regime is not terminated naturally within the dynamical $\alpha$-quenching paradigm
(but may be subject to some other unknown quenching mechanism).
Our code reproduces qualitatively their results using the same expression for the VC flux,
but the solution becomes numerically unstable.
If we turn off the VC flux and increase $\xi$ we eventually transition from growing and saturating solutions 
to solutions that decay in the kinematic regime.
However, if we introduce \textit{artificially} a factor in front of the new $\xi$-dependent part of the $\Emf\cdot\meanv{B}$ term,
we find that as this factor is increased from unity, 
the time scale for the solution to saturate in the non-linear regime becomes larger,
and that for large enough values of this artificial factor ($\sim3$ for Model~B parameter values and $\xi=0.2$) 
we find qualitatively similar behaviour to the VC flux case, as would be expected.
Similarly, if we reduce artificially the VC flux term (with $q-1$ replaced with unity to enable comparison with \citealt{Sur+07b})
by setting $C\VC$ to be less than unity, we obtain saturated solutions (for $C\VC\lesssim0.1$ for Model~B parameter values).
Such a reduction could also be achieved by including the full form 
of the term which is ~$\propto q-1$ (see equation~\eqref{dynamical_quenching_modified}) and assuming $q-1$ to be small.

We also consider a case similar to that presented in the top row of Figure~\ref{fig:Btime_log_ss},
that is with full dynamical quenching, but with $R_\kappa=0$ instead of $1$, 
so that the diffusive flux of mean small-scale helicity density vanishes.
We show this case in Figure~\ref{fig:Btime_Rkappa0_log_ss}.
The behaviour is qualitatively the same for each of the models A, B and C.
As expected, in the absence of small-scale helicity fluxes, the field is catastrophically quenched \citepalias{Brandenburg+Subramanian05a}
for the case $\xi$=0 (solid lines).
This remains true when $\xi$ is made to be finite, but like the growth rate, the decay rate reduces with $\xi$.
The magnetic R\"{a}dler effect prolongs the survival of the field in the catastrophic quenching regime 
and increases the time at which the field peaks, but the peak value reached by the field decreases with $\xi$.
Note that the growth and decay rates are comparable for each of the curves.
This means that for Model~A with $\xi=0$, for example, the field has a smaller decay rate and peak value than for Model~B with $\xi=0$.
We see then that reducing the growth rate, whether by changing the galaxy parameter values in going from one model to another,
or by increasing $\xi$, leads to qualitatively the same effects in the catastrophic quenching case.
Put succinctly, increasing $\xi$ weakens the dynamo, which results in a `flatter' temporal evolution of the field strength.

\subsection{Effect on the pitch angle of the mean magnetic field}
\label{sec:pitch}

\begin{figure*}
  \includegraphics[width=58mm,clip=true,trim=50 55  35 0]{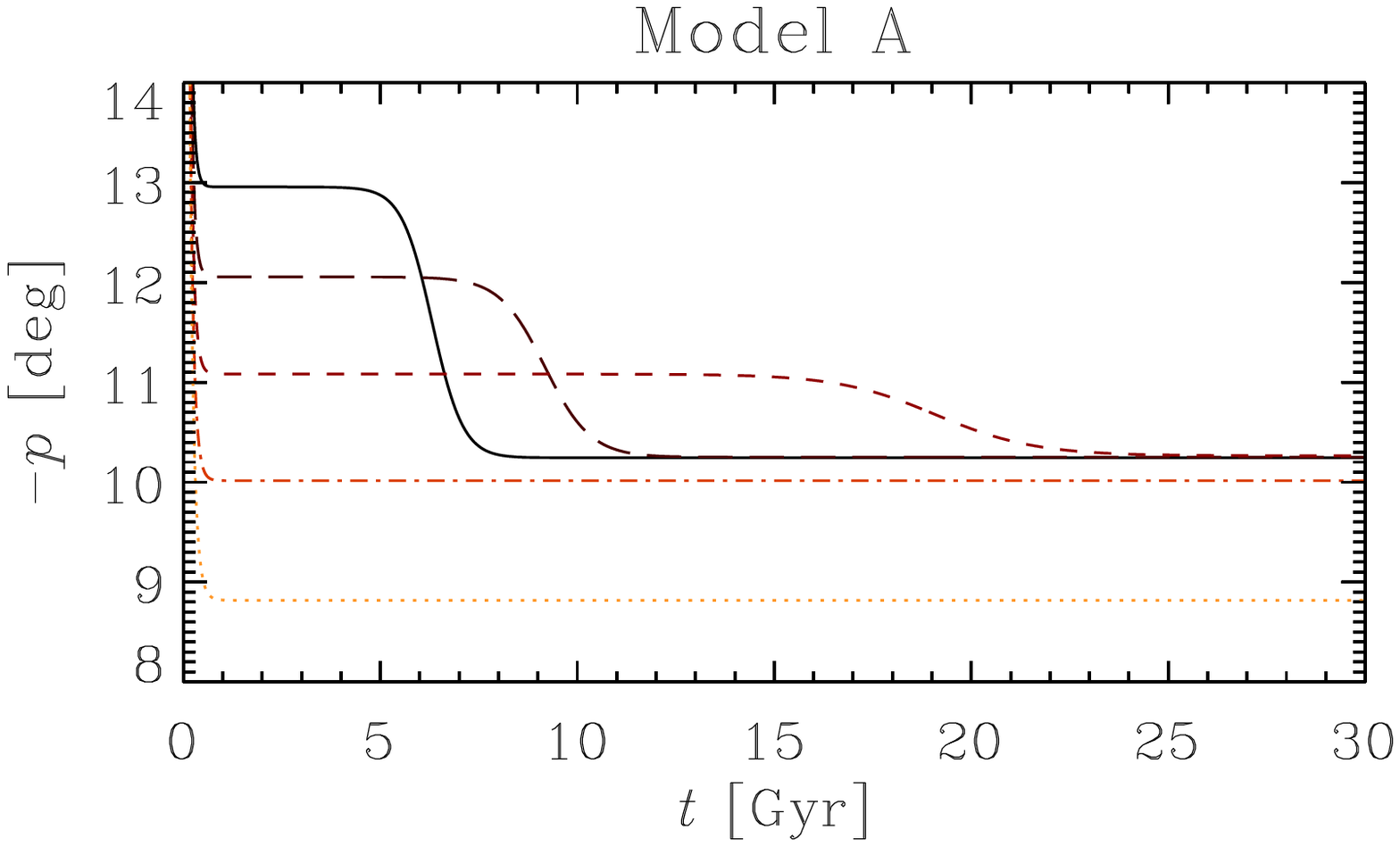}
  \includegraphics[width=58mm,clip=true,trim=50 55  35 0]{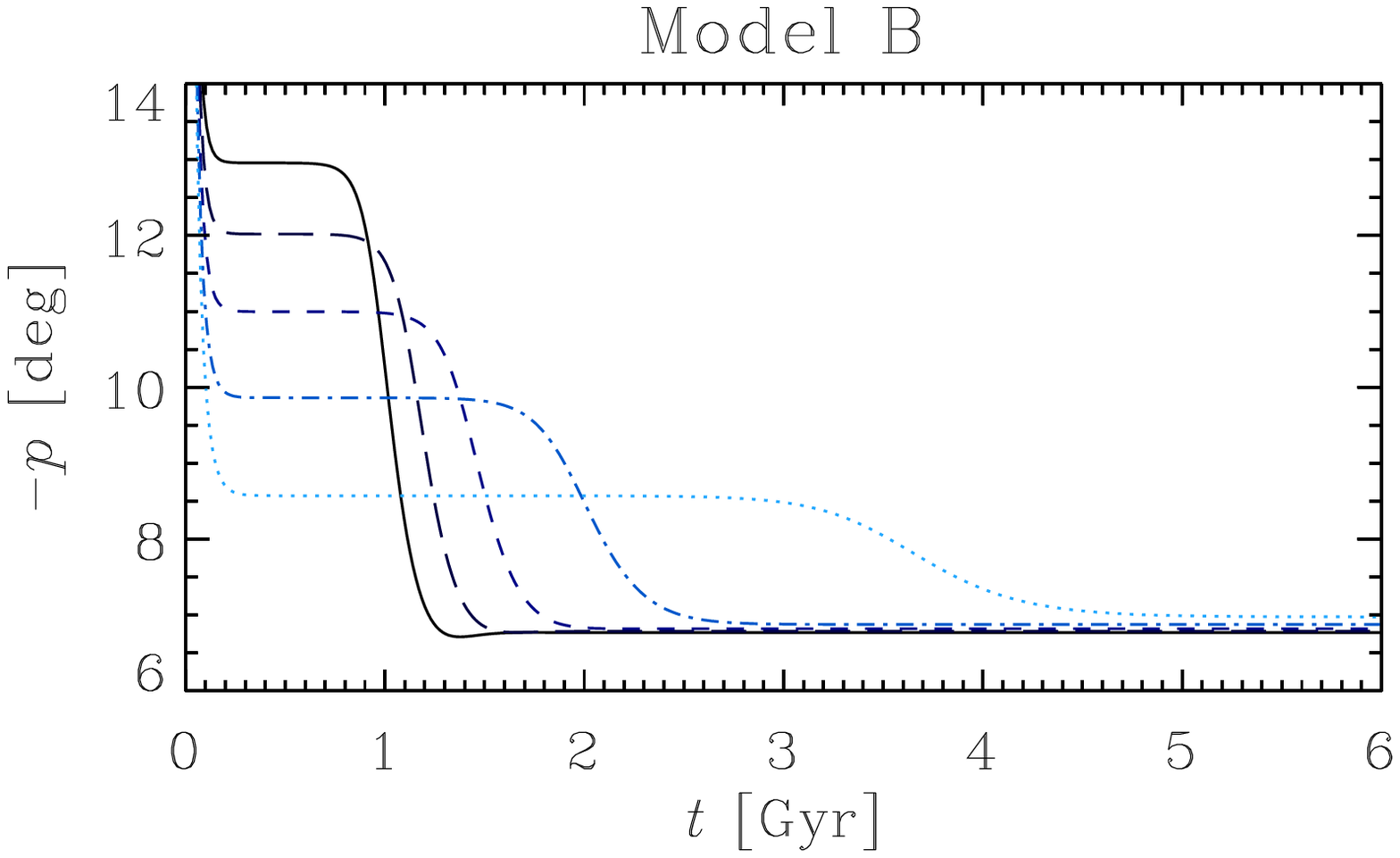}
  \includegraphics[width=58mm,clip=true,trim=50 55  35 0]{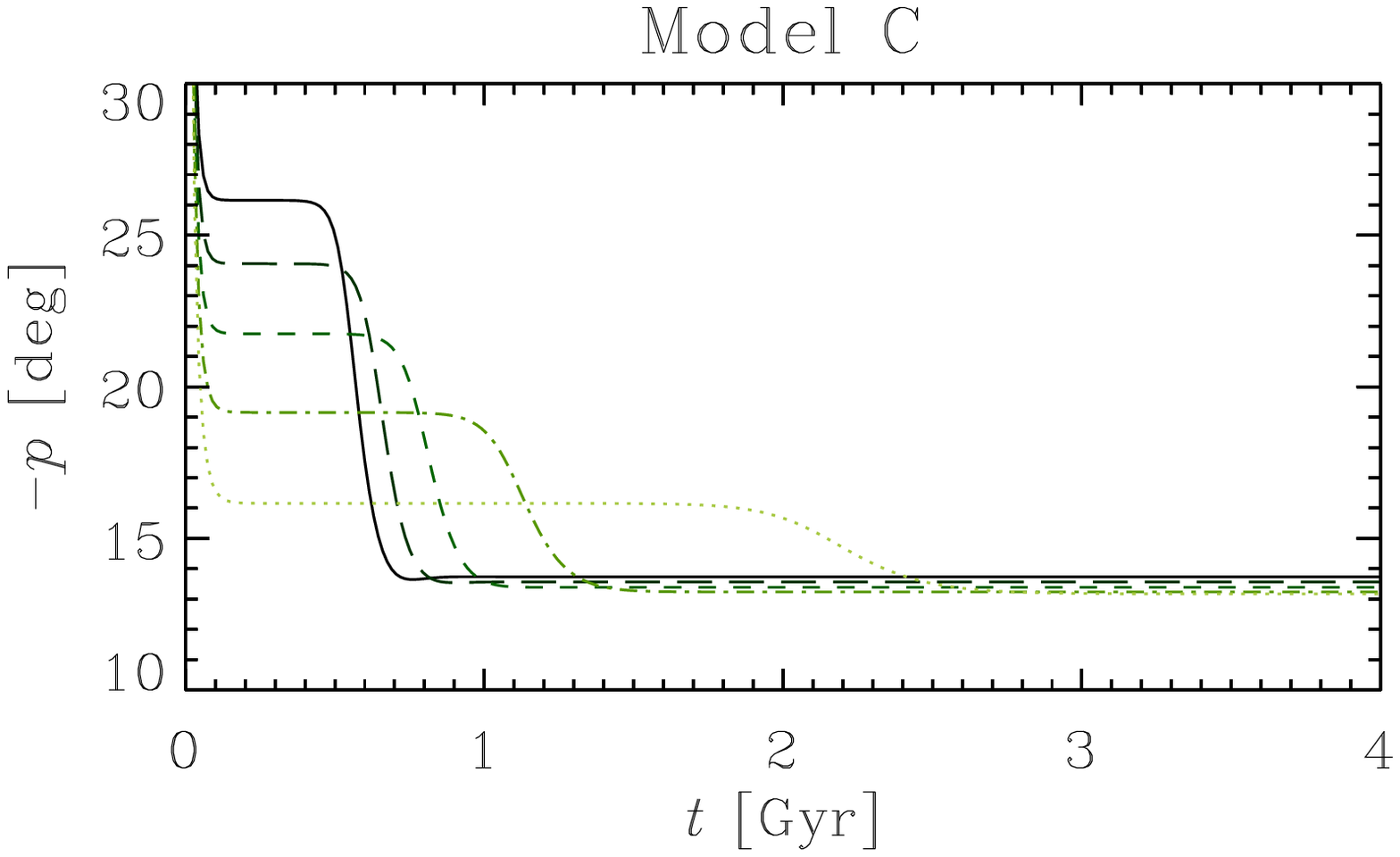}\\
  \includegraphics[width=58mm,clip=true,trim=50 55  35 0]{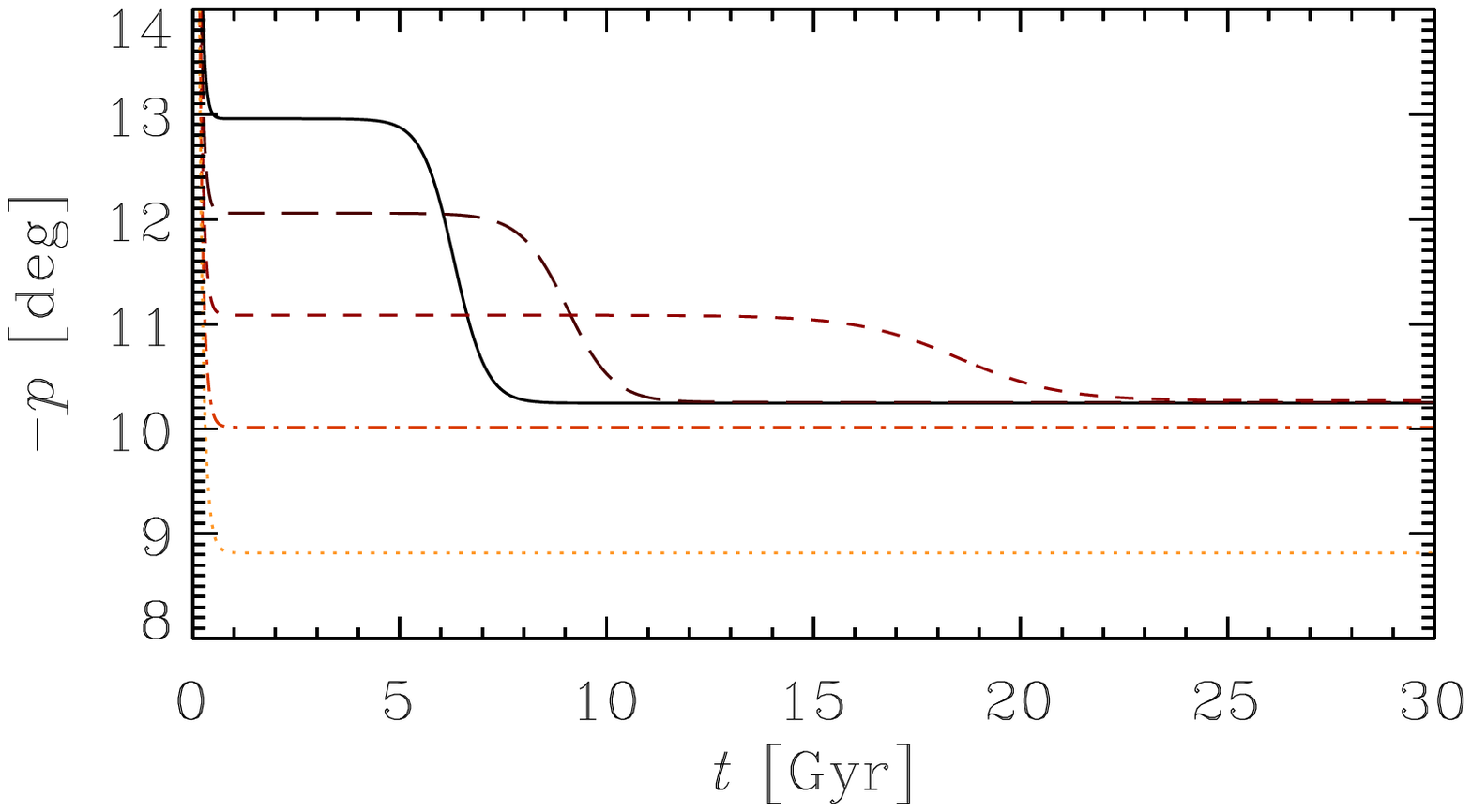}
  \includegraphics[width=58mm,clip=true,trim=50 55  35 0]{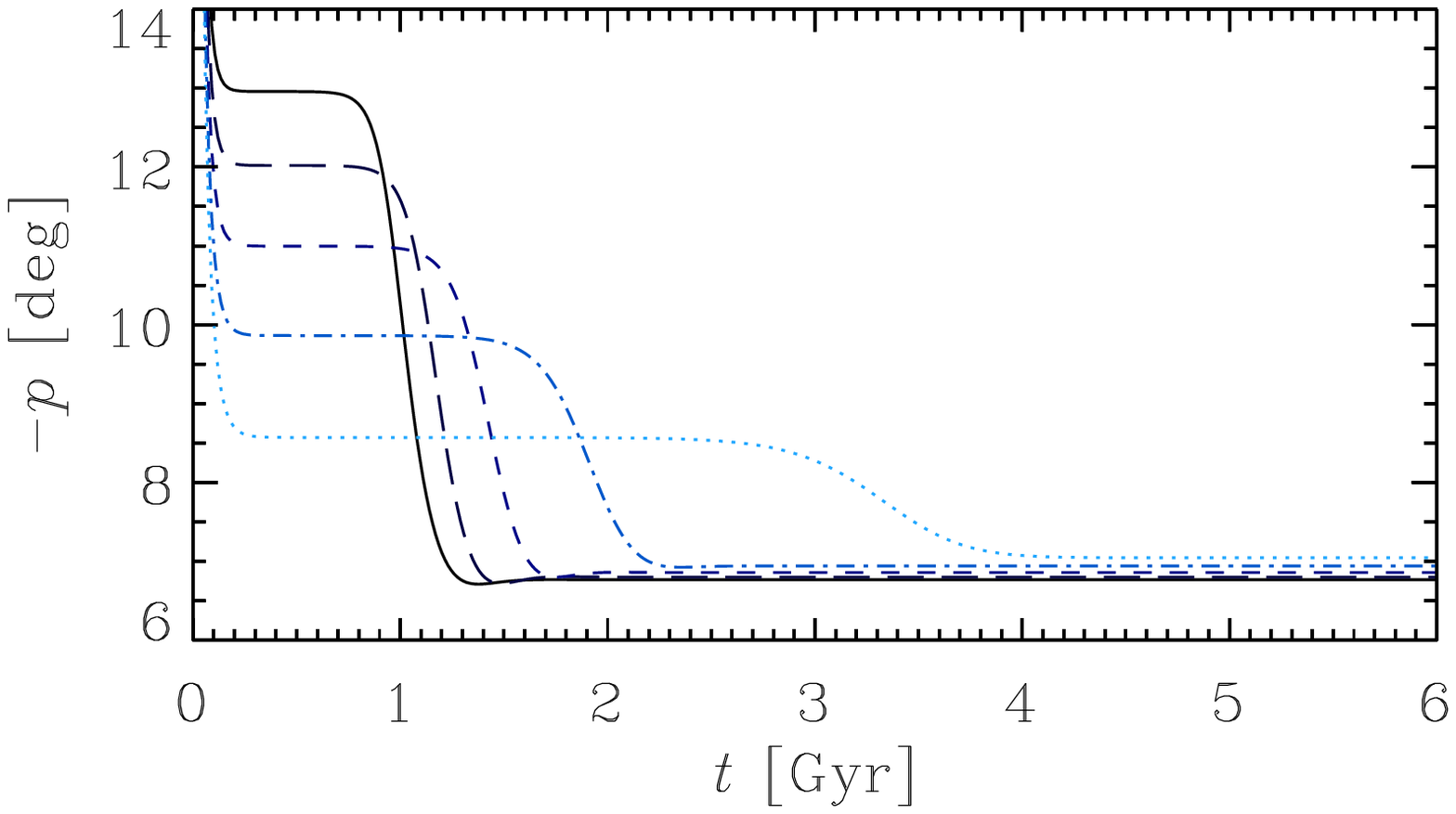}
  \includegraphics[width=58mm,clip=true,trim=50 55  35 0]{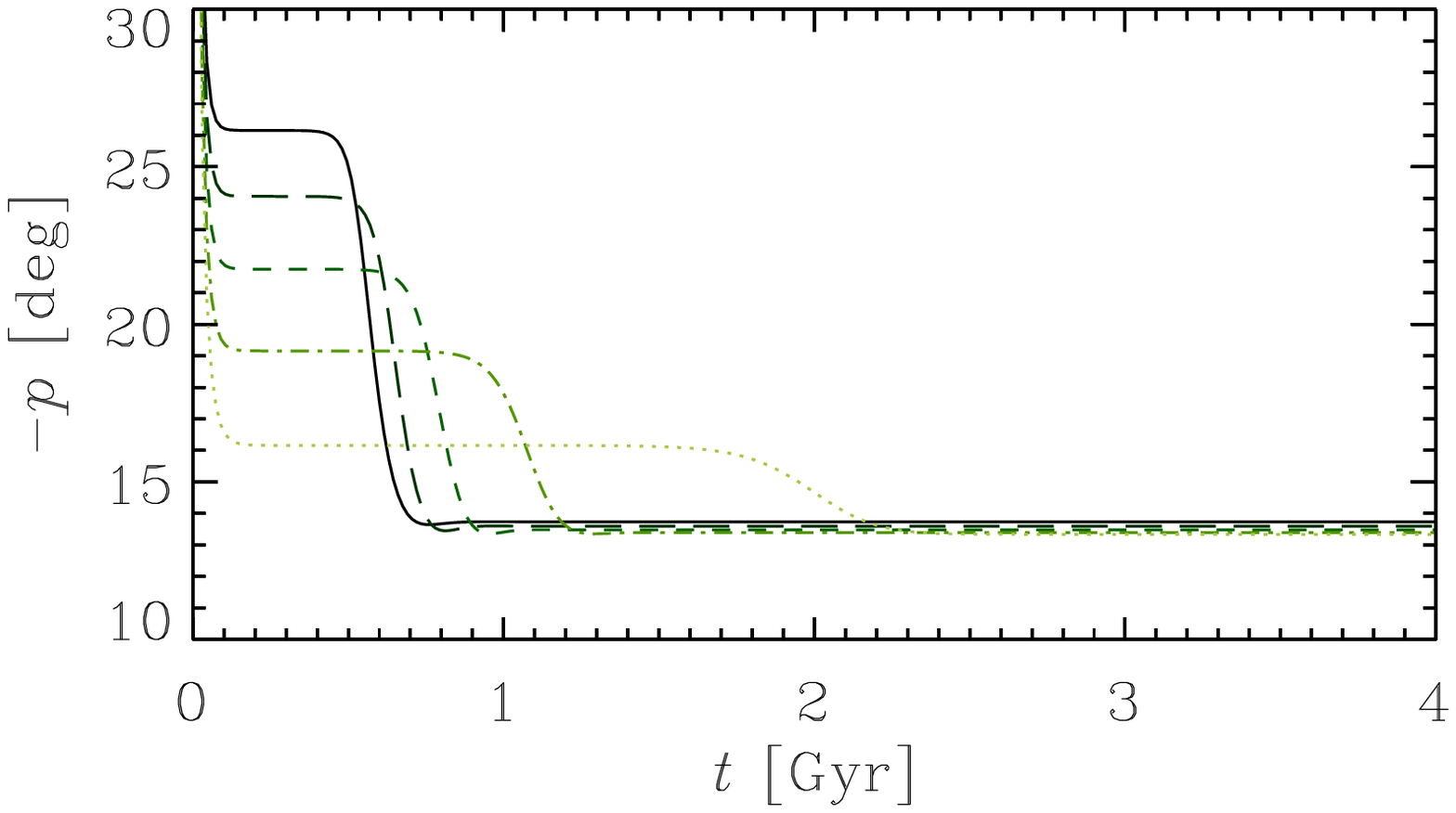}\\
  \includegraphics[width=58mm,clip=true,trim=50  0  35 0]{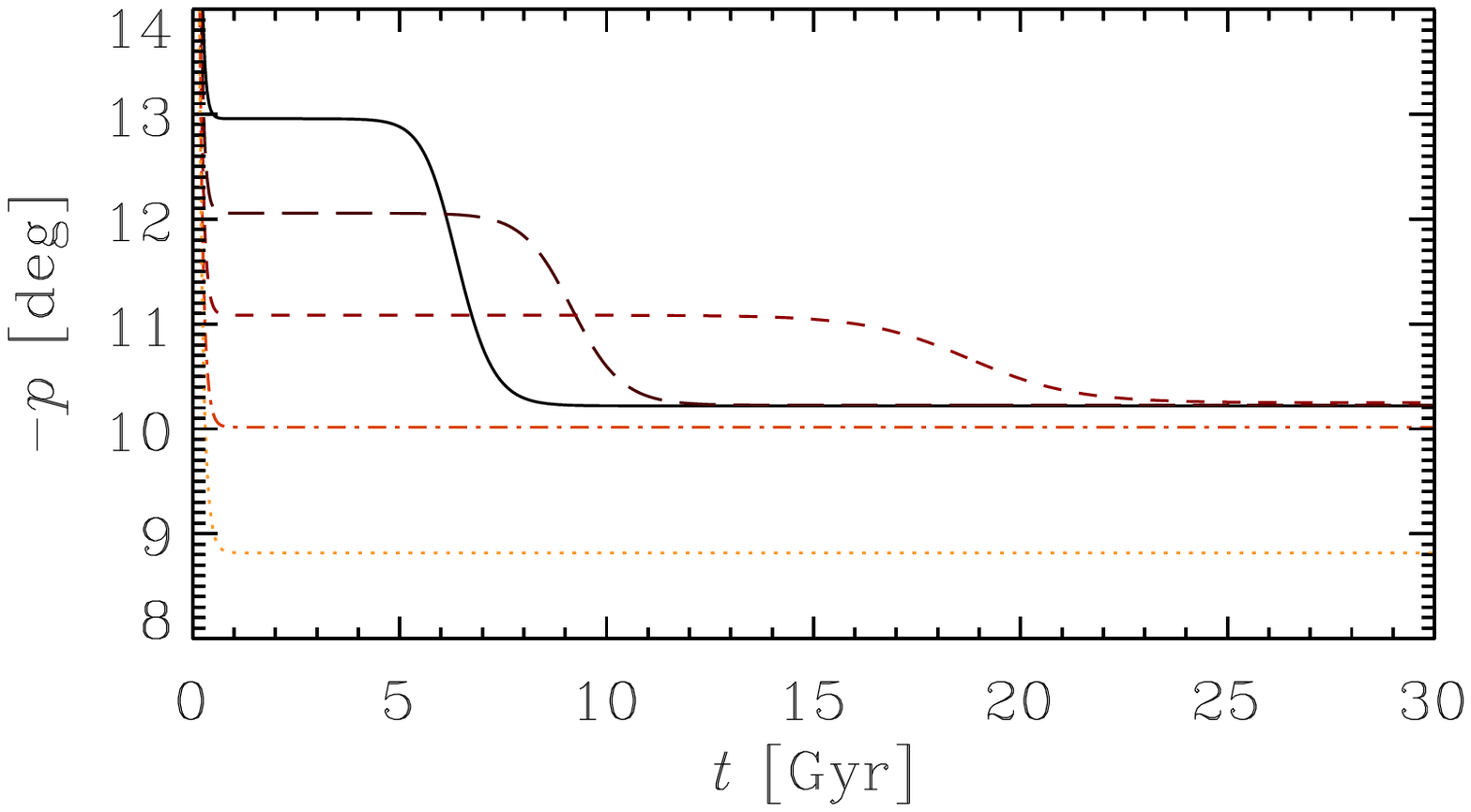}
  \includegraphics[width=58mm,clip=true,trim=50  0  35 0]{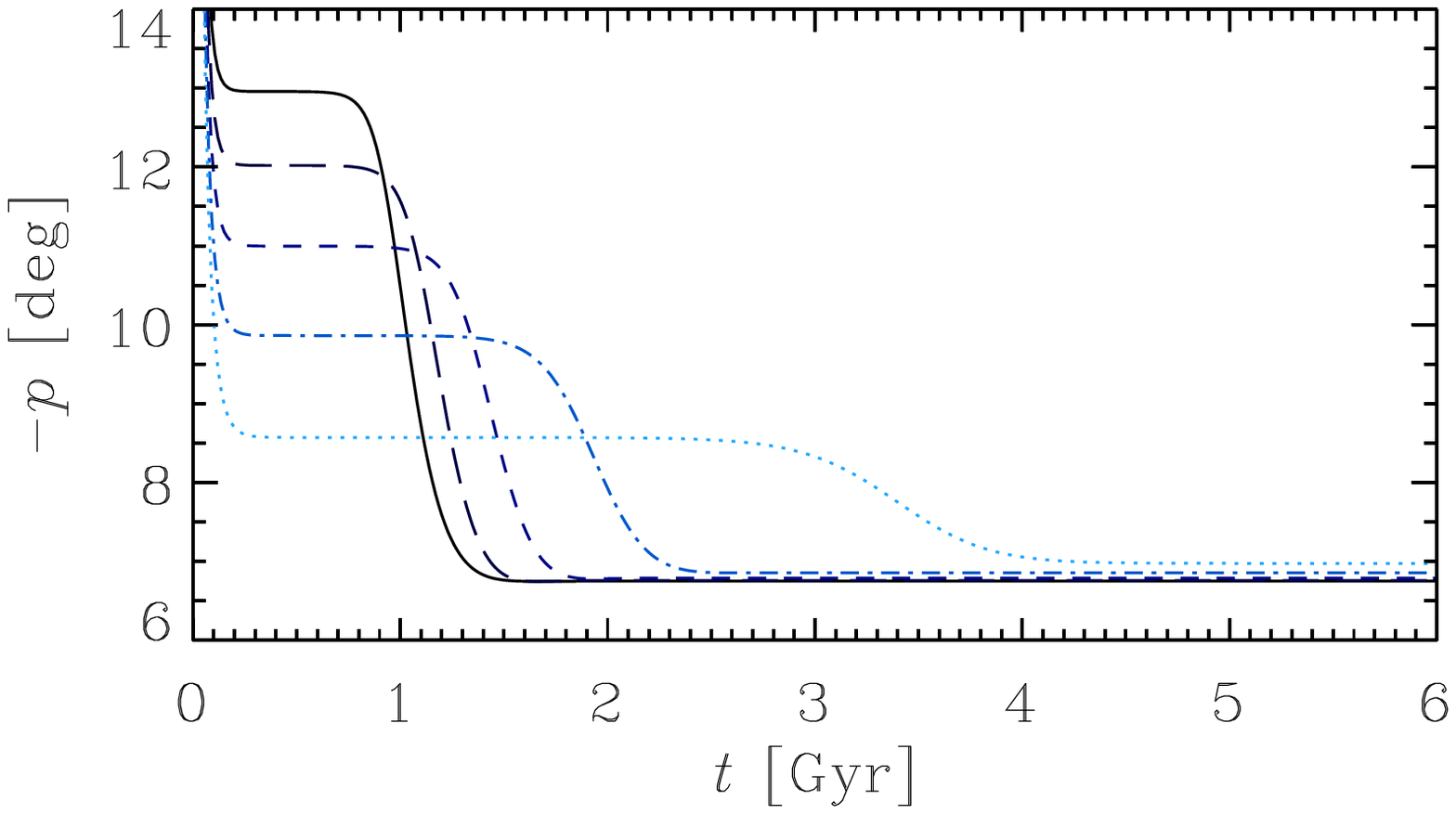}
  \includegraphics[width=58mm,clip=true,trim=50  0  35 0]{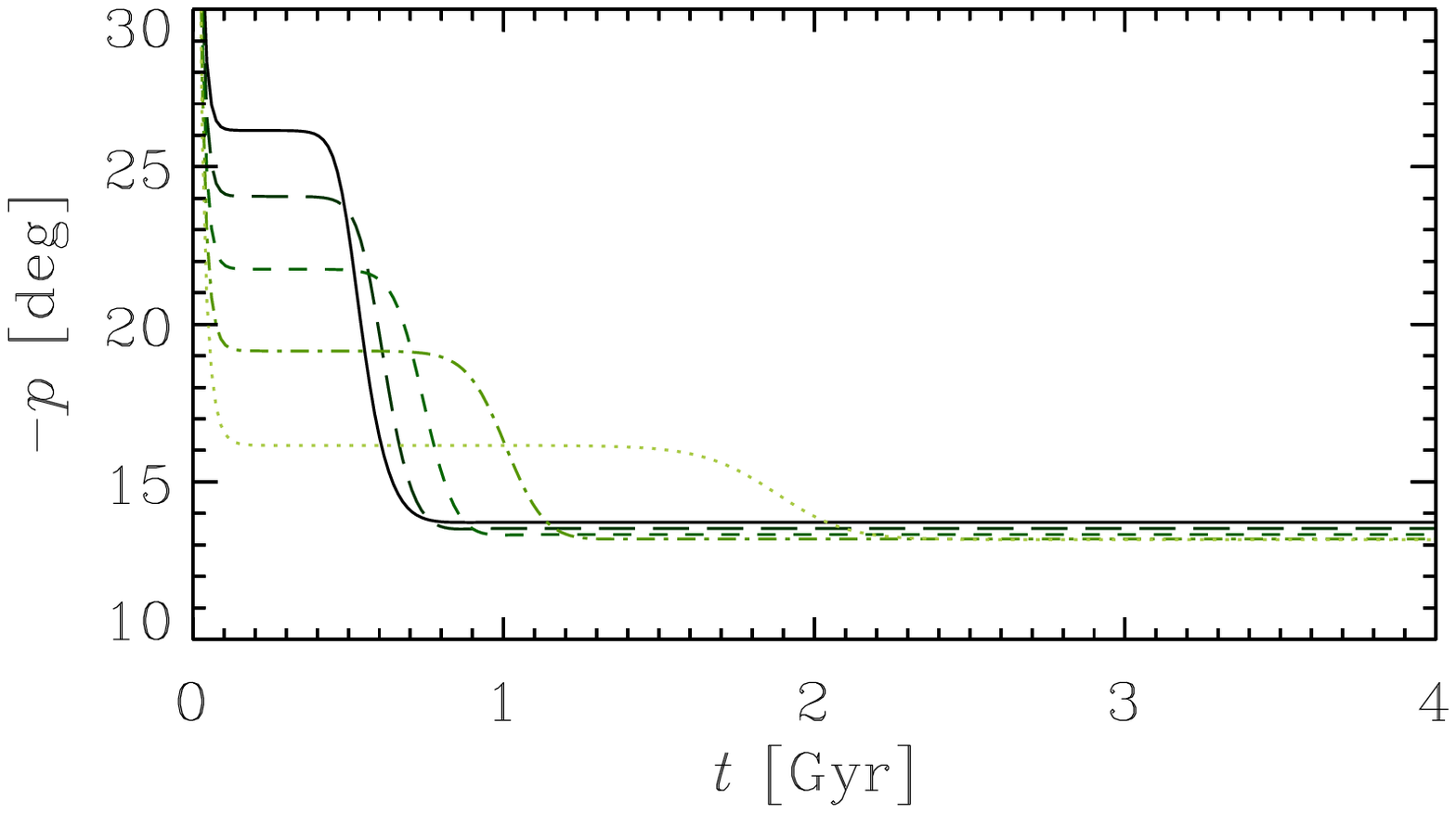}\\
  \caption{As Figure~\ref{fig:Btime_log_ss} but now showing the evolution of the pitch angle of $\meanv{B}$.
           \label{fig:ptime_ss}
          }            
\end{figure*}

In Figure~\ref{fig:ptime_ss} we present results for the time evolution of the weighted average of the pitch angle \citep{Chamandy+16}, 
\begin{equation}
  \label{pitch}
  p= \frac{\displaystyle\int_{-h}^h\arctan[\mbr(z)/\mbp(z)] B^2(z)dz}{\displaystyle\int_{-h}^h B^2(z)dz},
\end{equation}
with $-\pi/2<p\le\pi/2$.
The format of the figure is the same as for Figure~\ref{fig:Btime_log_ss}.
In the runs that result in dynamo growth of the field,
the magnitude of the pitch angle first settles to a fairly steady value in the kinematic regime, 
before decreasing again and settling to a smaller steady or almost steady value in the non-linear regime, 
as expected \citep[e.g.][]{Chamandy+Taylor15}.
For the runs which are subcritical to dynamo growth ($\xi=0.3$ and $\xi=0.4$ cases of Model~A), 
the pitch angle never enters the non-linear regime and retains a steady kinematic value.

Our alternate prescriptions for the non-linearity lead to almost identical results for the pitch angle.
This can be seen by the remarkable similarity between the different rows of Figure~\ref{fig:ptime_ss}, 
and especially in the values of the pitch angles in the saturated state.
For the cases shown, these values are within $1^\circ$ of one another for the each of the three models.
The uncertainty in observational estimates of large-scale magnetic field pitch angles 
in external galaxies is typically a few degrees \citep[][and references therein]{Chamandy+16}. 
Thus, the Magnetic R\"{a}dler effect has a rather negligible effect
on the pitch angle, at least with respect to current observational precision.

For saturated solutions, insensitivity of the pitch angle to the details of the non-linearity, and to $\xi$ more generally,
is expected because $p$ can be estimated simply by solving equations~\eqref{Br} and \eqref{Bp} for a steady state.
Under the `no-$z$' approximation (\citealt{Subramanian+Mestel93,Moss95,Phillips01}; \citetalias{Chamandy+14b}; \citealt{Chamandy16}), 
commonly used for galactic dynamos, the term in equation~\eqref{Bp} involving $\delta'$ can be combined with the term involving $\alpha$
resulting in an effective $\alpha$, or $\alphatilde$ effect.
This results in approximate analytical solutions that agree reasonably well 
with numerical solutions of the full equations in the kinematic regime \citepalias{Chamandy+Singh17}.
The steady state condition then leads to $\alphatilde=\alpha\crit$, 
which is the critical $\alpha$ needed for marginal dynamo growth for the case $\xi=0$.
Under the $\alphatilde\Omega$ approximation,
it is more convenient to consider the dimensionless control parameter $\Dtilde=R_{\alphatilde} R_\Omega$.
In such a steady state, one expects $\Dtilde=D\crit$, where \citepalias{Chamandy+Singh17},
\begin{equation}
  \Dtilde\sim D\left(1-\frac{\xi}{\xi\f}\right),
\end{equation}
with $\xi\f\approx20/\pi^3$,
and with $D\crit\approx -(\pi/2)^5$ in the no-$z$ approximation \citepalias{Chamandy+14b}.
One then obtains the remarkably simple result \citepalias{Chamandy+14b}
\begin{equation}
  \tan p\sim \frac{\pi^2}{4R_\Omega} 
        \sim -\frac{\pi^2}{12q}
               \left(\frac{\tau u}{h}\right)^2\frac{1}{\Omega\tau},
\end{equation}
which is independent of $\xi$ (see also Appendix~\ref{sec:analytic}).
This simple formula yields $p\sim-10^\circ$, $-7^\circ$ and $-13^\circ$ for Models~A, B and C, respectively,
which is very close to the values obtained numerically for the full equations, shown in Figure~\ref{fig:ptime_ss}.

\section{A new dynamo non-linearity}
\label{sec:dynamical_xi}
In reality, $\xi$ would not remain constant as the energy density of the large-scale field approaches that of the turbulence.
In this non-linear regime of $\meanv{B}$, 
$\xi$ would be expected to increase as a result of small-scale turbulent tangling of the large-scale magnetic field.
This provides a negative feedback loop between $\xi$ on the one hand and $\meanv{B}$ on the other,
and should result in the saturation of $\meanv{B}$.
Below we develop the formalism to explore this idea, and then incorporate this formalism into our dynamo model to obtain numerical solutions.

\subsection{Modeling non-linear feedback through tangling}
\label{sec:tangling}
The goal is thus to obtain an equation for $\xi$ as a function of $\meanv{B}$,
and we turn to the literature for an estimate.
\citet{Rogachevskii+Kleeorin07} (hereafter \citetalias{Rogachevskii+Kleeorin07}) derive the following expression 
for the small-scale magnetic field,%
\footnote{These authors make use of the spectral $\tau$ approximation, as in \citetalias{Radler+03},
and obtain results for large fluid and magnetic Reynolds numbers, as is appropriate here.
They do not include rotation (or shear) in the theory, but we do not expect that this would lead to drastic differences.}
which for non-convective turbulence simplifies to (their equation~(A21) with $a_*=0$):
\begin{equation}
  \label{RK}
  \mean{\bfb^2}= \mean{\bfb^2}^{(0)} 
                +\frac{1}{12}\left(\mean{\bfu^2}^{(0)}-\mean{\bfb^2}^{(0)}\right)
                \left[6-3A_1^{(0)}(\beta) -A_2^{(0)}(\beta)\right],
\end{equation}
with
\begin{equation}
  \beta= 4\frac{B}{B\eq}= 4\frac{B}{u},
\end{equation}
and from their equations~(A35) and (A36), the functions $A_1^{(0)}(\beta)$ and $A_2^{(0)}(\beta)$ are given by
\begin{equation}
  \begin{split}
    A_1^{(0)}(\beta)= \frac{1}{5}\Bigg[&2+2\frac{\arctan\beta}{\beta^3}(3+5\beta^2)-\frac{6}{\beta^2}\\
                                       &-\beta^2\ln\Rm
                                       -2\beta^2\ln\left(\frac{1+\beta^2}{1+\beta^2\sqrt{\Rm}}\right)\Bigg],
  \end{split}
\end{equation}
\begin{equation}
  \begin{split}
    A_2^{(0)}(\beta)= \frac{2}{5}\Bigg[&2-\frac{\arctan\beta}{\beta^3}(9+5\beta^2)+\frac{9}{\beta^2}\\
                                       &-\beta^2\ln\Rm
                                       -2\beta^2\ln\left(\frac{1+\beta^2}{1+\beta^2\sqrt{\Rm}}\right)\Bigg].
  \end{split}
\end{equation}
Note that the superscript `$(0)$' refers to the initial background turbulence
without the effects of the mean magnetic field on the turbulence included.
In \citetalias{Chamandy+Singh17}, 
we had adopted the relevant expressions from \citetalias{Radler+03} and \citetalias{Brandenburg+Subramanian05a},
which also contained $\mean{\bfu^2}^{(0)}$ rather than $\mean{\bfu^2}$.
Thus, as we did there, we will identify $u^2$ with $\mean{\bfu^2}^{(0)}$,
and so neglect any effects of $\meanv{B}$ on $u$,
but it should be kept in mind that these quantities may in general be different.
Equation~\eqref{RK} has been shown to fit simulation results reasonably well 
\citep[][but note the typographical error in their equation~(9)]{Karak+Brandenburg16}.

\begin{figure}
  \includegraphics[width=1.0\columnwidth,clip=true,trim=20 0  0 0]{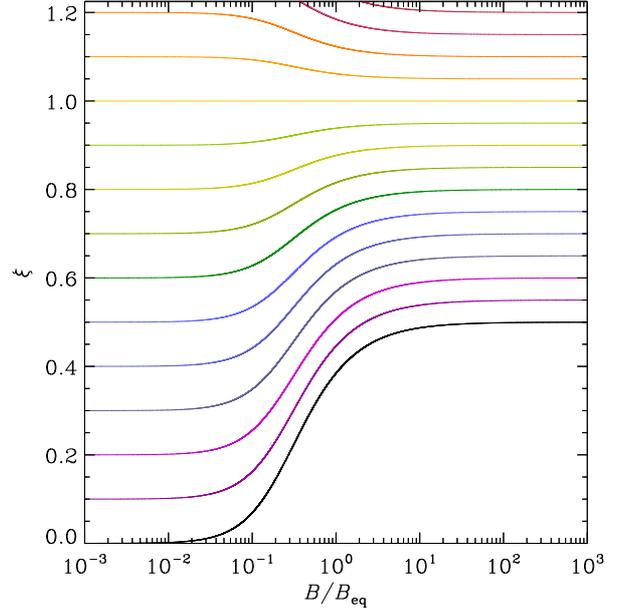}\\
  \caption{Graphical representation of equation~\eqref{dynamical_xi},
           showing how $\xi$ varies with $B/B\eq$ in the model of \citetalias{Rogachevskii+Kleeorin07}.
           Curves are for values of $\xi\f$ separated by $0.1$.
           \label{fig:xi}
          }
\end{figure}

Now, we are interested in the large $\Rm$ limit, since $\Rm\gg1$ in galaxies.
In this limit equation~\eqref{RK} simplifies to
\begin{equation}
  \begin{split}
    \mean{\bfb^2}=&\:\mean{\bfb^2}^{(0)} +\frac{1}{3}\left(\mean{\bfu^2}^{(0)}-\mean{\bfb^2}^{(0)}\right)
                   \Bigg[ 1-\frac{\arctan \beta}{\beta}\\
                         &+\frac{1}{4}\beta^2\ln\left(1+\frac{2}{\beta^2}+\frac{1}{\beta^4}\right)\Bigg].
  \end{split}
\end{equation}
Dividing by $u^2=\mean{\bfu^2}^{(0)}$, 
and using the subscript `$0$' in place of the superscript `$(0)$' on $\xi$,
we obtain
\begin{equation}
  \label{dynamical_xi}
  \xi= \xi\f +\frac{1}{3}\left(1-\xi\f\right)
                   \Bigg[ 1-\frac{\arctan \beta}{\beta}
                         +\frac{1}{4}\beta^2\ln\left(1+\frac{2}{\beta^2}+\frac{1}{\beta^4}\right)\Bigg],
\end{equation}
where $\xi\f$ is the value of $b^2/u^2$ in the kinematic regime of mean-field dynamo action.
This expression has the limit $\xi\rightarrow\xi\f$ as $\beta\rightarrow0$, as required,
since the second term in the square brackets $\rightarrow1$ and the third term $\rightarrow0$.
It is also instructive, from a mathematical point of view, to write down the large $\beta$ asymptotic behaviour, where
\begin{equation}
  \xi \rightarrow \frac{1}{2}\left(1+\xi\f\right)\;\; \mathrm{as}\;\; \beta\rightarrow\infty
\end{equation}
since $\arctan(\beta)/\beta\rightarrow0$ and $\beta^2\ln(1+2/\beta^2+1/\beta^4)\rightarrow2$ as $\beta\rightarrow\infty$.
Moreover, for $\xi\f<1$, $\xi$ increases with $B$,
for $\xi\f=1$, $\xi$ remains constant and is independent of $B$,
and for $\xi\f>1$, $\xi$ decreases with $B$.
A plot is shown in Figure~\ref{fig:xi}.
It can be seen that for realistic values of $\xi\f$ between $0$ and $0.4$, 
$\xi$ can increase by $0.2$ to $0.4$ before $B$ reaches $B\eq$.
We know from \citetalias{Chamandy+Singh17} that the kinematic growth rate, at least,
is rather sensitive to the value of $\xi$, and that increasing $\xi$ by this much
can even render the dynamo subcritical.
Therefore, we expect that by including this new non-linear effect, we should obtain quenching of the dynamo,
leading to a smaller growth rate, and eventually saturation.

As terms involving $\del\delta'/\del z$ occur in equations~\eqref{Br} and \eqref{Bp},
it is necessary to write down the derivative of equation~\eqref{dynamical_xi}.
We obtain
\begin{equation}
  \frac{\del\xi}{\del z}= \frac{\del\xi}{\del \beta}\frac{\del \beta}{\del z}
                        = \beta\frac{\del\xi}{\del \beta}\frac{1}{B^2}\left(\mbr\frac{\del\mbr}{\del z} +\mbp\frac{\del\mbp}{\del z}\right),
\end{equation}
where 
\begin{equation}
  \label{betadxidbeta}
  \beta\frac{\del\xi}{\del \beta}= \frac{1}{3}(1-\xi\f)\left[\frac{\arctan(\beta)}{\beta} +\beta^2\ln\left(1+\frac{1}{\beta^2}\right) -1\right].
\end{equation}
Expression~\eqref{betadxidbeta} has the limits $\beta \del\xi/\del \beta\rightarrow0$ as $\beta\rightarrow0$ and $\beta \del\xi/\del \beta\rightarrow0$ as $\beta\rightarrow\infty$.

\section{Results when tangling is included}
\label{sec:tangling_results}

\begin{figure*}
  \includegraphics[width=58mm,clip=true,trim=31 55 32 0]{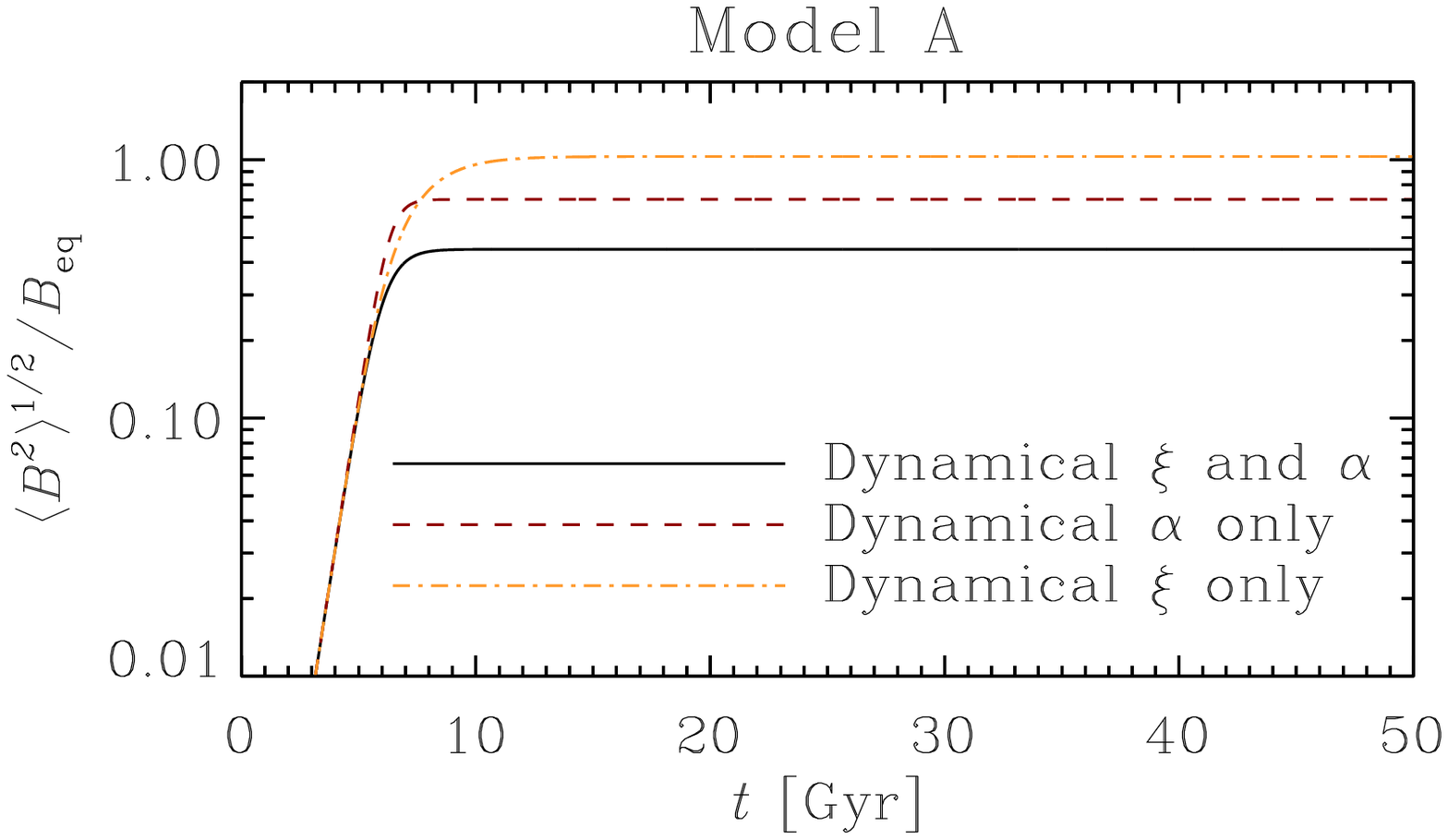}
  \includegraphics[width=58mm,clip=true,trim=31 55 32 0]{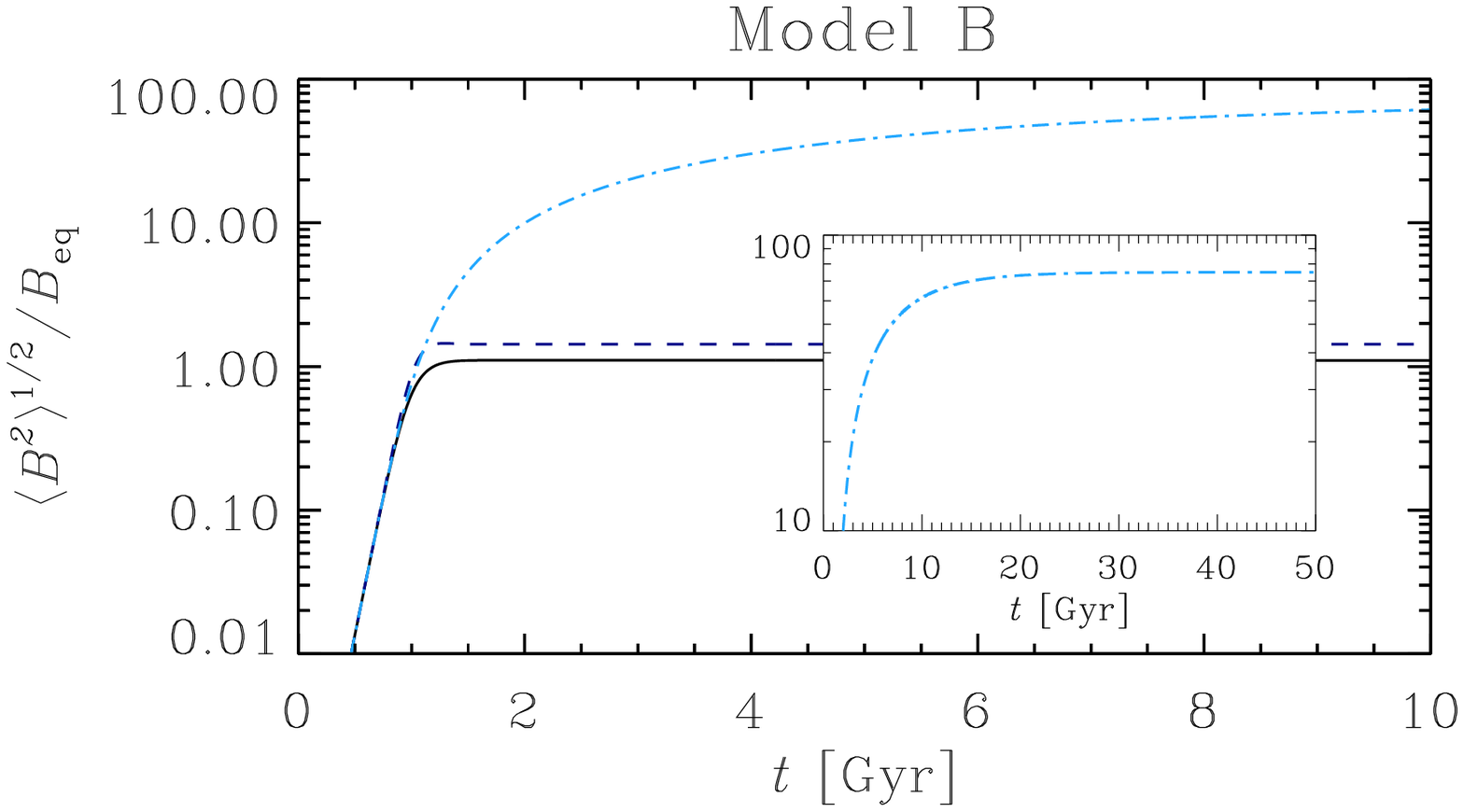}
  \includegraphics[width=58mm,clip=true,trim=31 55 32 0]{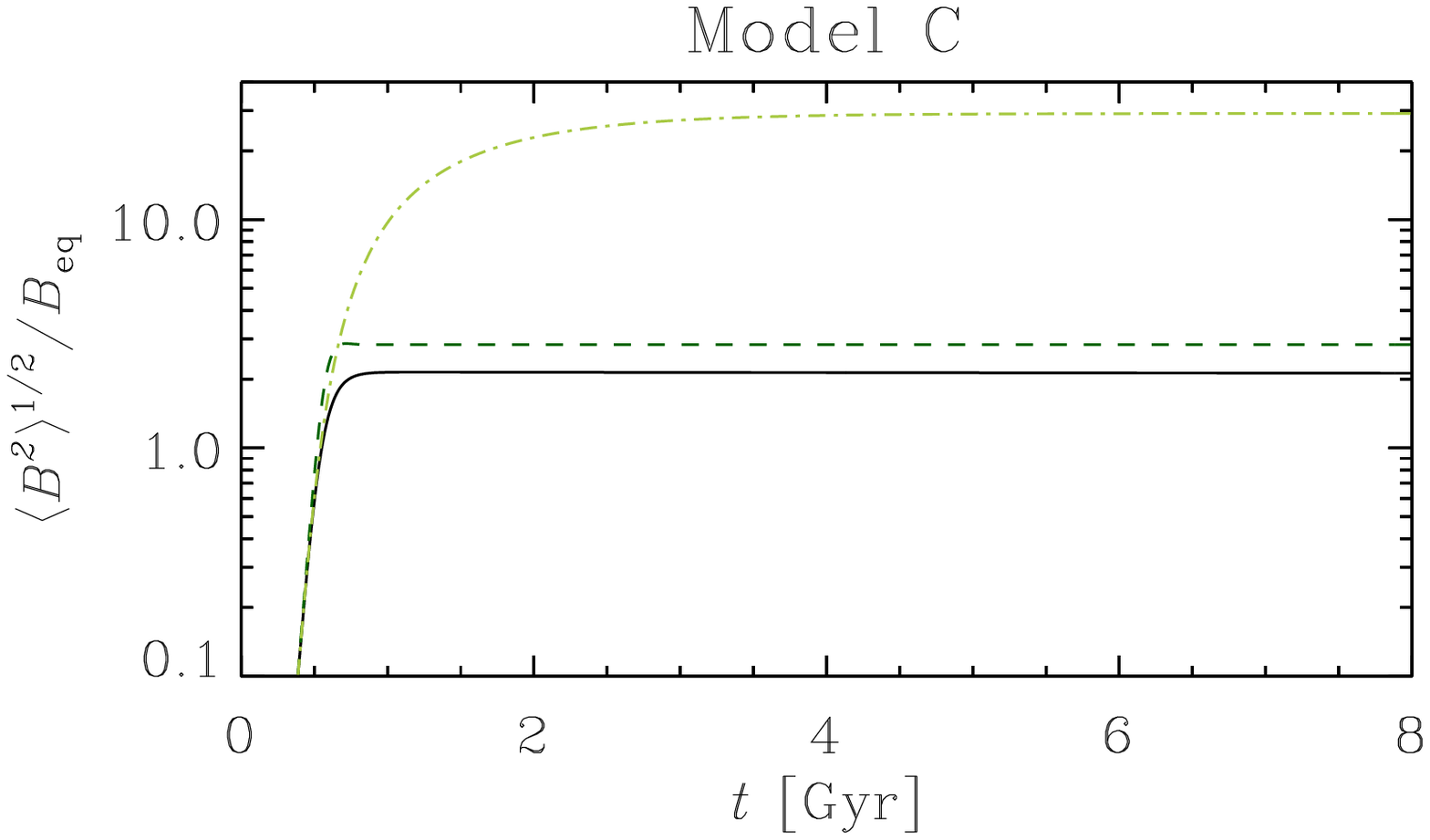}\\
  \includegraphics[width=58mm,clip=true,trim=31 55 32 0]{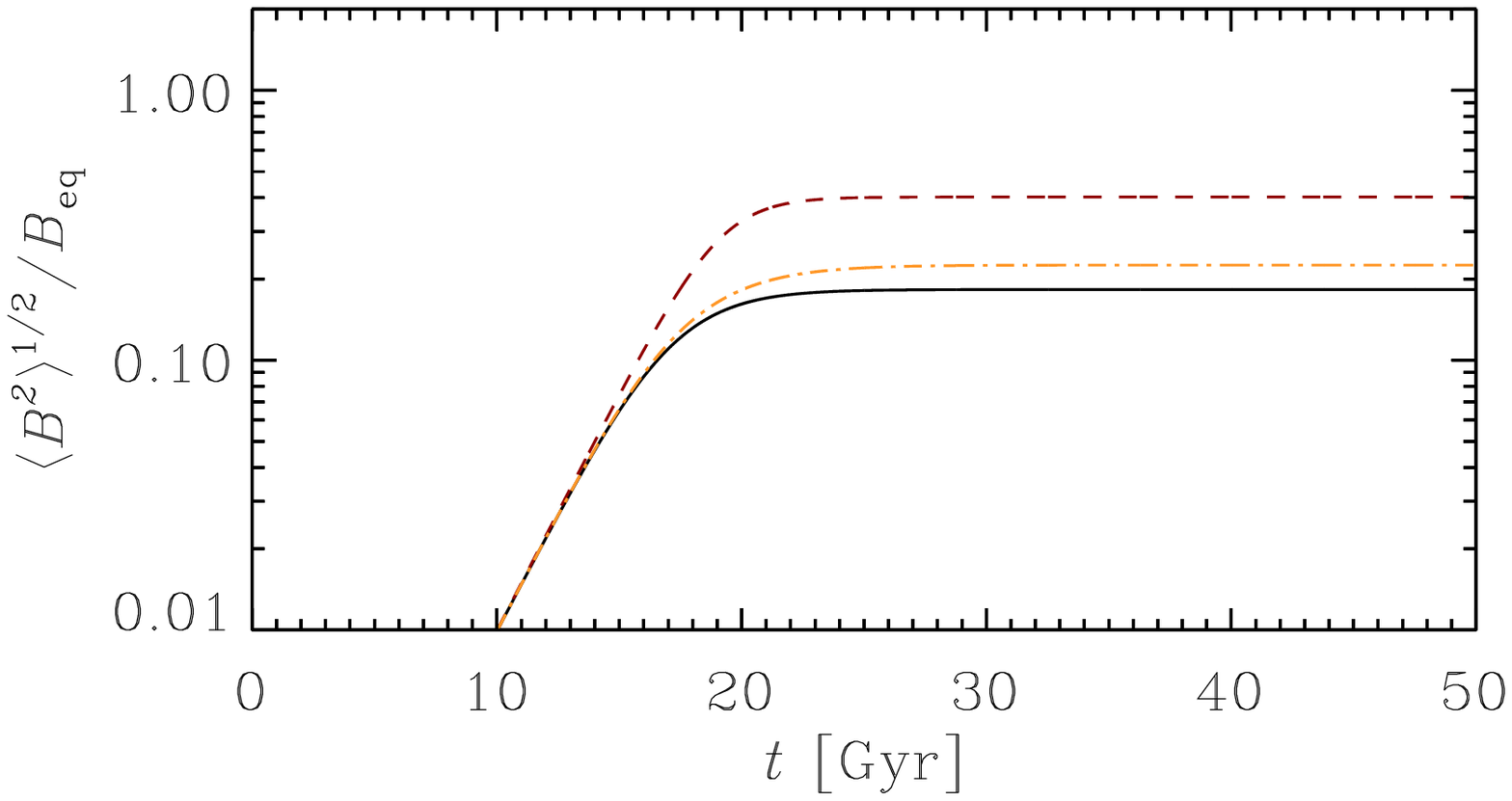}
  \includegraphics[width=58mm,clip=true,trim=31 55 32 0]{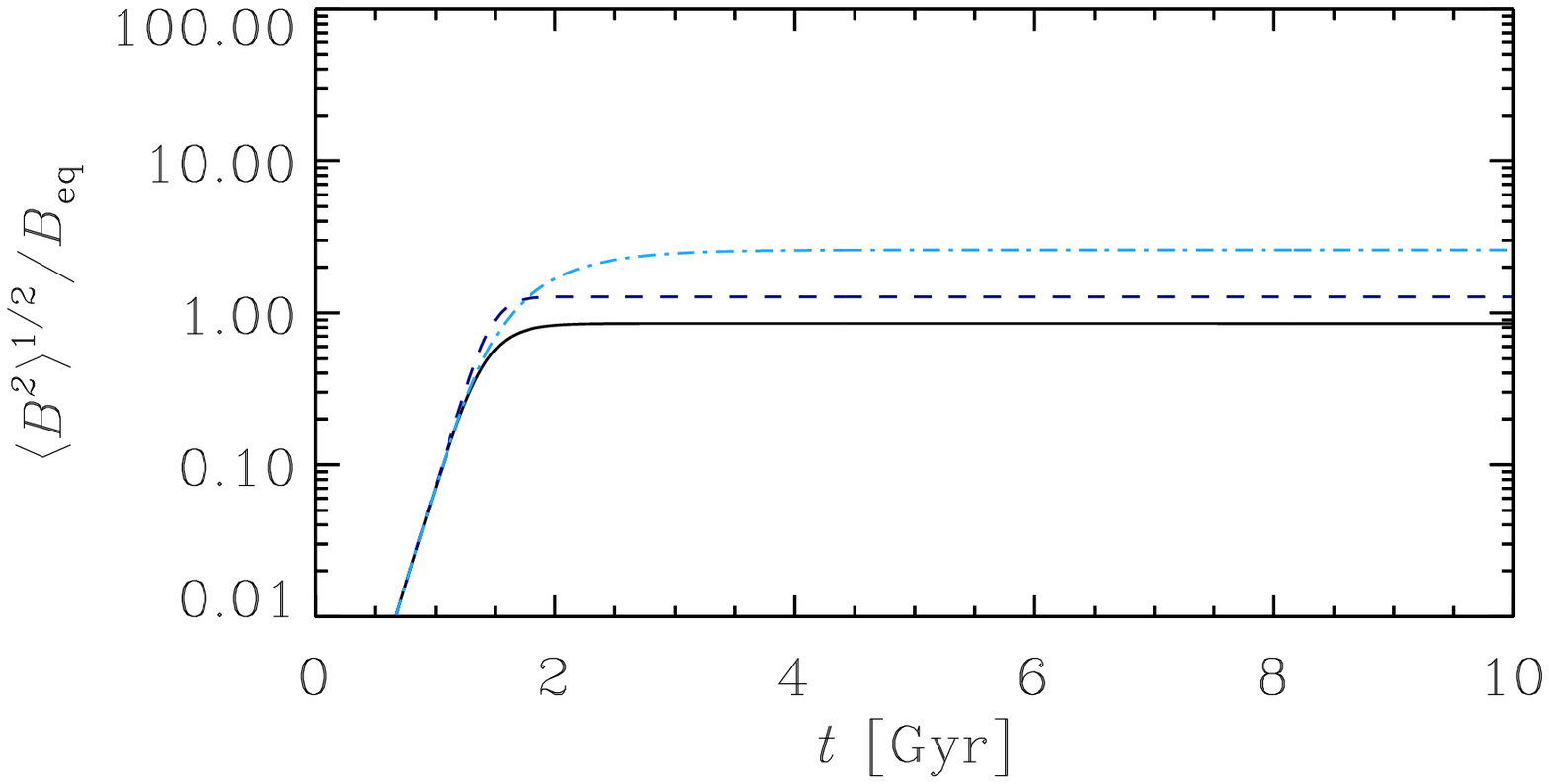}
  \includegraphics[width=58mm,clip=true,trim=31 55 32 0]{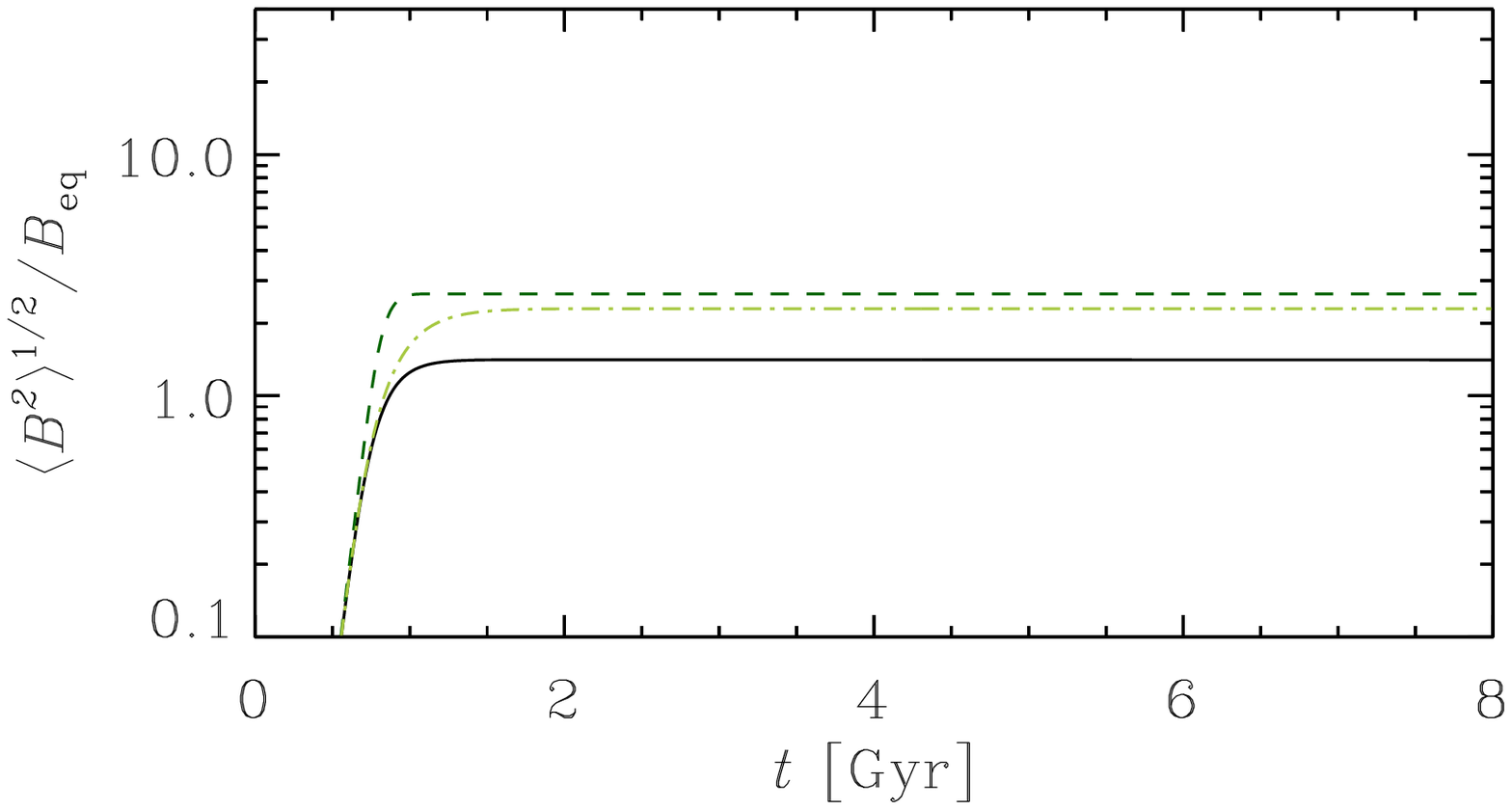}\\
  \includegraphics[width=58mm,clip=true,trim=31 0  32 0]{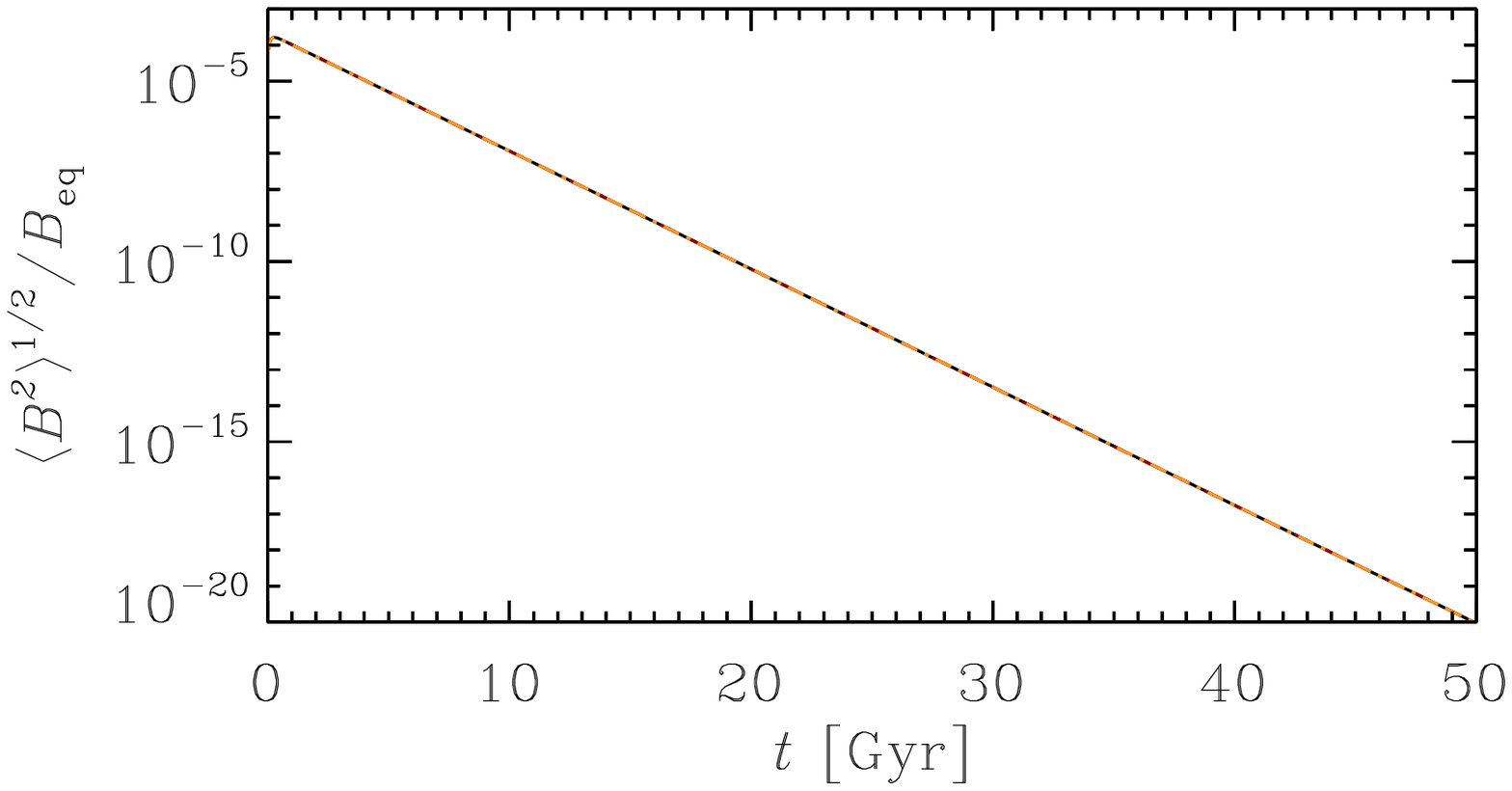}
  \includegraphics[width=58mm,clip=true,trim=31 0  32 0]{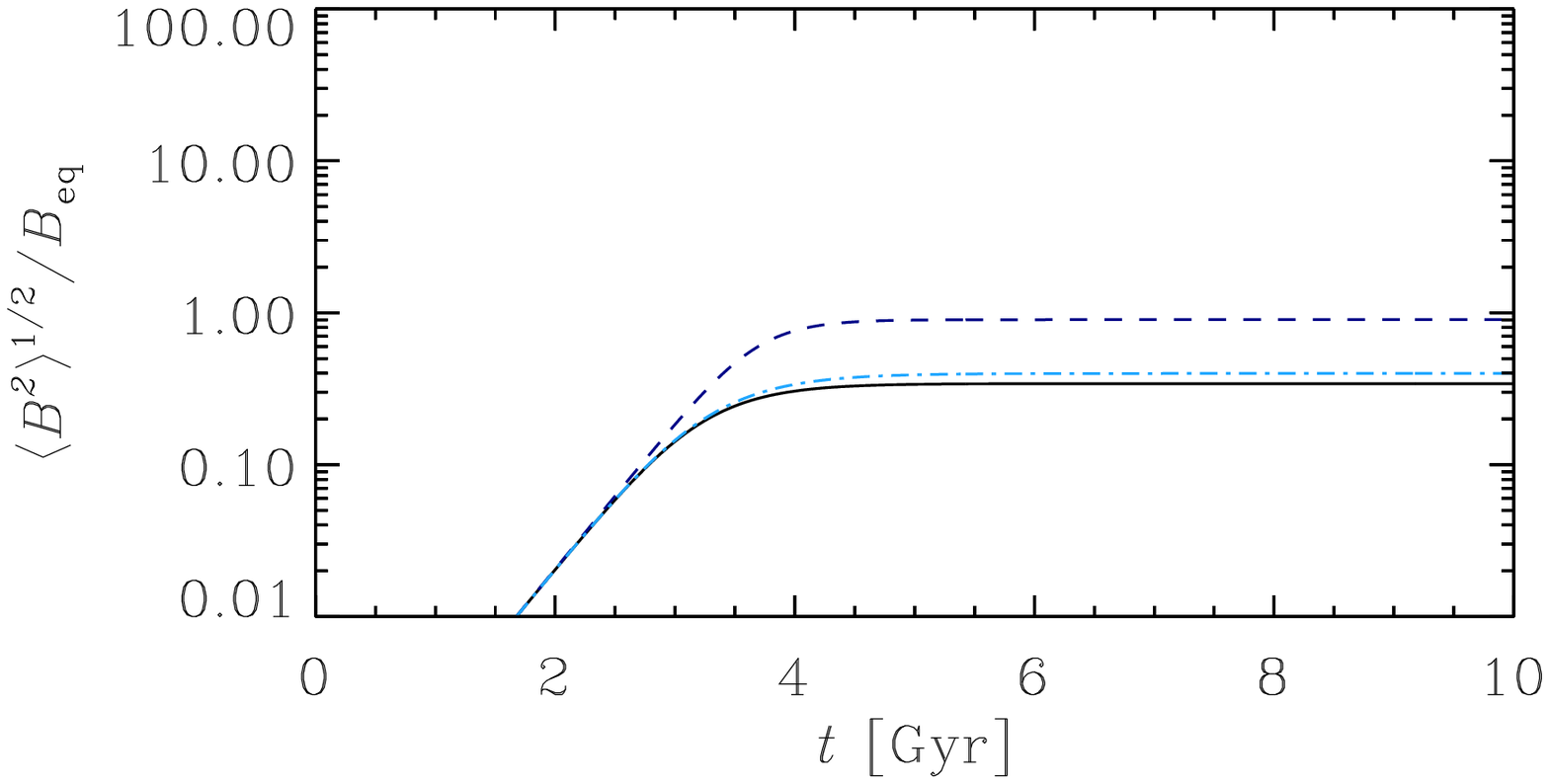}
  \includegraphics[width=58mm,clip=true,trim=31 0  32 0]{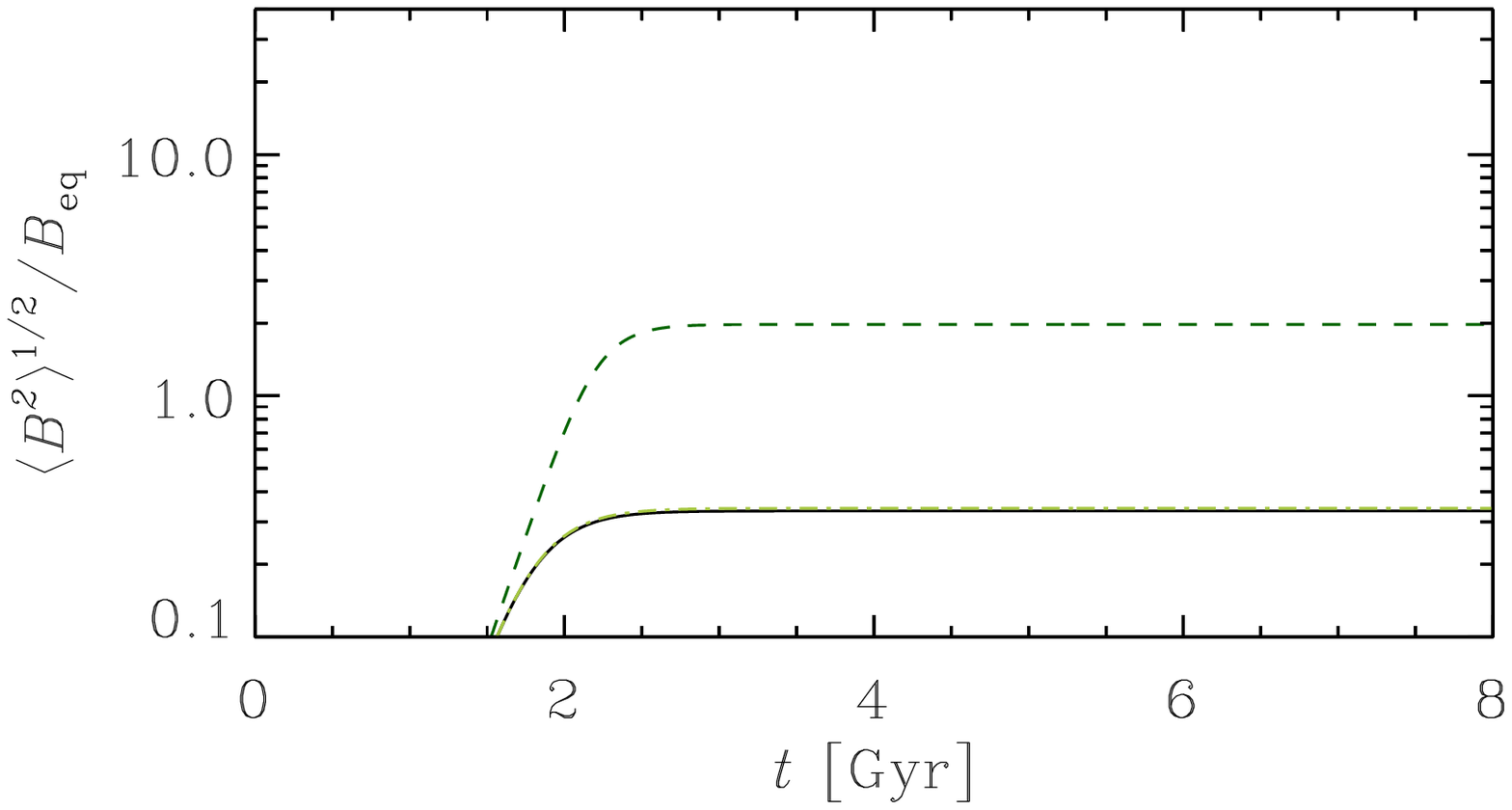}\\
  \caption{Mean magnetic field strength evolution for Model~A (leftmost column), Model~B (middle column) and Model~C (rightmost column).
           For each model three different curves are shown corresponding to different types of quenching:
           dynamical $\alpha$ and dynamical $\xi$   (solid);
           dynamical $\xi$ only (dash-dotted);
           dynamical $\alpha$ only (dashed).
           In some cases the solid and dash-dotted curves coincide with one another.
           Top row: $\xi\f=0$. Inset in the middle panel shows the evolution of the dynamical $\xi$ only case up to late times.
           Middle row: $\xi\f=0.2$
           Bottom row: $\xi\f=0.4$.
           \label{fig:Btime_tangling_log_ss}
          }            
\end{figure*}

\begin{figure*}
  \includegraphics[width=58mm,clip=true,trim=31 55  32 0]{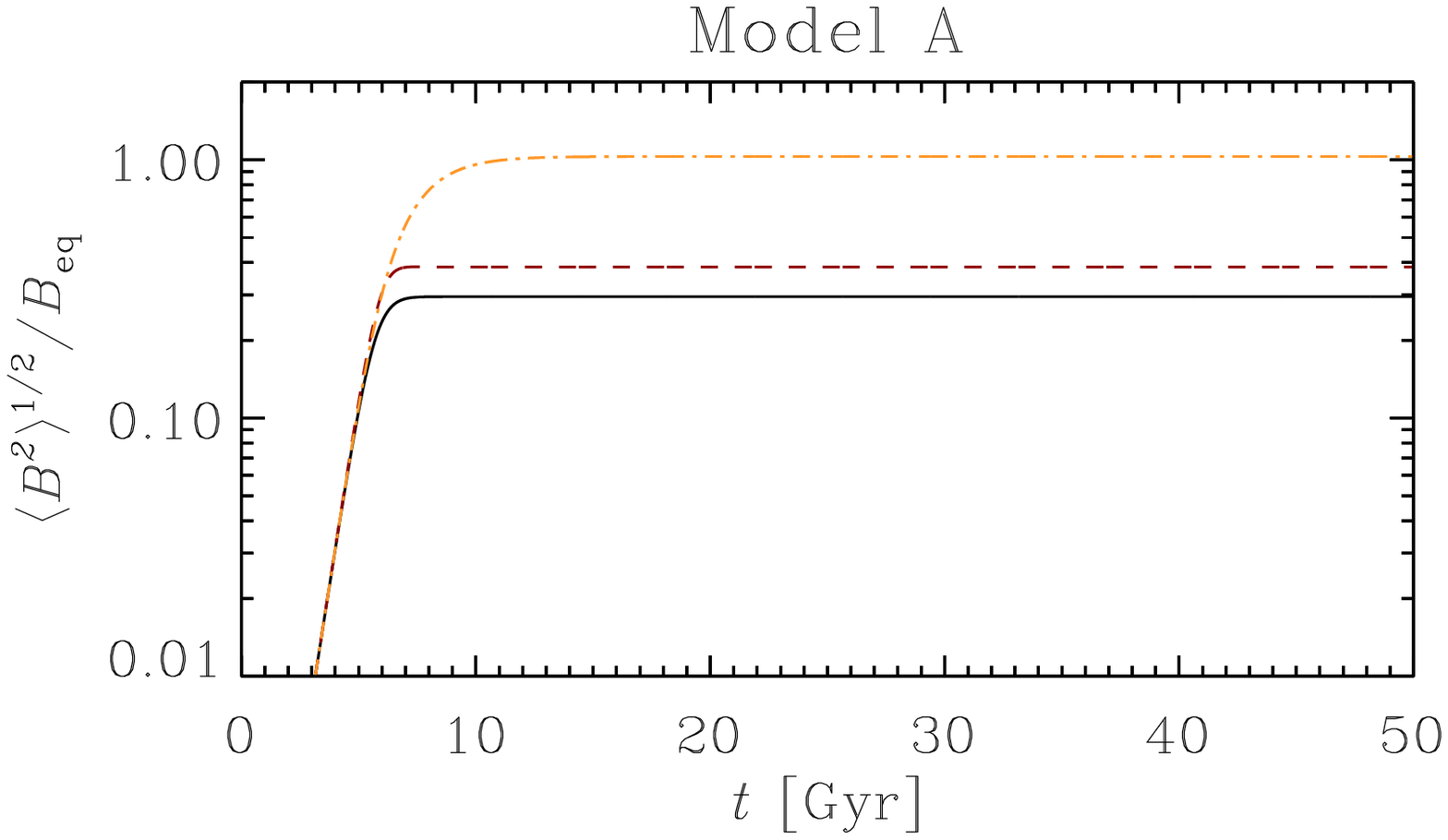}
  \includegraphics[width=58mm,clip=true,trim=31 55  32 0]{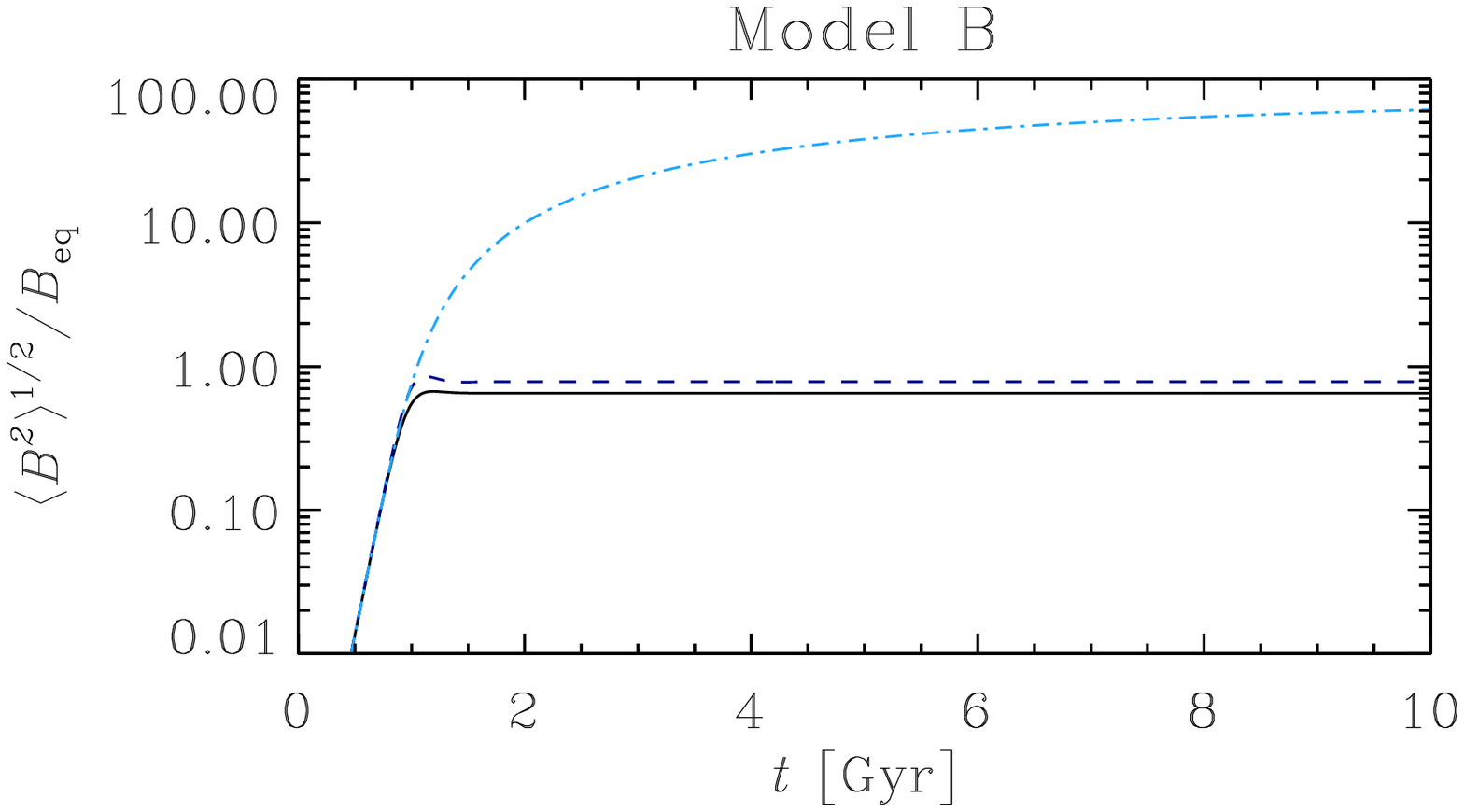}
  \includegraphics[width=58mm,clip=true,trim=31 55  32 0]{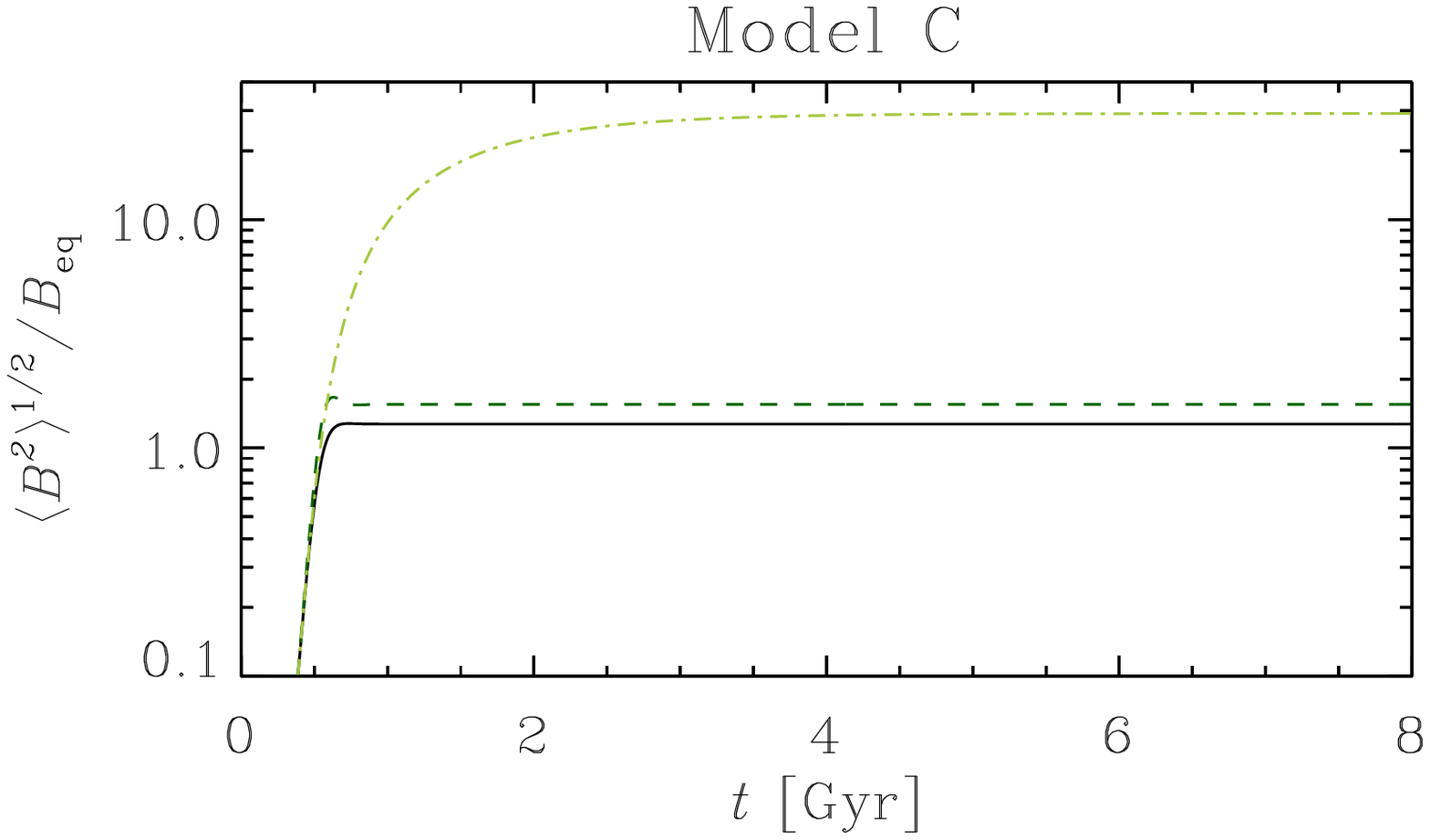}\\
  \includegraphics[width=58mm,clip=true,trim=31 55  32 0]{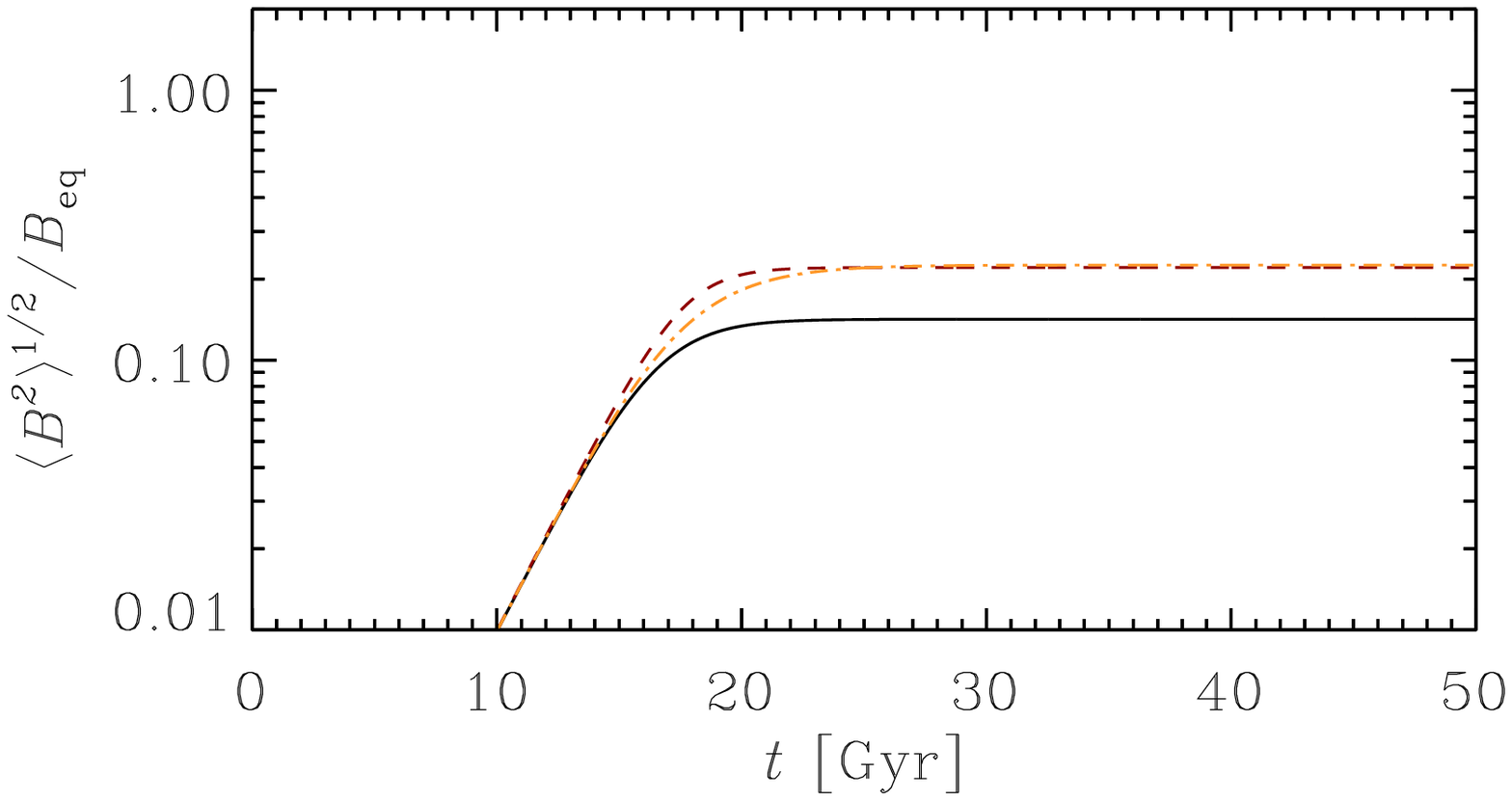}
  \includegraphics[width=58mm,clip=true,trim=31 55  32 0]{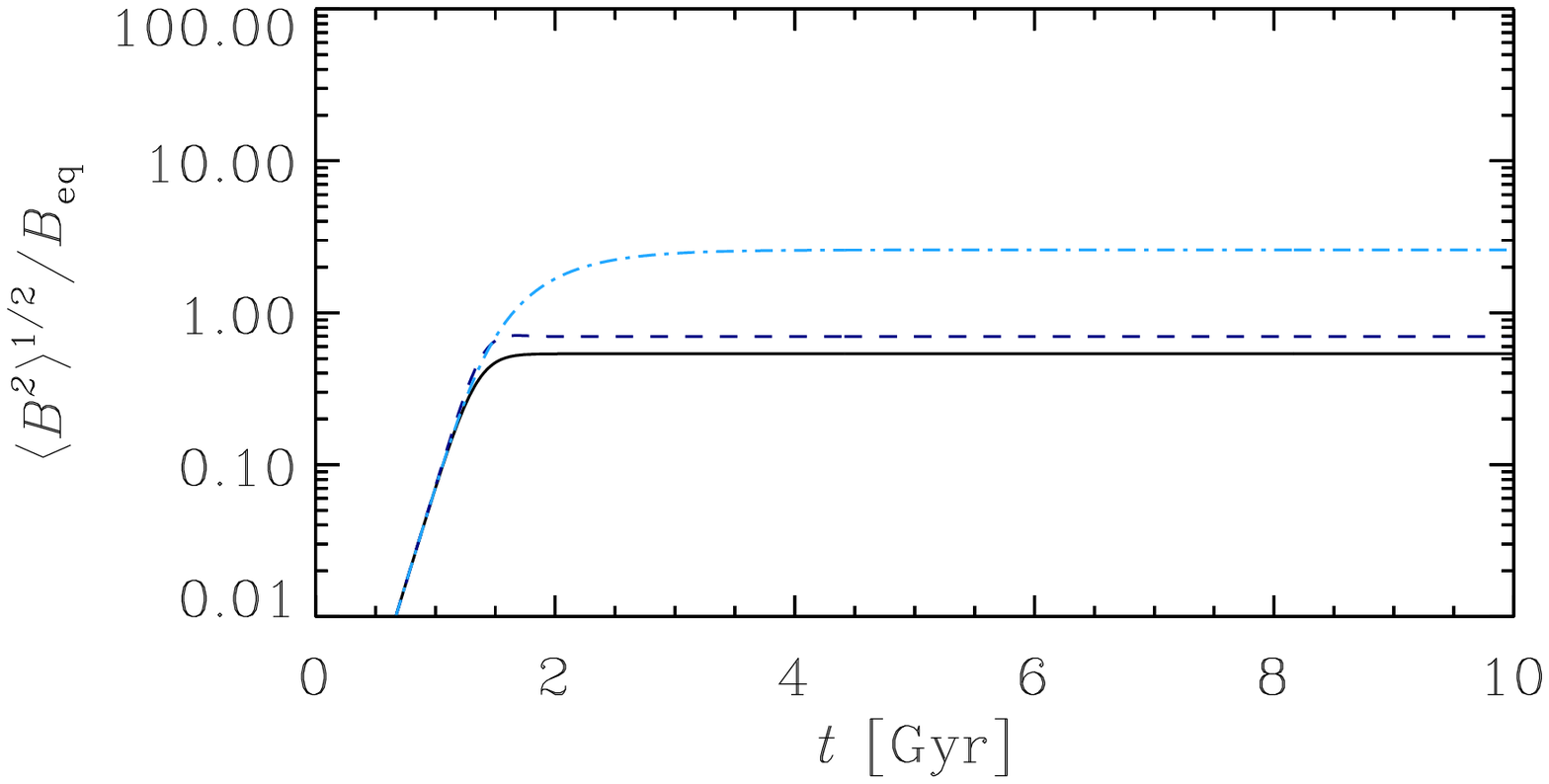}
  \includegraphics[width=58mm,clip=true,trim=31 55  32 0]{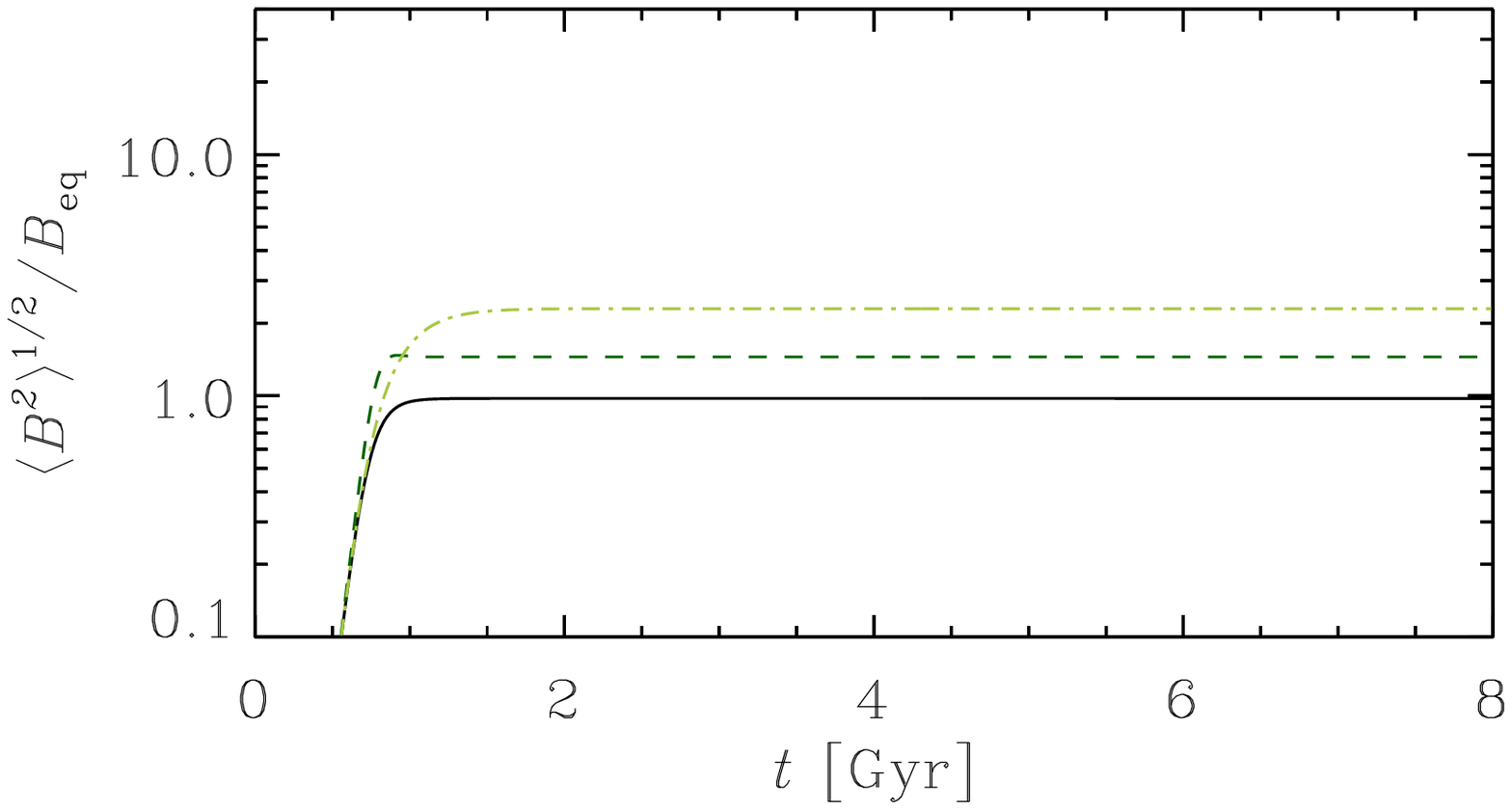}\\
  \includegraphics[width=58mm,clip=true,trim=31 0   32 0]{B_time_modelA_xi0_0o4_ss.eps}
  \includegraphics[width=58mm,clip=true,trim=31 0   32 0]{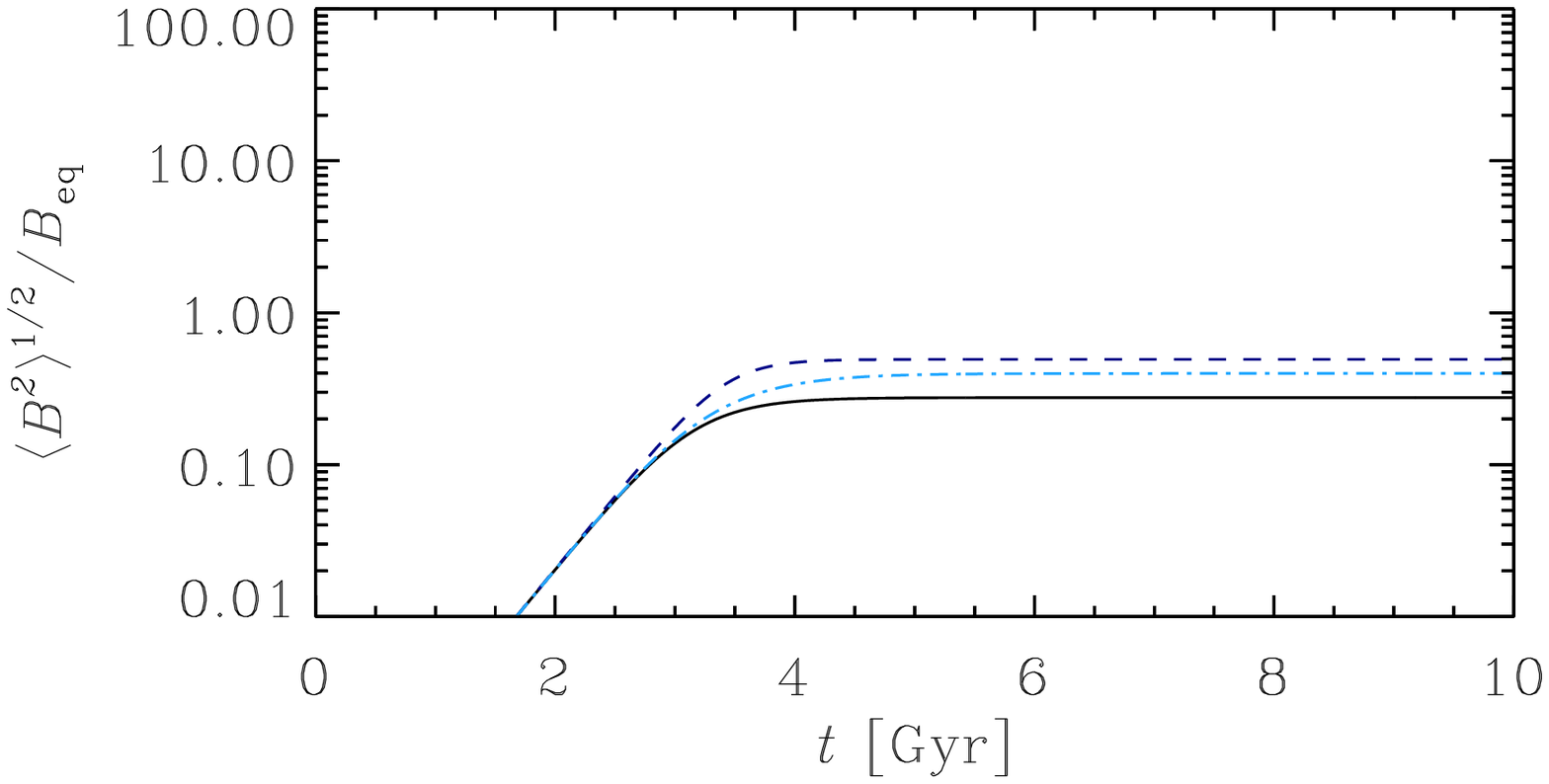}
  \includegraphics[width=58mm,clip=true,trim=31 0   32 0]{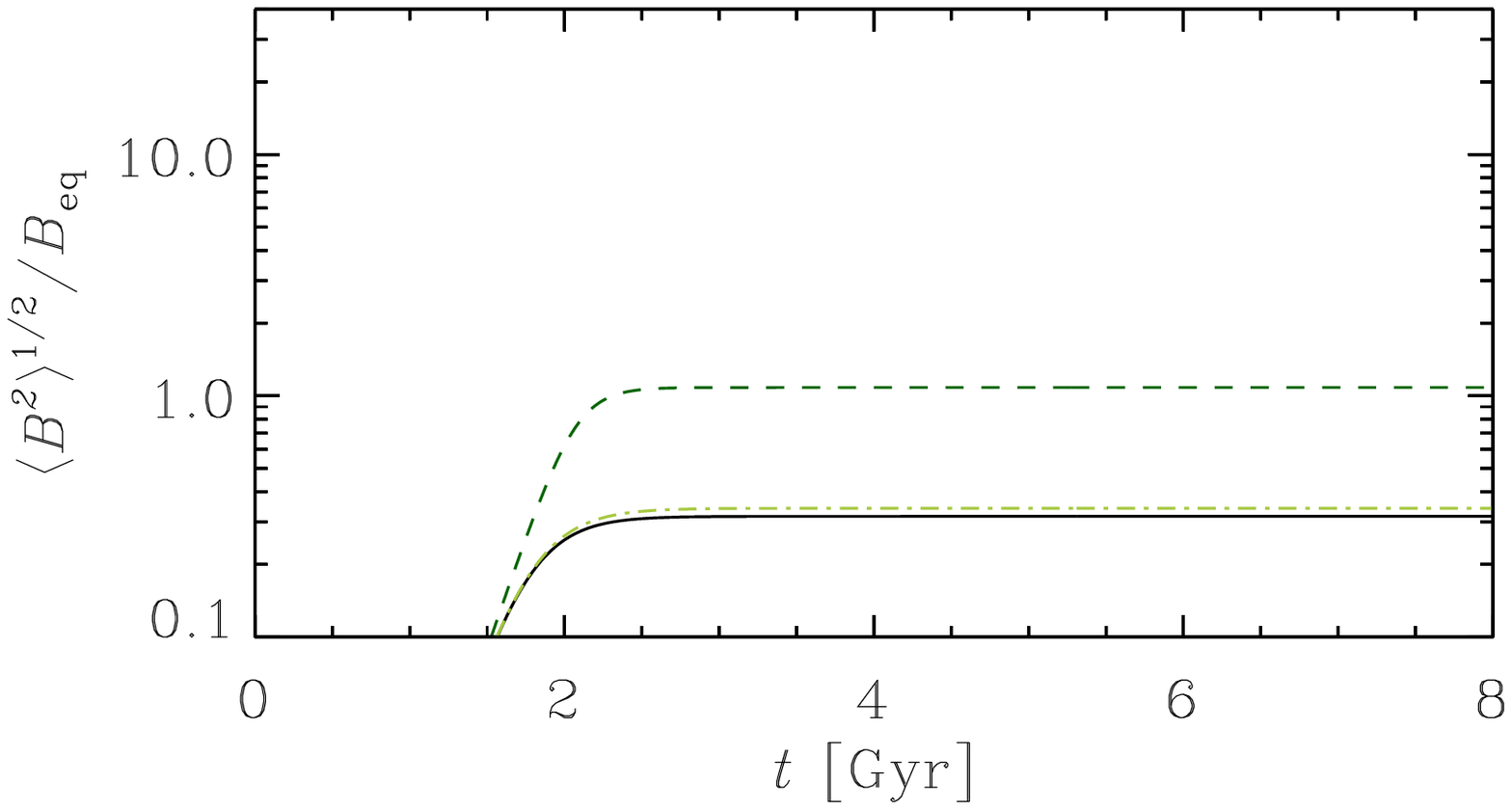}\\
  \caption{As Figure~\ref{fig:Btime_tangling_log_ss} but with $R_\kappa=0.3$ instead of $1$.
           The dash-dotted curves (no dynamical $\alpha$-quenching) are the same as in Figure~\ref{fig:Btime_tangling_log_ss}.
           \label{fig:Btime_tangling_Rkappa0o3_log_ss}
          }            
\end{figure*}

Here we explore what happens when $\xi$ is allowed to change dynamically through its coupling with the turbulent tangling of the large scale field.
Results for the rms field strength averaged across the disc are shown in Figure~\ref{fig:Btime_tangling_log_ss}.
Each panel shows three curves, each corresponding to a different assumed non-linearity.
The dashed curves show the case of dynamical $\alpha$-quenching with $\xi=\xi\f$,
and are identical to the results of Section~\ref{sec:saturation}.
The dash-dotted curves show the case that includes a dynamical $\xi$ which responds to tangling of $\meanv{B}$, 
but with $\alpha\magn=0$, so that $\alpha=\alpha\kin$.
Finally, the solid curves show the case when both the dynamical $\xi$ and dynamical $\alpha$ non-linearities are included.
The columns again correspond to Models~A, B and C of Table~\ref{tab:models}, but the rows now correspond to values of $\xi\f=\xi|_{t=0}$.
The top row shows $\xi\f=0$, the middle row $\xi\f=0.2$ and the bottom row $\xi\f=0.4$.

In all cases, except for Model~A with $\xi\f=0.4$ (bottom-left panel) where the dynamo is subcritical, 
the field eventually saturates, reaching a steady solution with even symmetry about the midplane.
(We remind the reader that the large growth and saturation times for Model~A are an artifact of our model being local;
global solutions would result in much faster growth at the radius meant to be represented by Model~A,
but its saturated solution is expected to resemble closely the local saturated solution; see Sections~\ref{sec:emf} and \ref{sec:saturation}.)

\subsection{Relative importance of the new quenching mechanism}
Figure~\ref{fig:Btime_tangling_log_ss} shows that the two prescriptions for quenching are generally competitive.
In some cases, the dynamical $\xi$-quenching is stronger than the dynamical $\alpha$-quenching,
while in other cases the reverse is true.
This can be seen by comparing the saturation levels for the dynamical $\xi$ with $\alpha\magn=0$ case (dash-dotted)
and the dynamical $\alpha$ with $\xi=\xi\f$ case (dashed) in each panel.
In some panels $\xi$-quenching leads to a smaller saturation strength, implying that this form of quenching is strongest,
while in other panels the reverse is true.
In the bottom row (Models~B and C, middle and right-most panels), where $\xi\f=0.4$, 
the solid and dash-dotted lines almost coincide, 
which implies that dynamical $\xi$-quenching is the dominant quenching mechanism.
On the other hand, when $\xi\f=0$ for Models~B and C, $\xi$-quenching alone without $\alpha$-quenching
leads to field strengths that are probably unreasonably large, and saturation times that are also very large,
so the results for those cases seem to be less physical.
For incompressible turbulence, we expect $\xi\f\sim0.3$ from saturation of the small-scale dynamo at early times when the large-scale field is still weak.%
\footnote{This number is expected to be different for compressible turbulence, 
and likely also depends on other factors such as the nature of the turbulence driving
\citepalias[see][and references therein]{Chamandy+Singh17}.}
Thus, this new non-linear feedback mechanism involving $\xi$, $\delta'$ and turbulent tangling
can lead to quenching and saturation of the large scale magnetic field at values near equipartition for realistic parameter values,
completely independently of the dynamical $\alpha$-quenching mechanism and small-scale magnetic helicity.
When both effects are included, they are generally competitive with one another in strength, 
and reinforce one another. 
Further, the dynamical $\xi$-quenching actually dominates over dynamical $\alpha$-quenching in some cases.

\begin{figure*}
  \includegraphics[width=58mm,clip=true,trim=31 55  32 0]{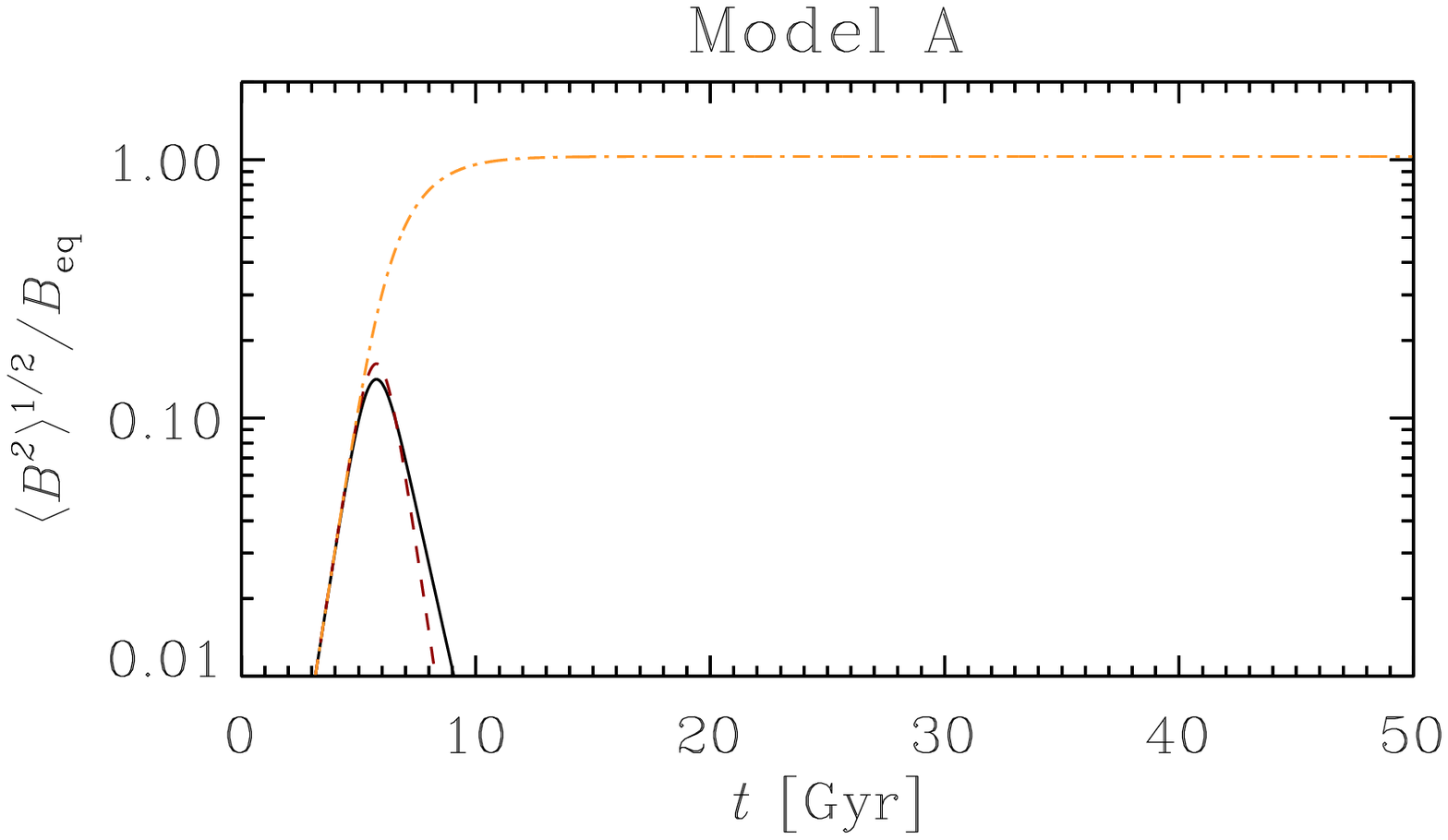}
  \includegraphics[width=58mm,clip=true,trim=31 55  32 0]{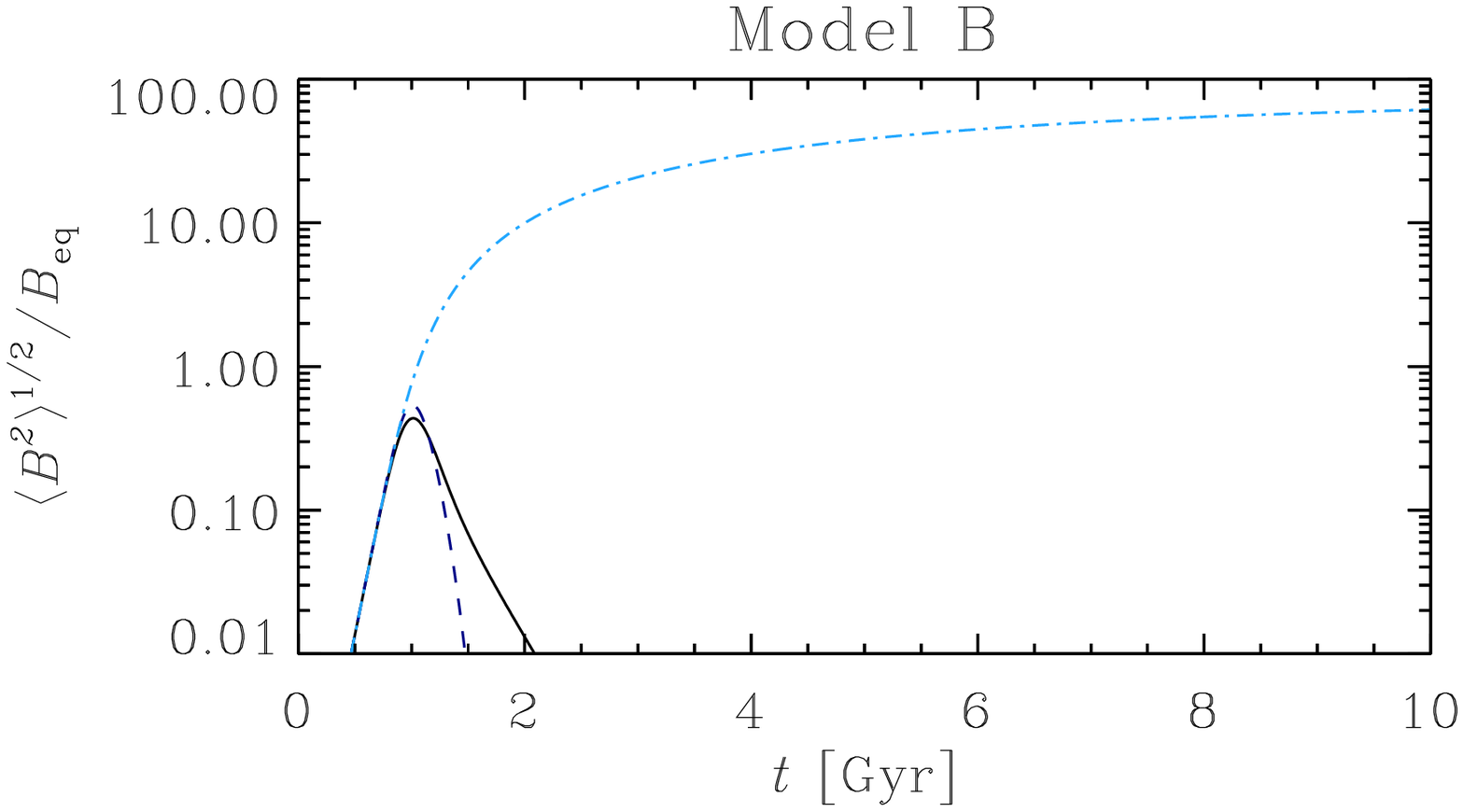}
  \includegraphics[width=58mm,clip=true,trim=31 55  32 0]{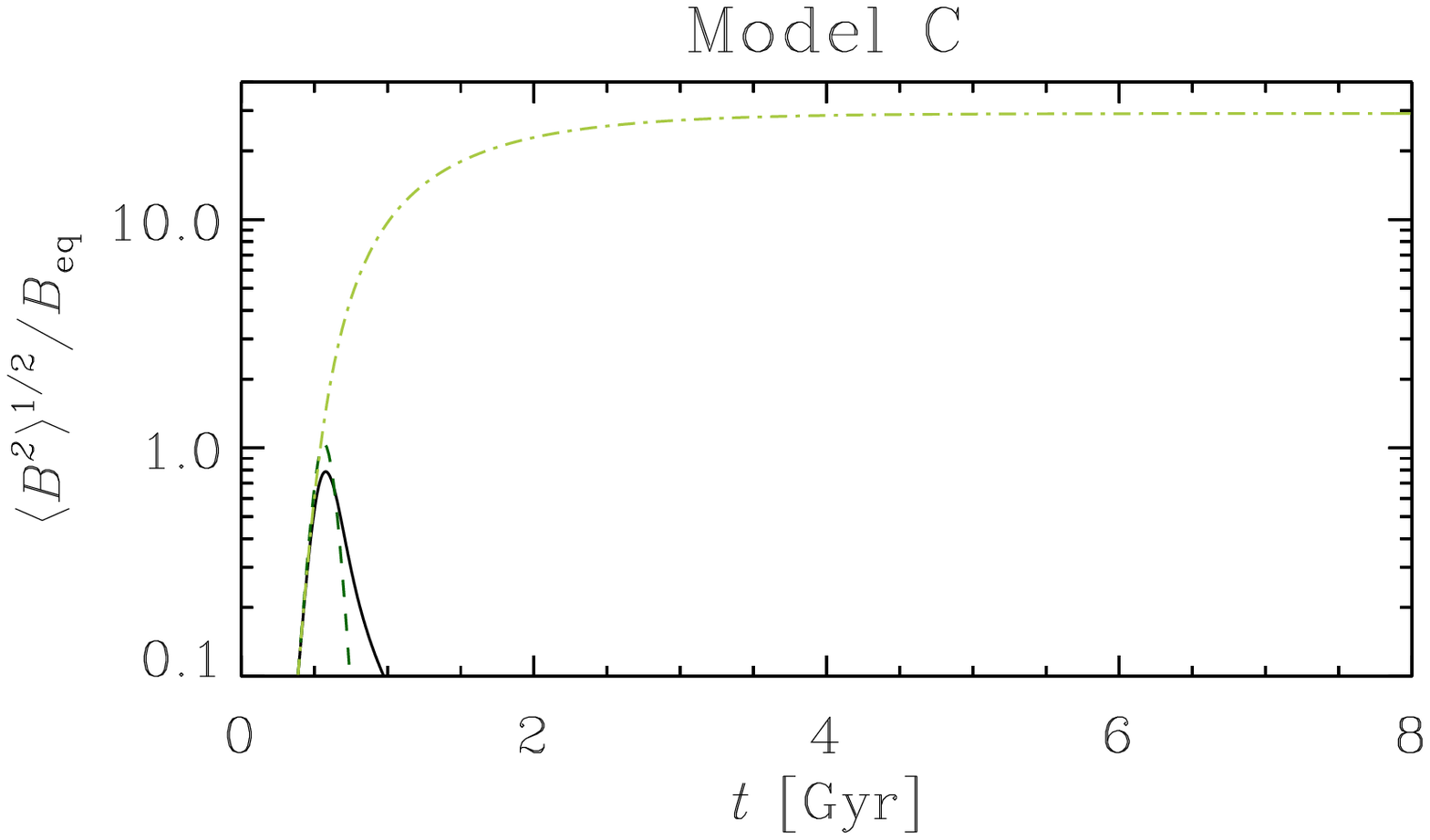}\\
  \includegraphics[width=58mm,clip=true,trim=31 55  32 0]{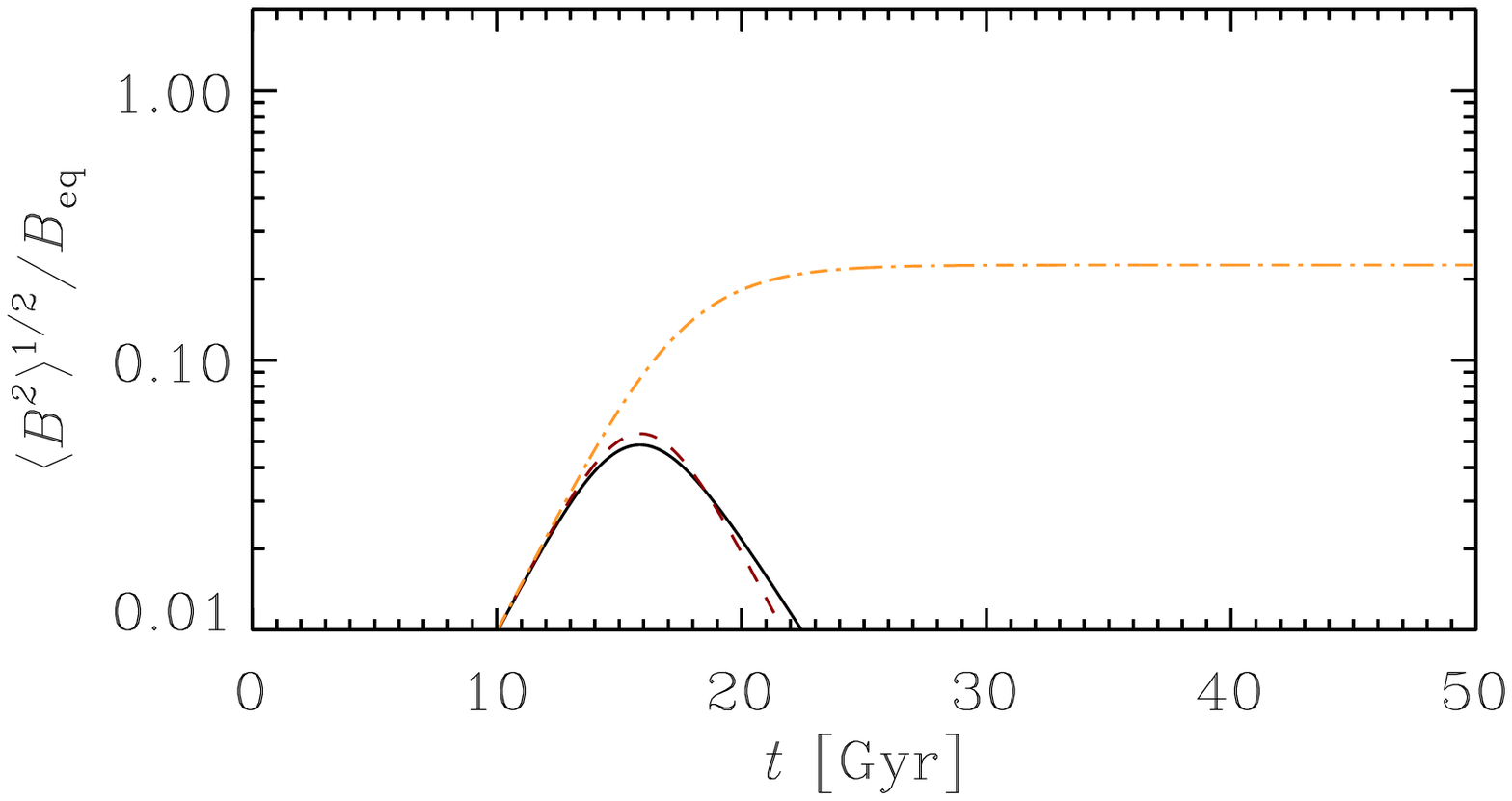}
  \includegraphics[width=58mm,clip=true,trim=31 55  32 0]{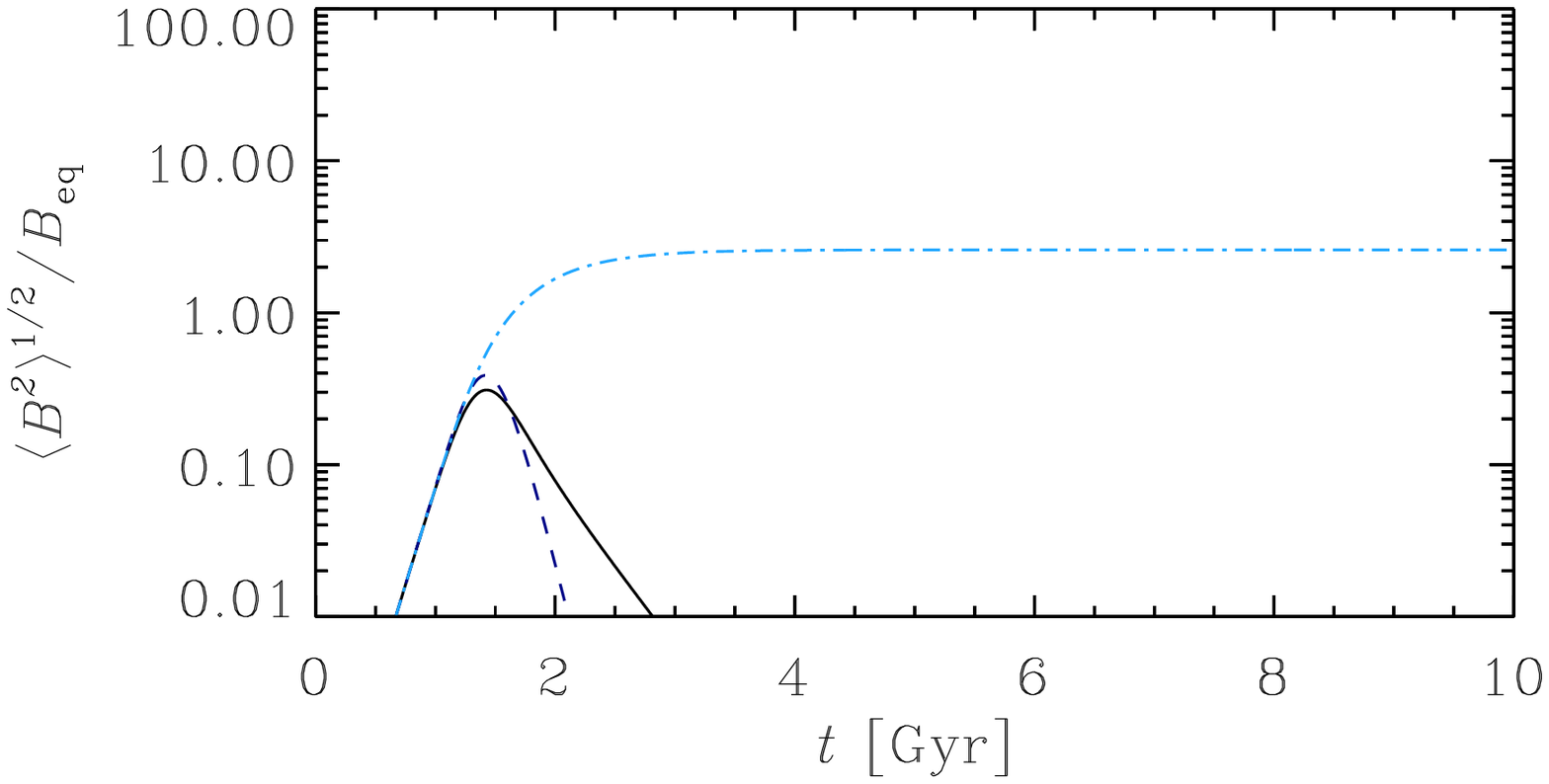}
  \includegraphics[width=58mm,clip=true,trim=31 55  32 0]{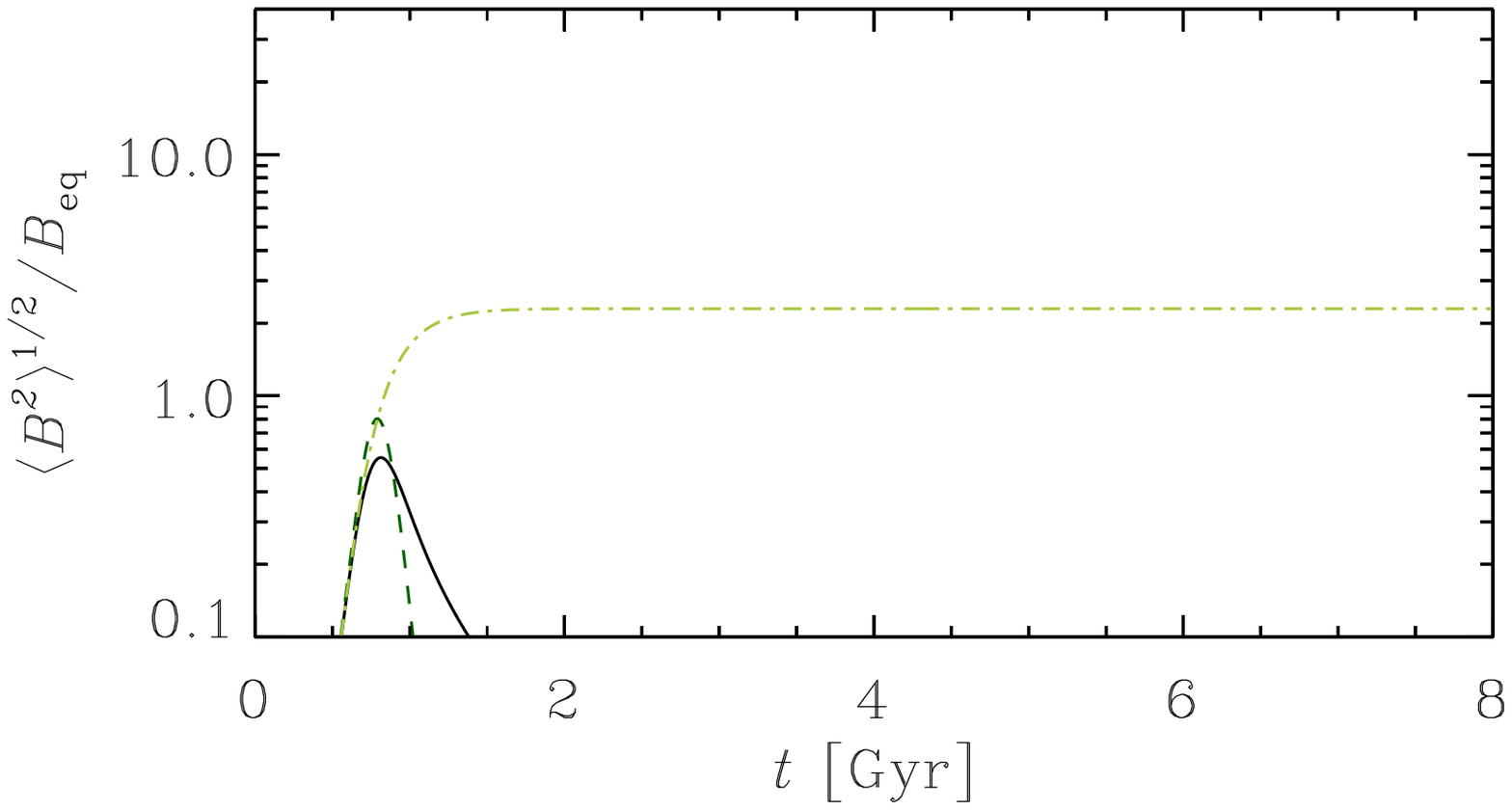}\\
  \includegraphics[width=58mm,clip=true,trim=31 0   32 0]{B_time_modelA_xi0_0o4_ss.eps}
  \includegraphics[width=58mm,clip=true,trim=31 0   32 0]{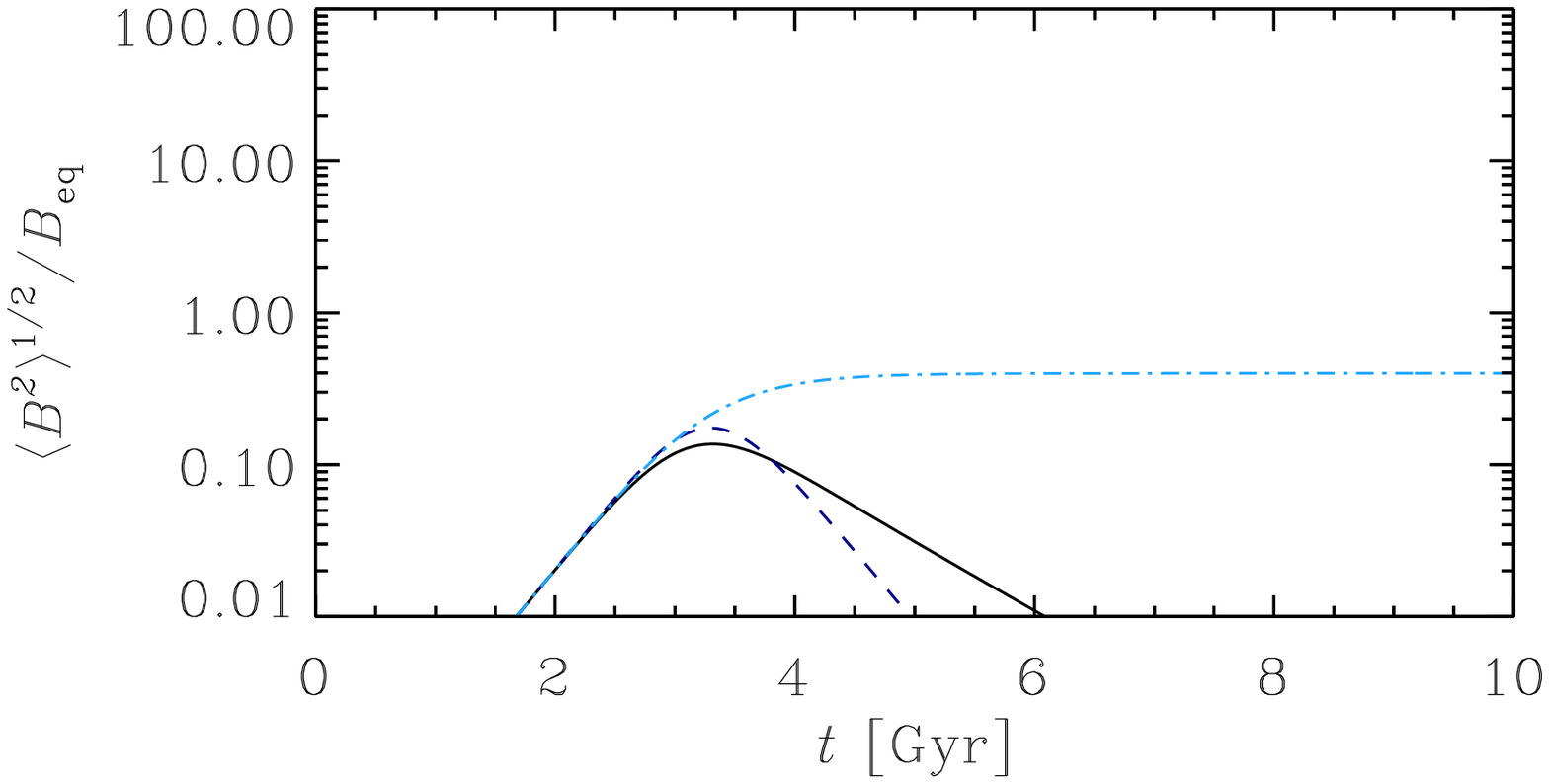}
  \includegraphics[width=58mm,clip=true,trim=31 0   32 0]{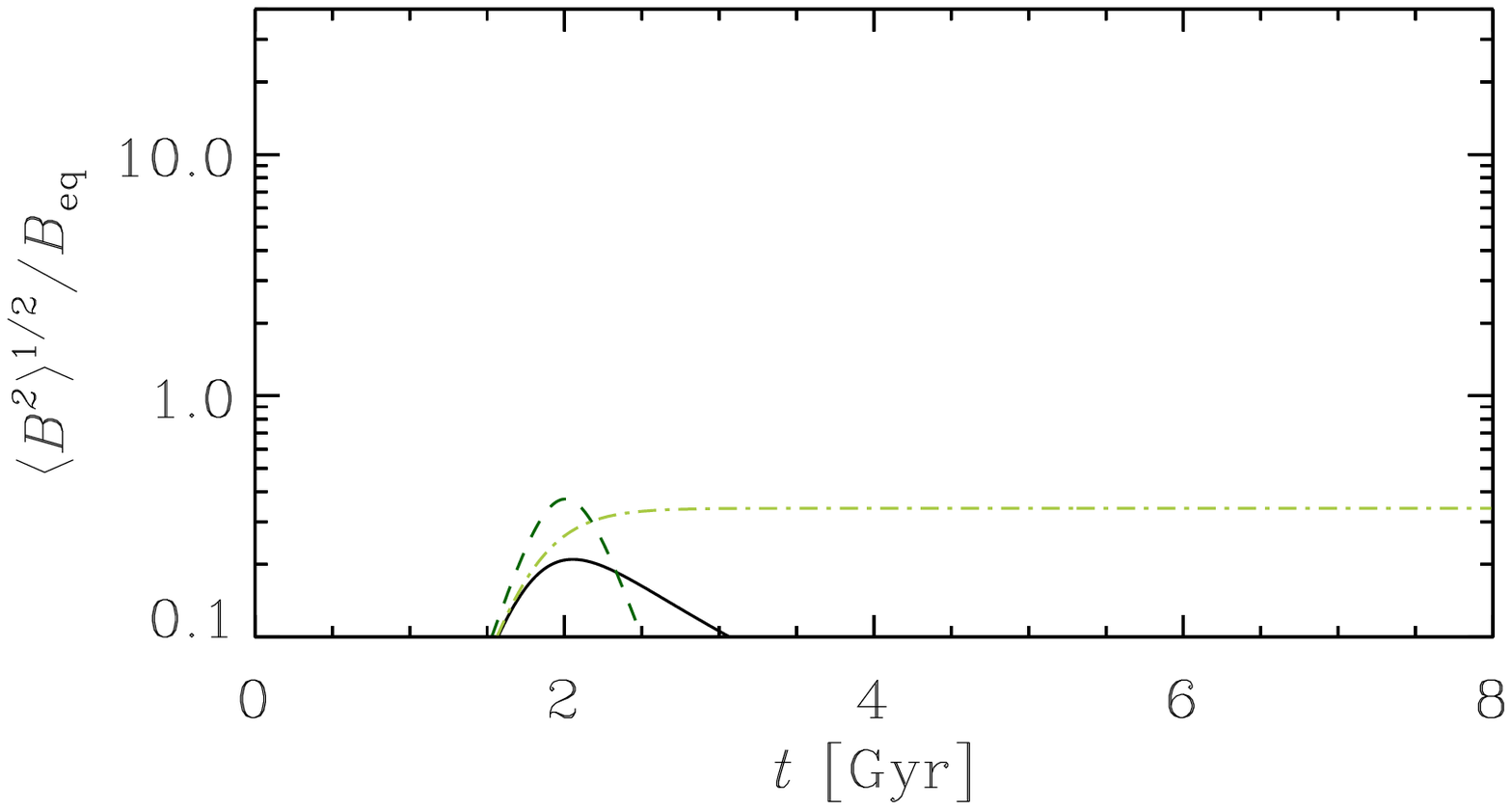}\\
  \caption{As Figure~\ref{fig:Btime_tangling_log_ss} but with $R_\kappa=0$ instead of $1$.
           Catastrophic quenching is not prevented (but is partly alleviated) by including the dynamical $\xi$ non-linearity.
           \label{fig:Btime_tangling_Rkappa0_log_ss}
          }            
\end{figure*}

We note that there are some caveats to these conclusions.  
First, we have assumed the most natural value $R_\kappa=1$.
However, DNS suggest that $R_\kappa\approx0.3$ \citep{Mitra+10}, which implies a smaller flux of $\alpha\magn$.
The saturation strength of $\meanv{B}$, assuming dynamical $\alpha$-quenching, 
is approximately proportional to this flux, and thus to $R_\kappa$ \citepalias{Chamandy+14b}.
Therefore, dynamical $\alpha$-quenching with $R_\kappa=0.3$ is stronger than for $R_\kappa=1$, 
and is thus expected to be more competitive with dynamical $\xi$-quenching.
This expectation is borne out in our simulations, and the results are illustrated in Figure~\ref{fig:Btime_tangling_Rkappa0o3_log_ss}.
However, results are remarkably similar to those with $R_\kappa=1$, which means that the dynamical $\xi$ continues to play almost
as large a role in the quenching of $\meanv{B}$ as it did for the $R_\kappa=1$ case.
However, if we instead chose to use a variant of equation~\eqref{dynamical_quenching} with $l^{-2}$ replaced by $k\f^2=(2\pi/l)^2$ \citep{Shukurov+06},
then there would be extra factors of $(2\pi)^2$ multiplying terms other than the flux term on the right-hand-side of equation~\eqref{dynamical_quenching}.
We have confirmed that this still results in the expected qualitative behaviour.
However, it leads to a much smaller saturated field strength in the pure dynamical $\alpha$-quenching case,
which implies a stronger quenching. In that case, our new dynamical $\xi$-quenching is relatively less important.
Given this kind of uncertainty in the underlying theory, it is not possible to evaluate precisely the relative importance of the two quenching mechanisms.
Generally, however, it is clear that the dynamical $\xi$-quenching mechanism can be competitive with dynamical $\alpha$-quenching,
and may even dominate in some cases.

\begin{figure*}
  \includegraphics[width=58mm,clip=true,trim=50 55  35 0]{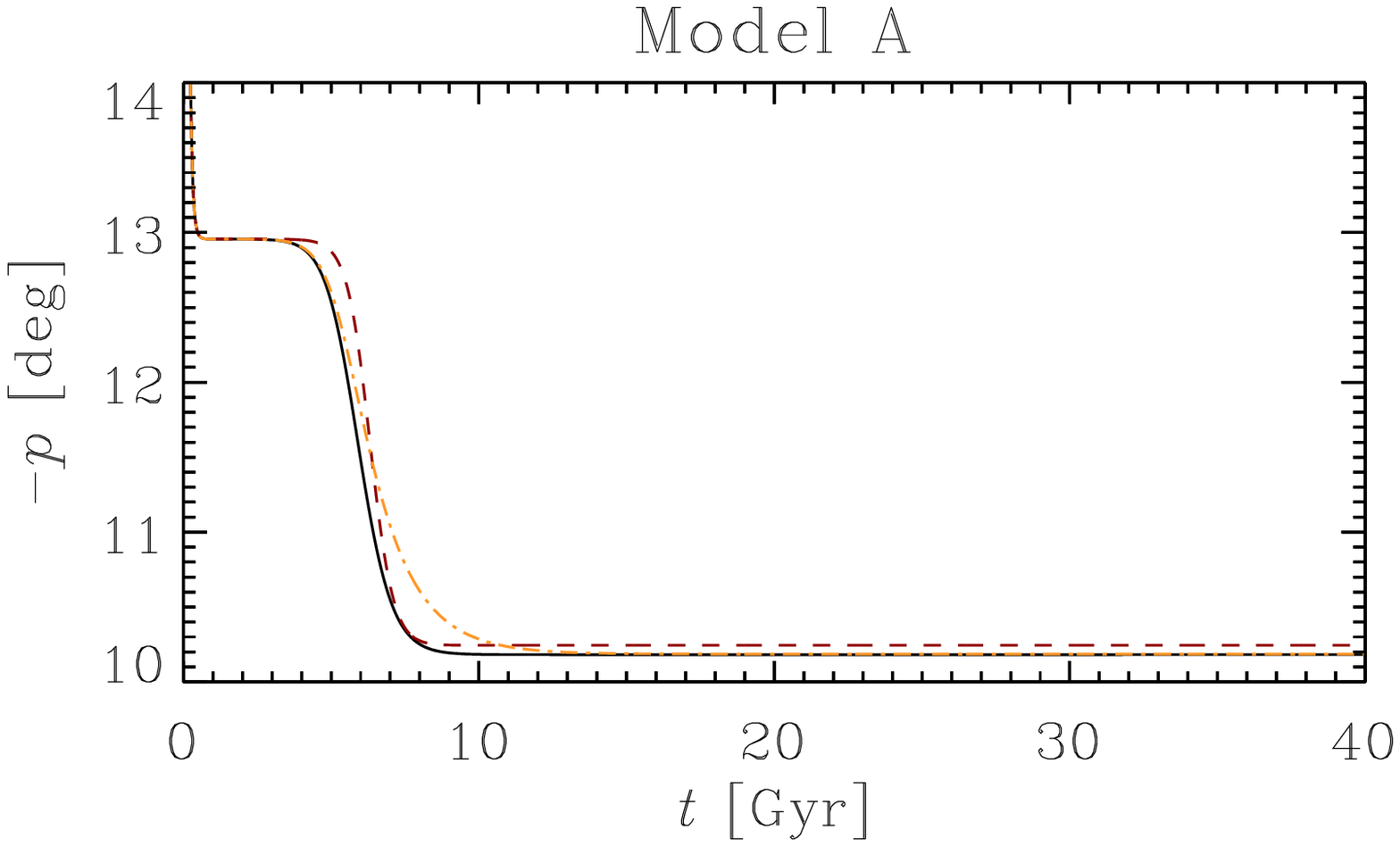}
  \includegraphics[width=58mm,clip=true,trim=50 55  35 0]{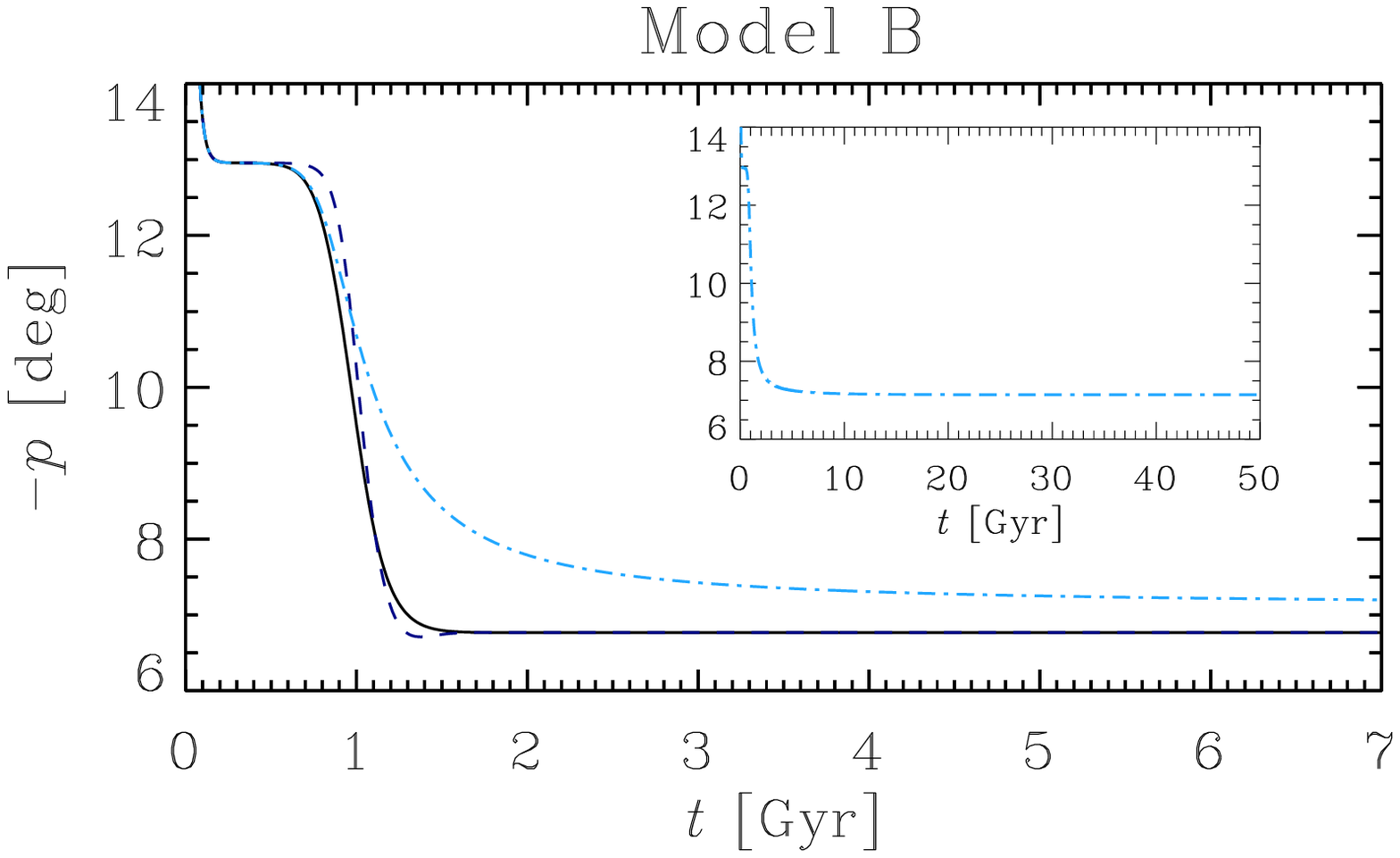}
  \includegraphics[width=58mm,clip=true,trim=50 55  35 0]{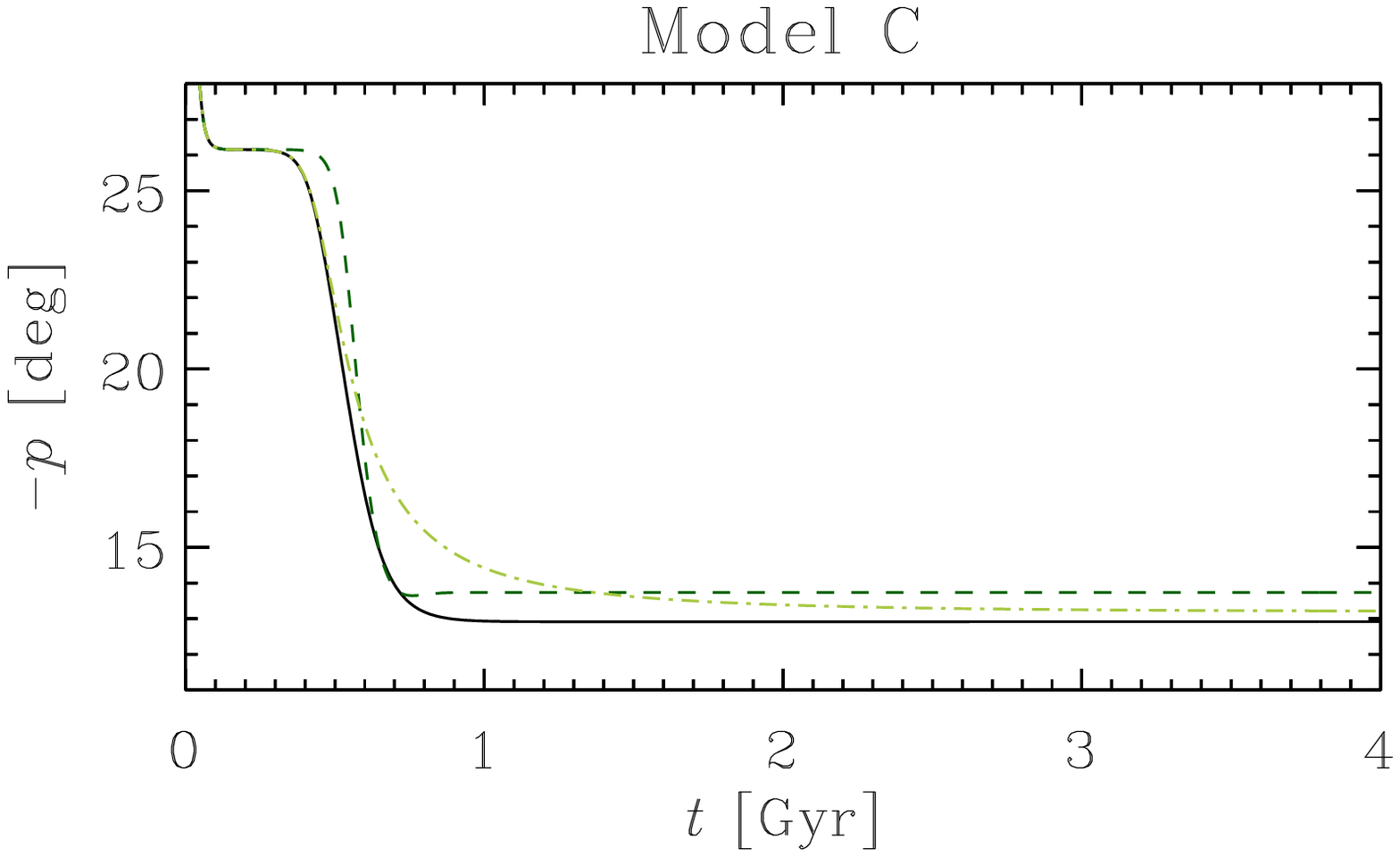}\\
  \includegraphics[width=58mm,clip=true,trim=50 55  35 0]{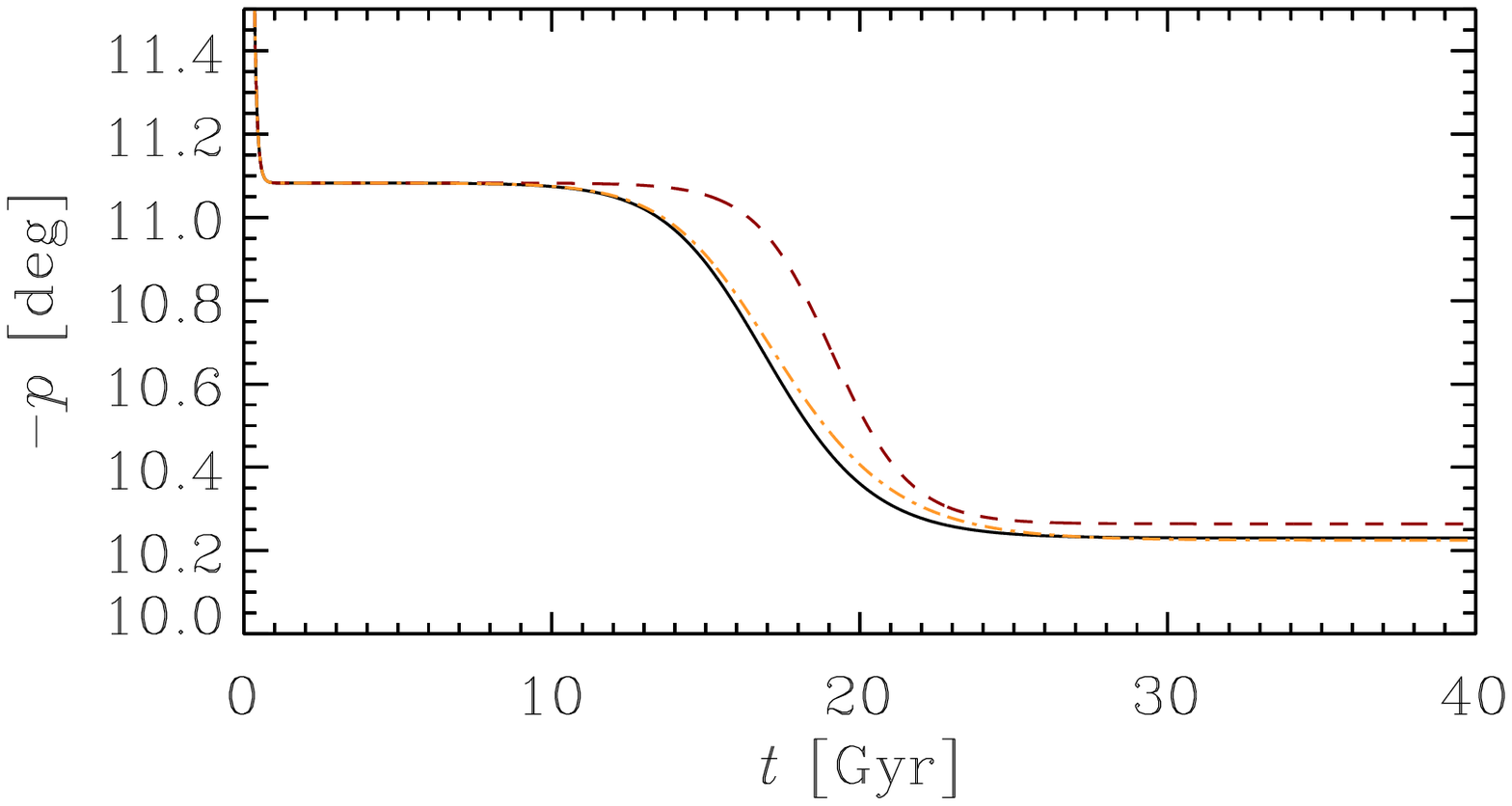}
  \includegraphics[width=58mm,clip=true,trim=50 55  35 0]{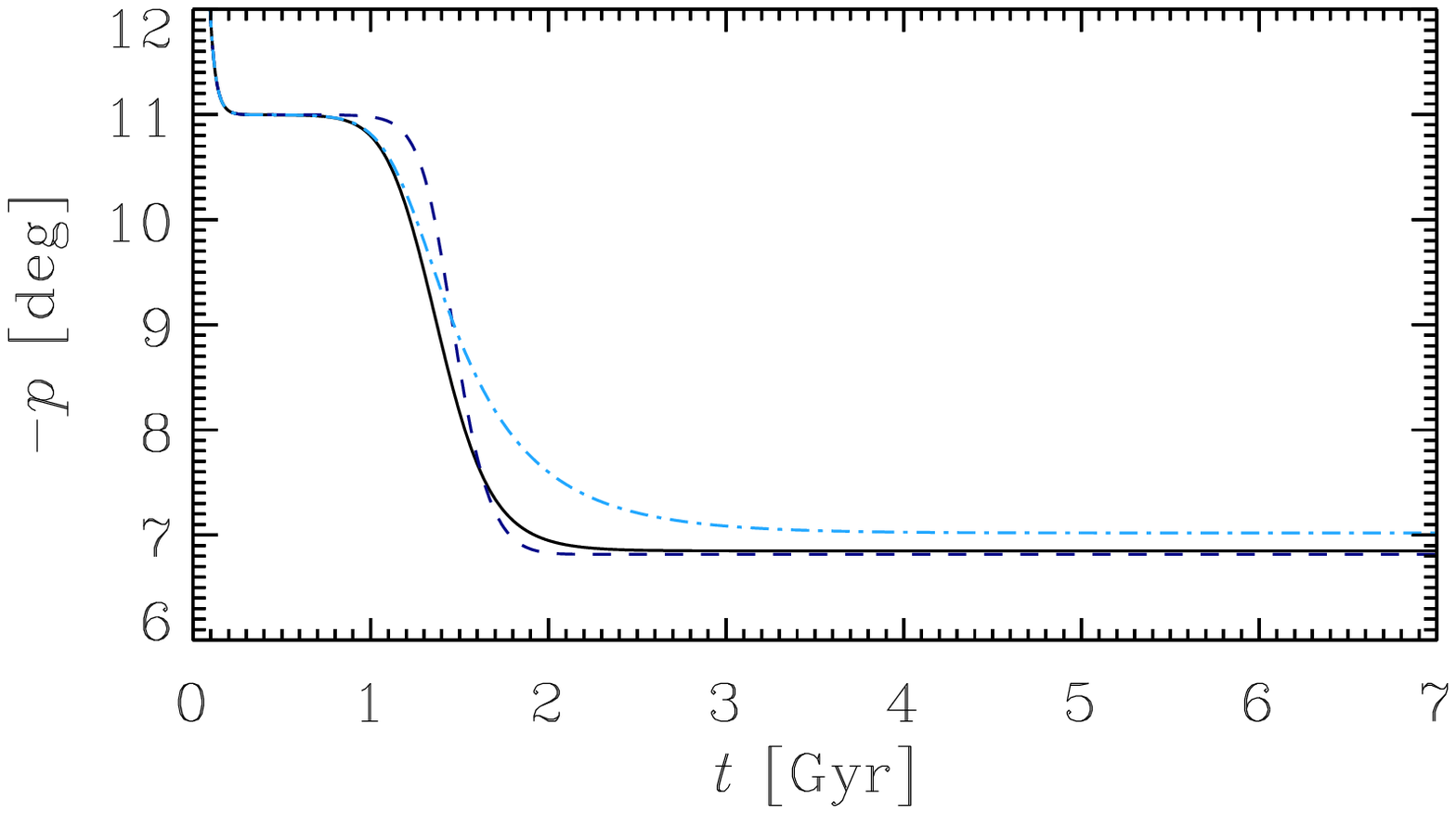}
  \includegraphics[width=58mm,clip=true,trim=50 55  35 0]{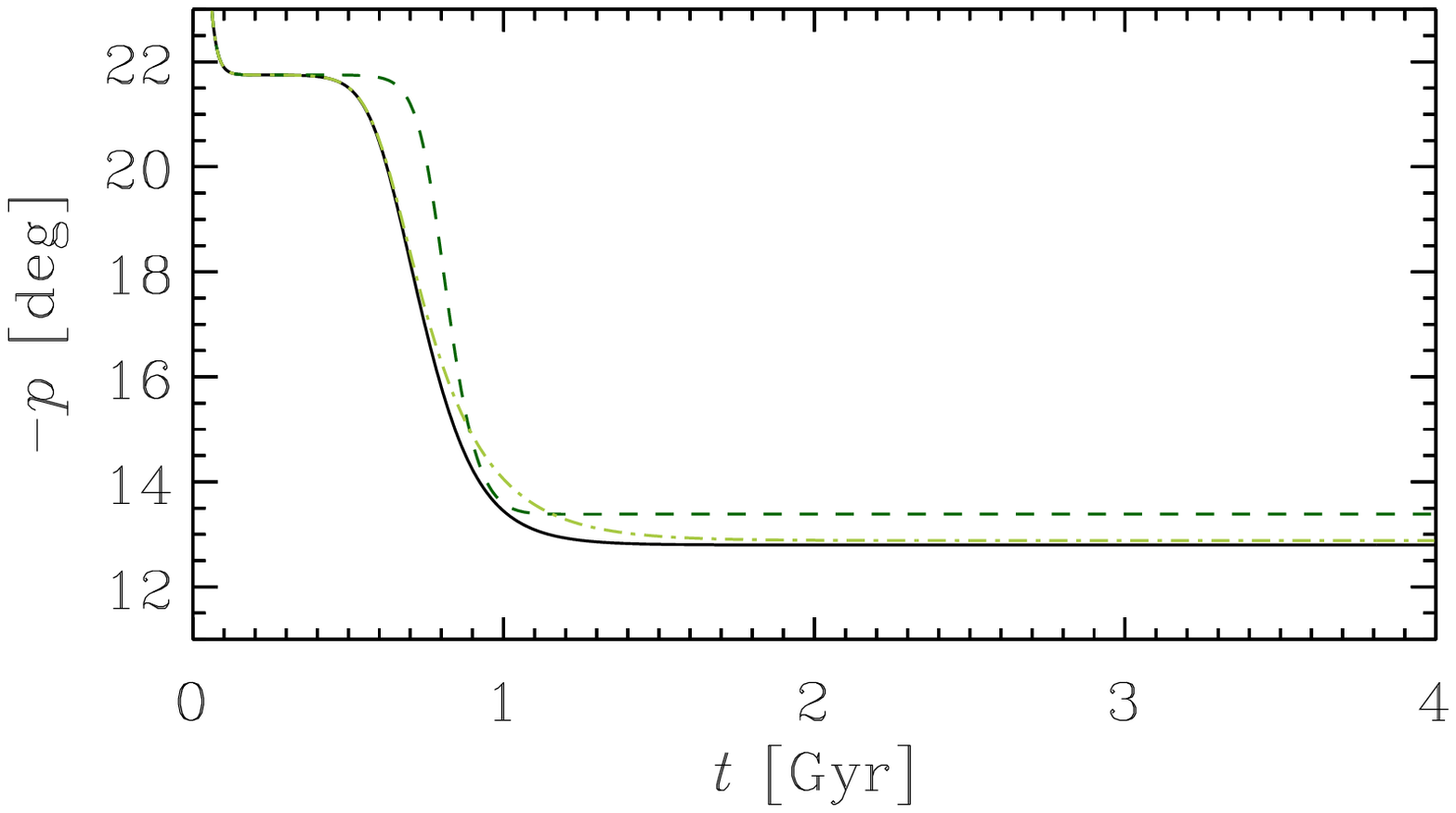}\\
  \includegraphics[width=58mm,clip=true,trim=50 0   35 0]{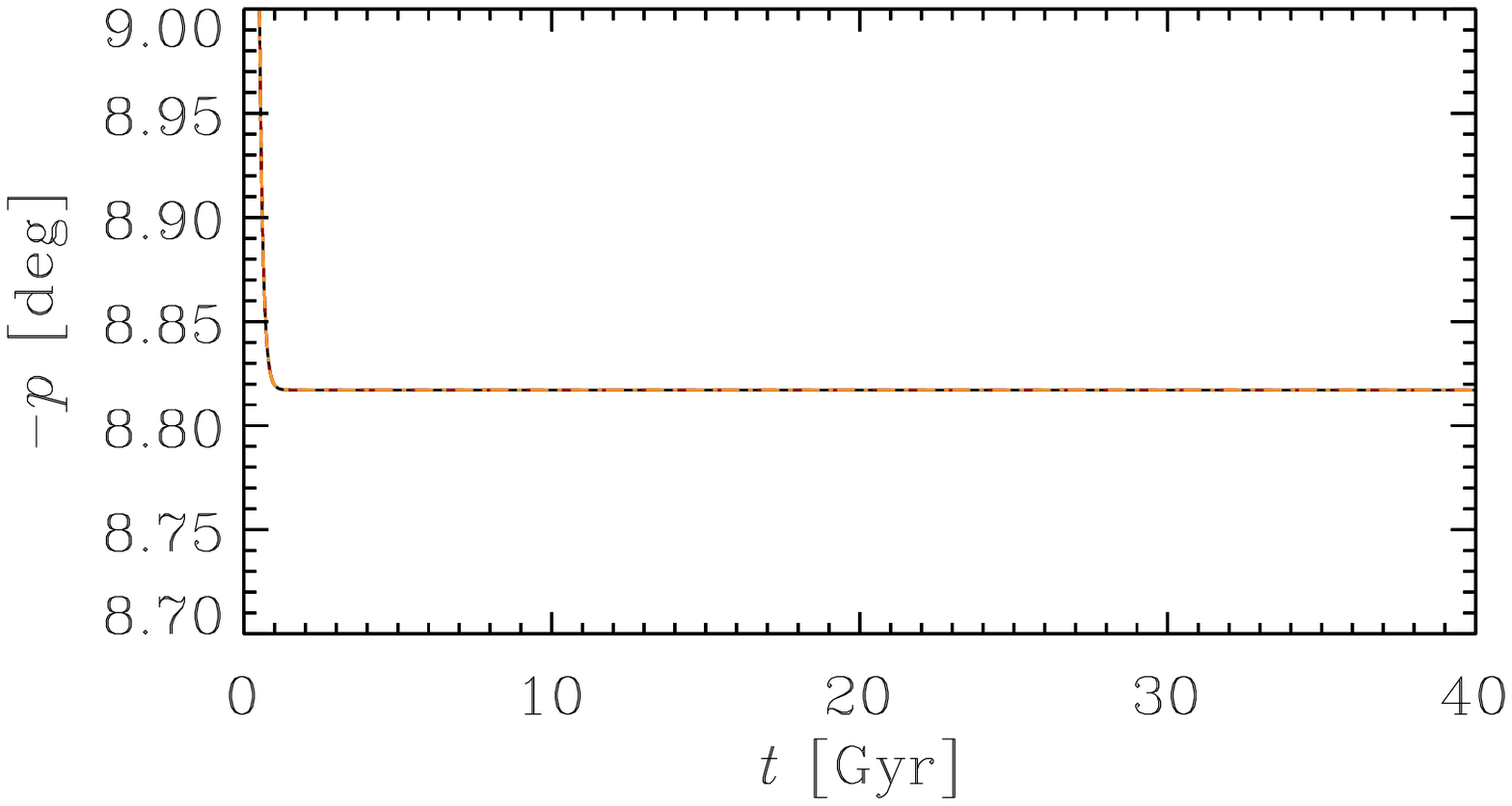}
  \includegraphics[width=58mm,clip=true,trim=50 0   35 0]{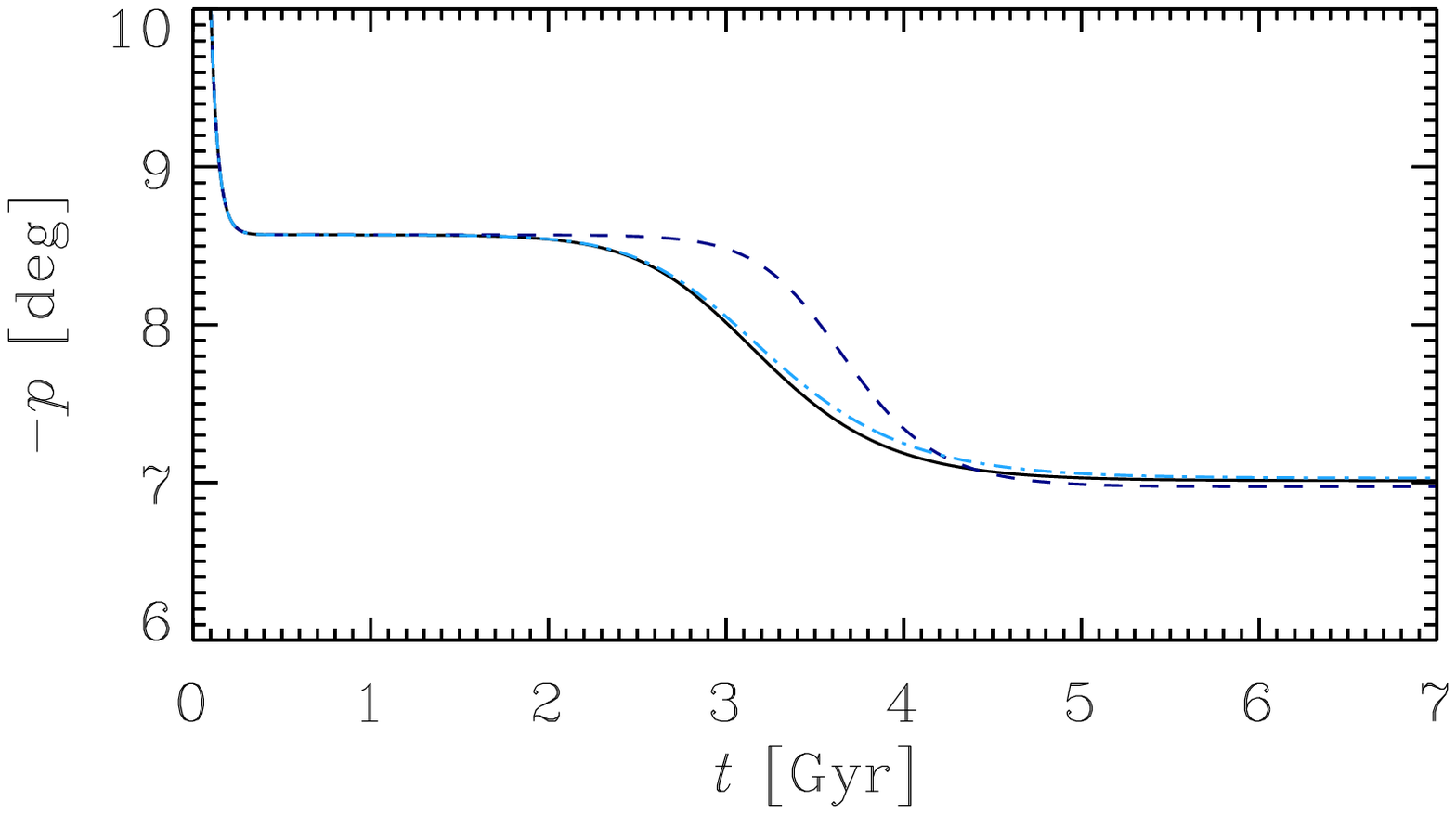}
  \includegraphics[width=58mm,clip=true,trim=50 0   35 0]{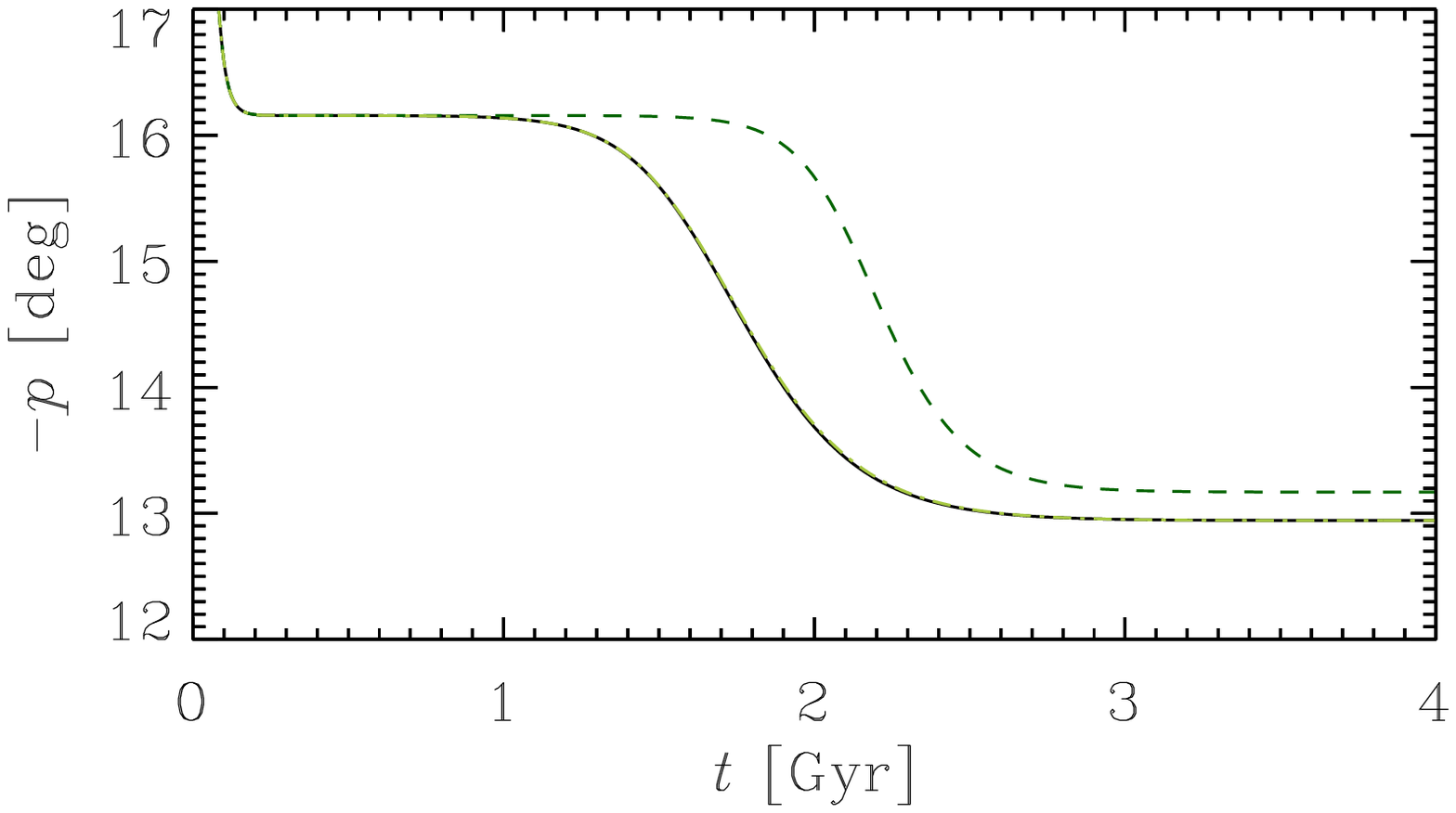}\\
  \caption{As Figure~\ref{fig:Btime_tangling_log_ss} but now showing the evolution of the 
           weighted average of the pitch angle of $\meanv{B}$ across the disc, defined in equation~\eqref{pitch}.
           \label{fig:ptime_tangling_ss}
          }            
\end{figure*}

\subsection{Implications for catastrophic quenching}
One could even be tempted to take the unconventional view that if this new effect leads to saturation when $\alpha$ is prevented from changing dynamically, 
then the  dynamical $\alpha$-quenching prescription may not be necessary.
The new effect is simpler than dynamical $\alpha$-quenching in at least one way,
namely the feedback stems from the presence of small-scale magnetic energy, rather than small-scale current (or magnetic) helicity.
By \textit{Occam's razor}, this would suggest that the new mechanism may be more appealing than dynamical $\alpha$-quenching.
However, we believe that such a point of view would be ill-conceived. 
Rather than being ad hoc, dynamical $\alpha$-quenching arises as a consequence of magnetic helicity conservation 
\citep[][see \citetalias{Brandenburg+Subramanian05a} for a review]{Field+Blackman02,Blackman+Brandenburg02,Subramanian02,Blackman+Field02}.
Therefore, it seems unavoidable. 
But could a dynamical $\xi$ somehow obviate the need for mean small-scale magnetic helicity fluxes, 
which in dynamical $\alpha$-quenching theory, are responsible for the alleviation of 
`catastrophic' quenching, allowing the mean field to saturate at near-equipartition levels?
In other words, by including the new dynamical $\xi$ effect along with dynamical $\alpha$-quenching, 
but setting the flux of $\alpha\magn$ to zero, can we still obtain saturation at near-equipartition levels?

To answer this question, we performed simulations similar to those discussed above, but we set $R_\kappa=0$.
Results are shown in Figure~\ref{fig:Btime_tangling_Rkappa0_log_ss}.
Clearly, catastrophic quenching of the dynamo is \textit{not} prevented, but it is partially alleviated,
in the sense that the dynamical $\xi$ effect leads to a longer decay time of the field in the non-linear regime (compare dashed and solid lines).
Dynamical $\xi$-quenching prevents the field strength from peaking at large values, for which catastrophic quenching is more effective,
as can be seen from equation~\eqref{dynamical_quenching} with the flux term set to zero and $\Rm\gg1$.
Thus, the magnetic R\"{a}dler effect along with turbulent tangling of the large-scale field 
can weaken the catastrophic quenching, but not prevent it.

\subsection{Observational significance}
\label{sec:observational}
It is also interesting to ask what would be the observable consequences of the new proposed mechanism.
The evolution of the field strength would be affected, 
but there is considerable theoretical as well as observational uncertainty in the determination of the field strength.
The magnetic pitch angle tends to be better constrained by observations and is also predicted with higher confidence by the theory,
compared to the field strength \citep{Chamandy+16}.
Thus, in Figure~\ref{fig:ptime_tangling_ss} we compare the evolution of the pitch angle, given by equation~\eqref{pitch},
for the various types of quenching.
Here we plot results for $R_\kappa=1$.  Panels and curves correspond to those of Figure~\ref{fig:Btime_tangling_log_ss}.
We see that for each set of parameter values, the different quenching prescriptions lead to values of the saturated pitch angle
that differ by less than $1^\circ$.
Since pitch angles observed in nearby galaxies tend to have uncertainties of a few degrees, 
pitch angle observations cannot distinguish between the different models for the dynamo non-linearity
(but see Section~\ref{sec:profiles} below, where we discuss a possible exception).
And what good is a new model, one might ask, if it does not make testable predictions that differentiate it from another model?
However, we see this apparent weakness as a strength, because it means that there exist base quantitative predictions of dynamo theory
that are robust to variations in what can at this stage aptly be described as \textit{details} of the models.
This provides a first order test of the theory, while other diagnostics such as the field strength could provide higher order tests in the future,
once the data and models have improved.
Given that quantitative evidence for mean-field galactic dynamo theory is still fragmentary and incomplete,
such first order tests are very valuable, especially when they can be carried out using existing observational technology and theoretical tools,
as is the case here.

\begin{figure*}
  \includegraphics[width=58mm,clip=true,trim=-5 55  30 0]{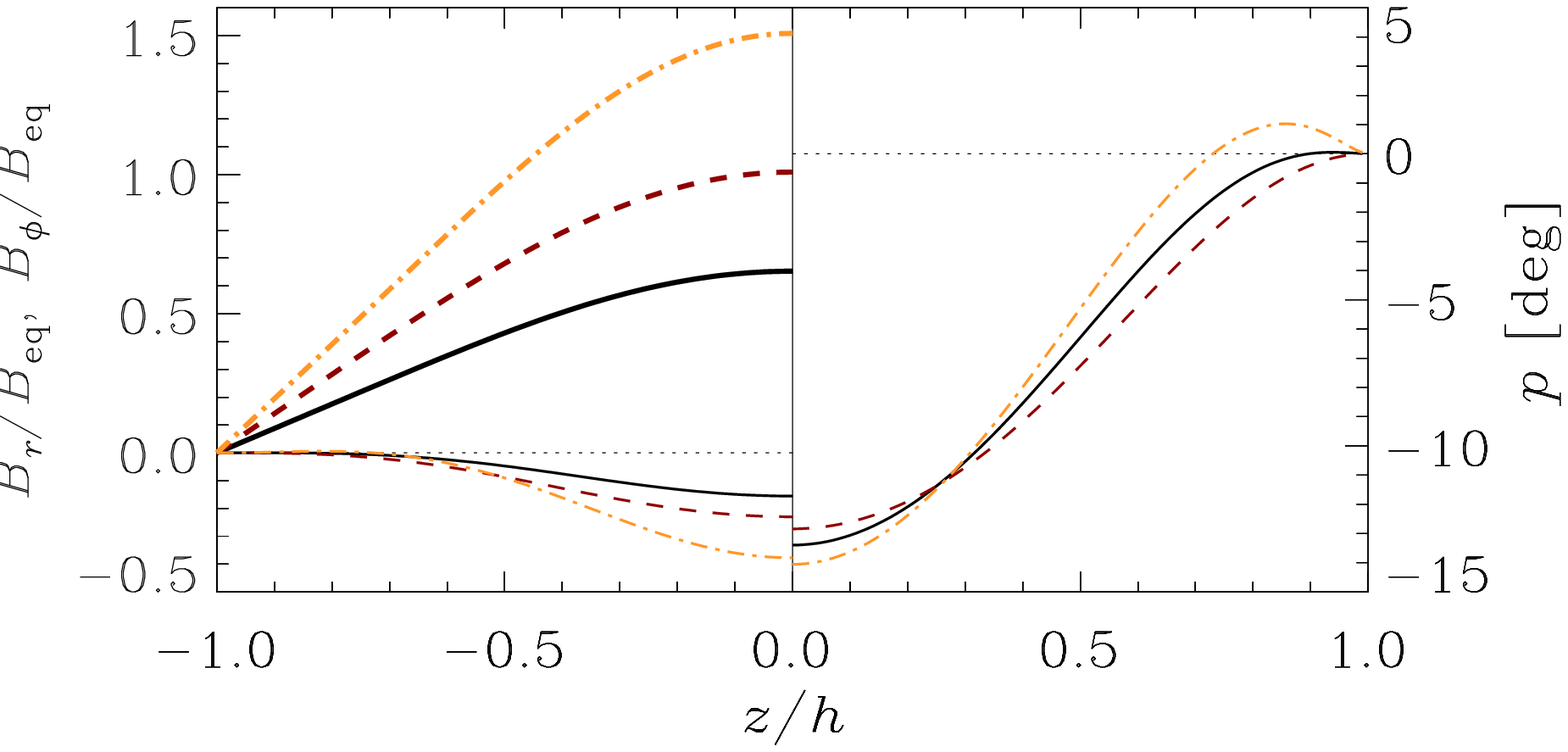}
  \includegraphics[width=58mm,clip=true,trim=-5 55  30 0]{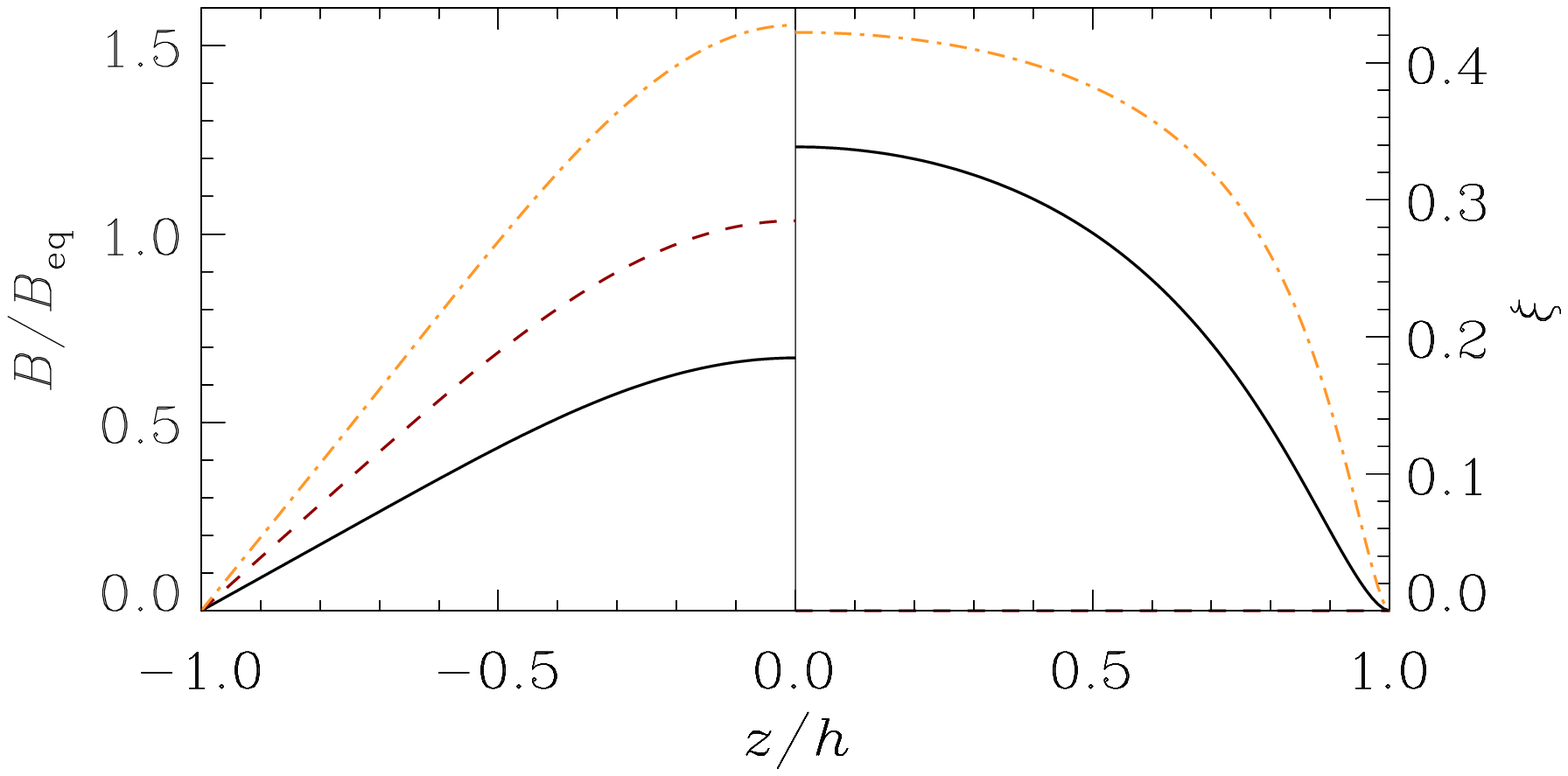}
  \includegraphics[width=58mm,clip=true,trim=-5 55  30 0]{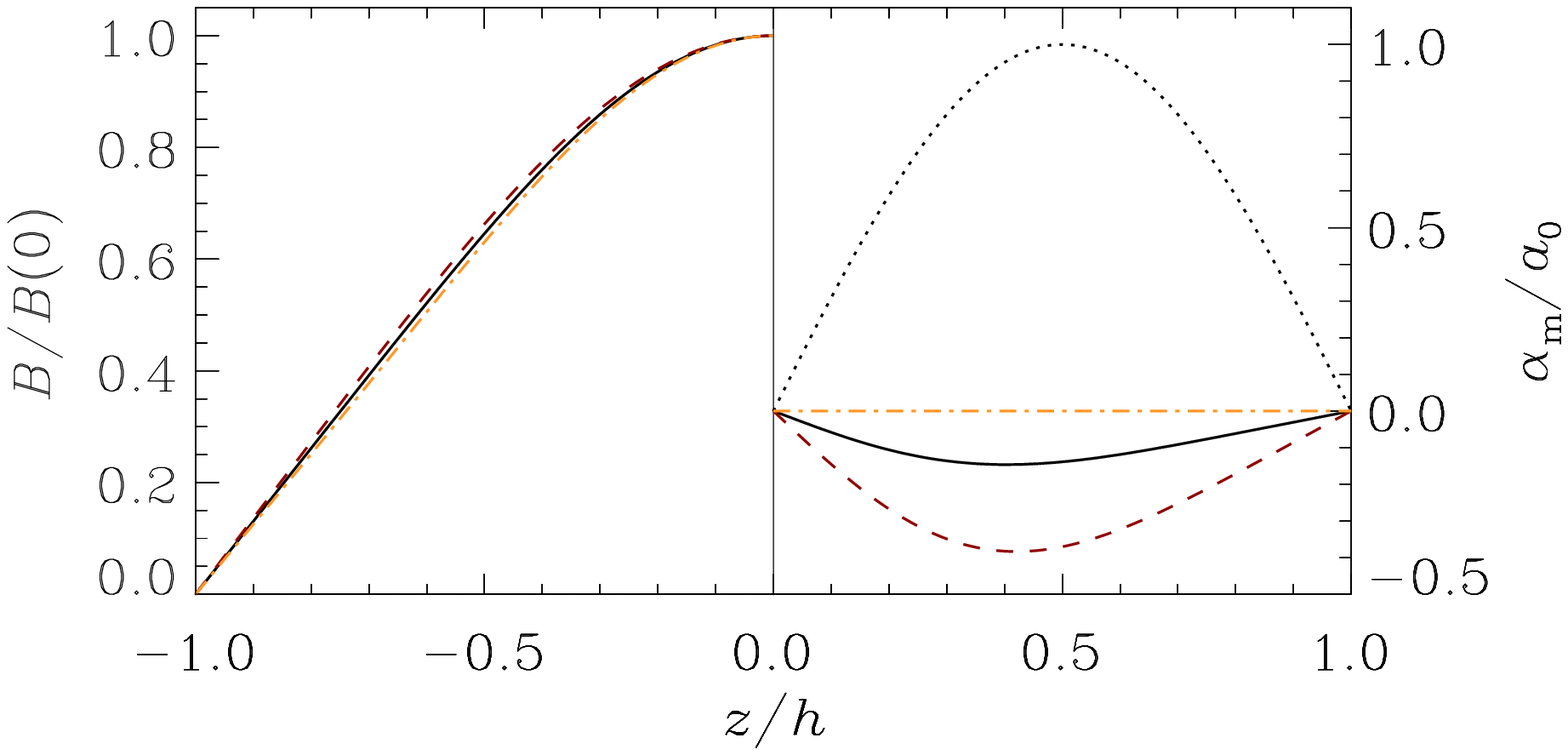}\\
  \includegraphics[width=58mm,clip=true,trim=-5  0  30 0]{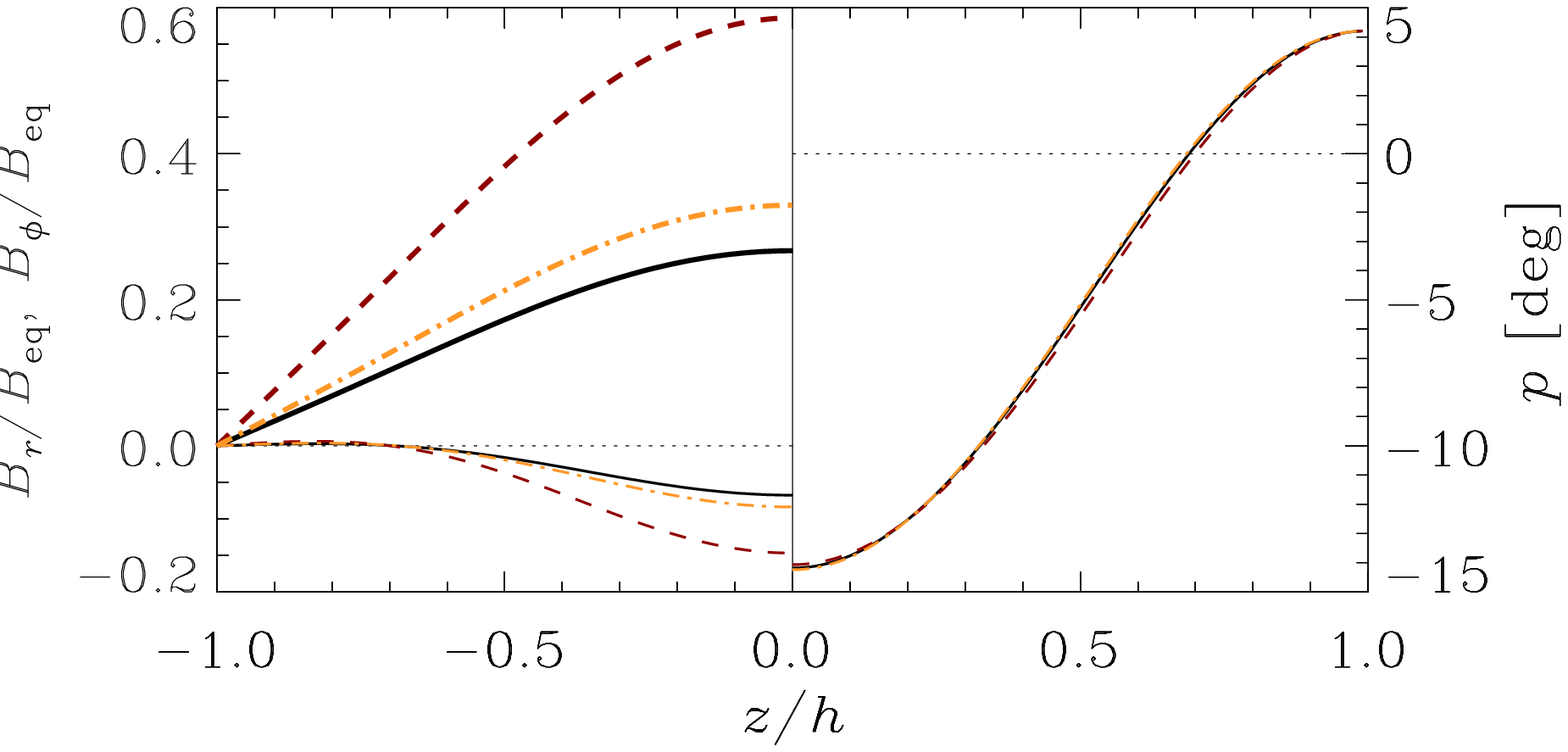}
  \includegraphics[width=58mm,clip=true,trim=-5  0  30 0]{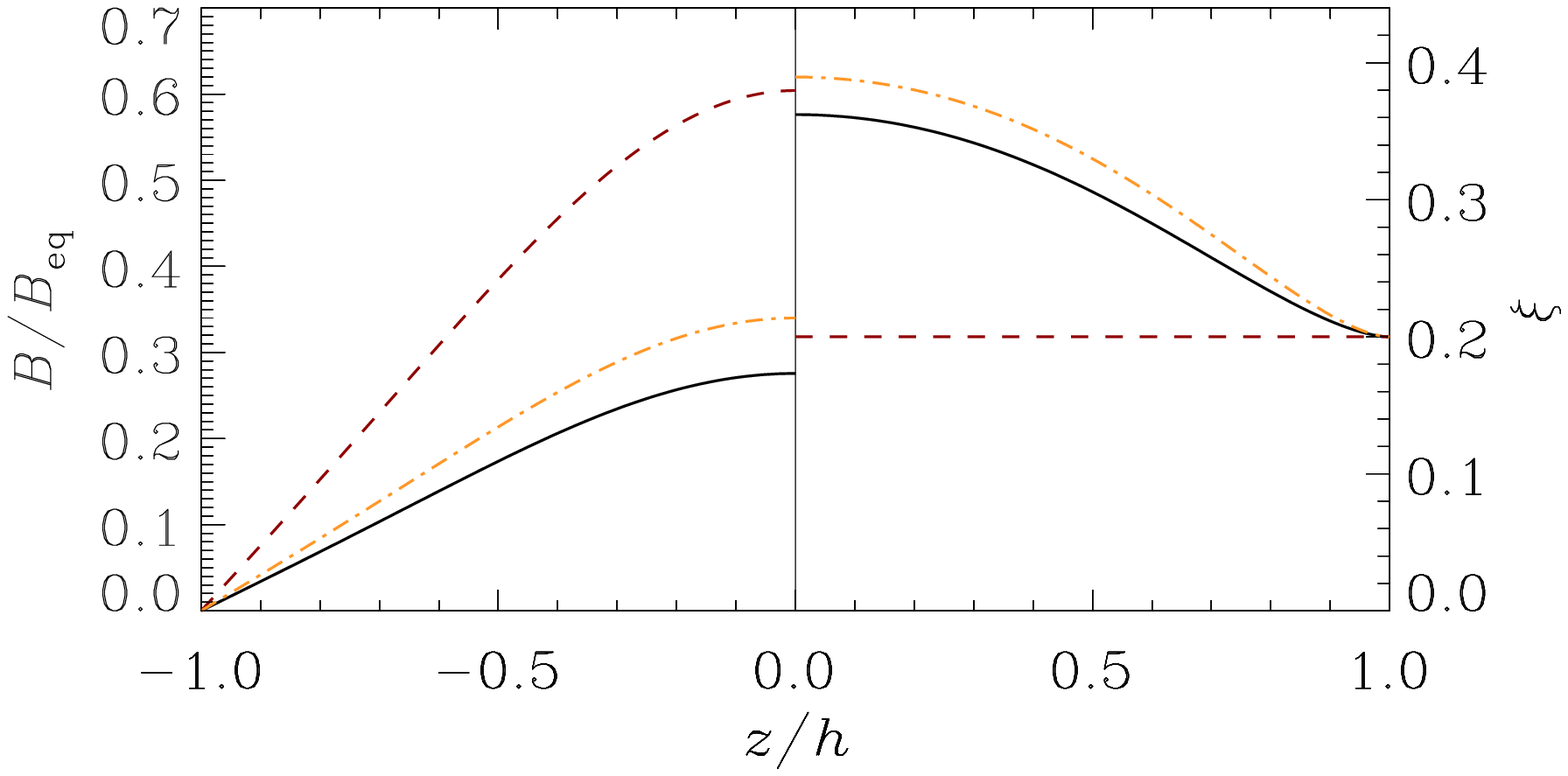}
  \includegraphics[width=58mm,clip=true,trim=-5  0  30 0]{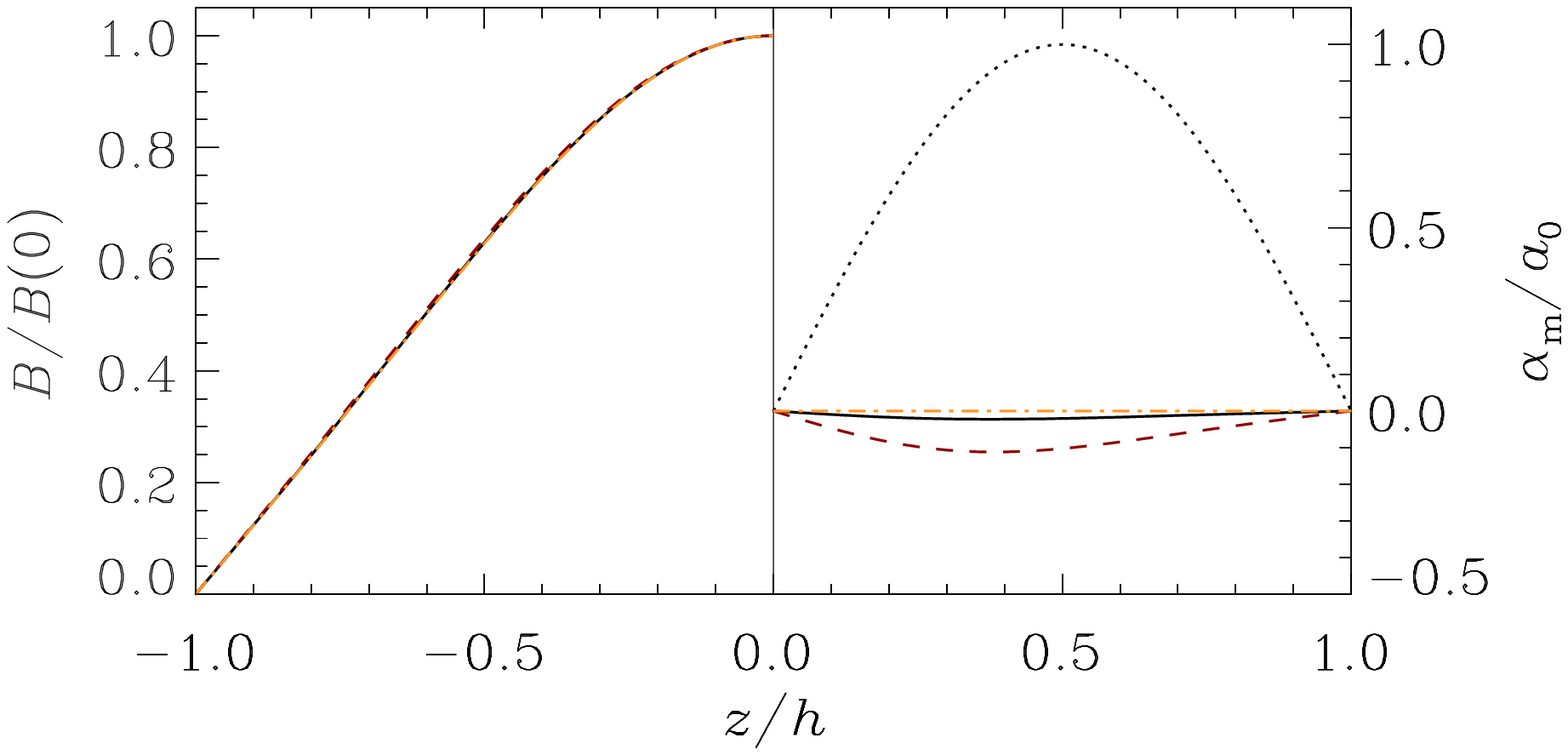}\\
  \caption{Profiles along $z$ for the saturated solutions of Model~A. 
           Profiles are shown for a single hemisphere to save space because all profiles have either even symmetry,
           or, for panels showing $\alpha\magn$ and $\alpha\kin$ only, odd symmetry about the midplane $z=0$.
           Line styles are the same as for the middle columns of Figures~\ref{fig:Btime_tangling_log_ss} to \ref{fig:ptime_tangling_ss}.
           From left to right, quantities plotted are
           (i) the components $\mbr$ (thin) and $\mbp$ (thick) normalized with respect to the equipartition strength $B\eq$;
           (ii) the magnetic pitch angle $p=\arctan(\mbr/\mbp)$ in degrees; 
           (iii) the field strength $\sqrt{\mbr^2+\mbp^2}$ normalized by $B\eq$;
           (iv) the quantity $\xi=b^2/u^2=(b/B\eq)^2$;
           (v) the field strength now normalized by its midplane value;
           (vi) the magnetic part of $\alpha$, denoted as $\alpha\magn$, normalized by $\alpha\f=\tau^2u^2\Omega/h$.
           In (vi) the kinetic part $\alpha\kin$ is denoted by a dotted line.
           Top row: $\xi\f=0$. 
           Bottom row: $\xi\f=0.2$
           \label{fig:profile_ss_modelA}
          }            
\end{figure*}

\begin{figure}
  \centering
  \includegraphics[width=75mm,clip=true,trim=-5 55  30 0]{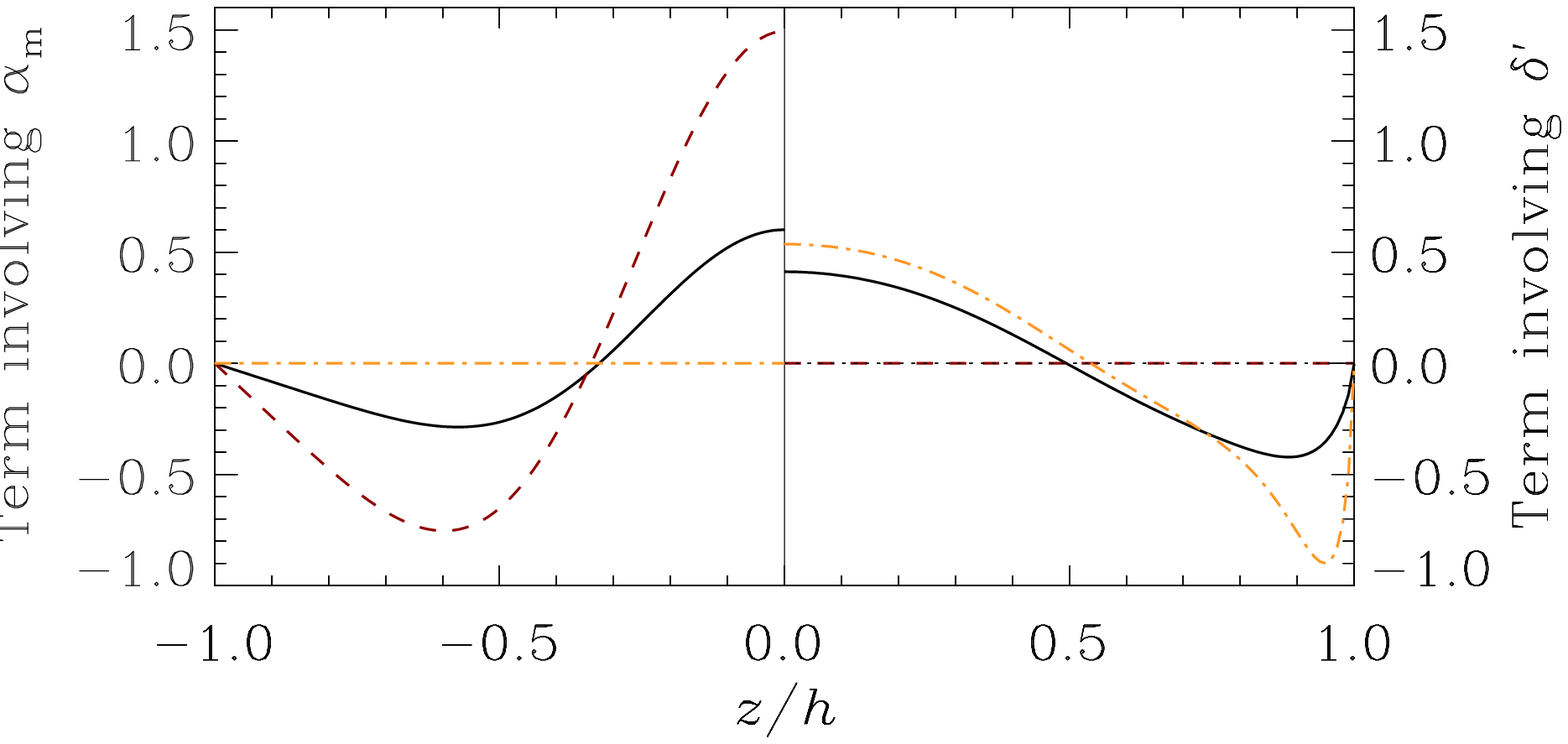}\\
  \includegraphics[width=75mm,clip=true,trim=-5  0  30 0]{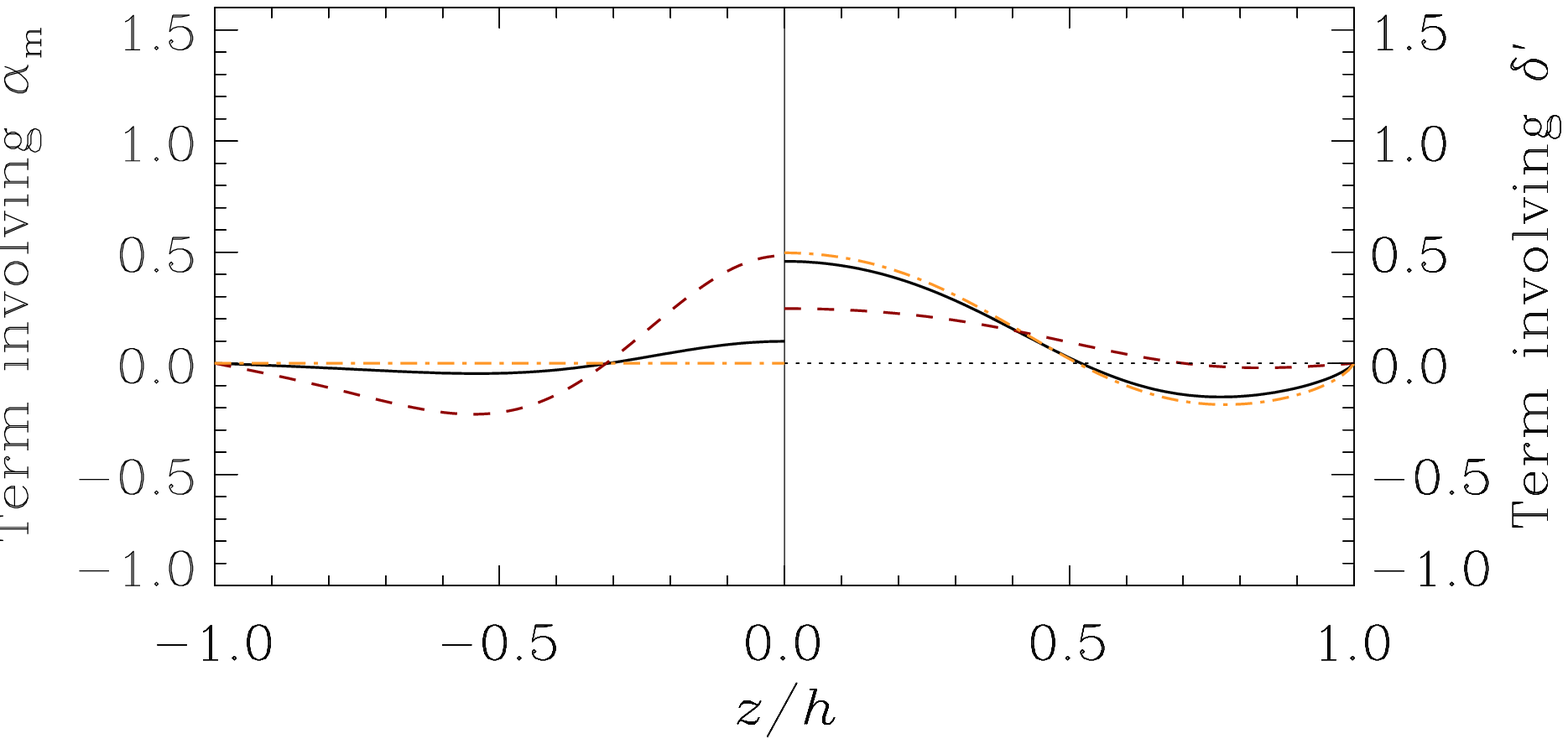}\\
  \caption{Similar to Figure~\ref{fig:profile_ss_modelA} but now comparing quenching terms $-\del(\alpha\magn\mbp)/\del z$ (left) 
           and $\del(\delta'\del\mbp/\del z)/\del z$ (right) of equation~\eqref{Br}. 
           Quantities are normalized by $\mbp(0)\Omega\tau^2u^2/h^2$.
           Top row: $\xi\f=0$. 
           Bottom row: $\xi\f=0.2$
           \label{fig:term_profile_ss_modelA}
          }            
\end{figure}

\subsection{A closer look at saturated solutions}
\label{sec:profiles}
In Figure~\ref{fig:profile_ss_modelA} we present profiles across the disc for various quantities of interest
to gain more insight into the solutions.
We do this for Model~A only both to avoid unnecessary repetition since results of other models are qualitatively very similar,
and for presentational convenience because the range of saturated field strengths is smaller in Model~A than in Models~B and C.
As above, line styles represent the three types of quenching, 
dashed for dynamical $\alpha$-quenching, dash-dotted for dynamical $\xi$-quenching, and solid for the combination of the two,
and rows from top to bottom show runs with $\xi\f=0$ and $0.2$, respectively.
Half-profiles are shown to save space as all have even symmetry about the midplane (or odd symmetry for $\alpha\kin$ and $\alpha\magn$).

In the left-hand column we show the components of the field $\mbr$ and $\mbp$ along with the magnetic pitch angle $p=\arctan(\mbr/\mbp)$.
Negative pitch angle means that magnetic field lines form a trailing spiral.
Although, as discussed in Section~\ref{sec:observational}, the \textit{average} value of the pitch angle given by equation~\eqref{pitch}
is insensitive to the type of quenching used, the shape of the profile differs significantly between the different cases.
Specifically, we see that a larger value of $\xi$ leads to a steeper profile with a larger negative value at $z=0$, 
increasing to a large positive value at the disc surface (where the field vanishes).
This is caused by $\mbr$ reversing sign at $z\approx\pm0.7h$, while $\mbp$ does not change sign.
In contrast, for the case with dynamical $\alpha$-quenching and $\xi=\xi\f=0$ (dashed line, top row),
$\mbr$ has the same sign for all $-h<z<h$.\footnote{Note that the equations do not favour solutions of one direction or its opposite,
so that replacing $\mbr$ by $-\mbr$ and $\mbp$ by $-\mbp$ would give an equally valid solution.
The symmetry is broken by the arbitrary seed field chosen.}
Even when $\xi\f$ is large and dynamical $\xi$ is turned off so that $\xi(z,t)=\xi\f$, 
we see a reversal in $\mbr$ (bottom panel in left-hand column).
This provides an observational prediction for the magnetic R\"{a}dler effect:
we would expect to see the field undergo a single reversal in $\mbr$ on either side of the midplane,
so that magnetic field lines transition from trailing to leading spirals as one moves away from the midplane.
However, it should be noted that other effects not included in our simple model, 
such as outflows and the presence of a thicker gaseous disc (halo) surrounding the disc are likely to affect this prediction.
Note that even for the pure dynamical $\alpha$-quenching case with $\xi\f=0$, 
such a reversal in $\mbr$ in the saturated solution occurs in the presence of an outflow, 
which suggests a possible degeneracy between vertical outflow speed and $\xi$ (\citetalias{Chamandy+14b}; \citealt{Chamandy+Taylor15}).
Nevertheless, the possibility of a clear potentially observable signature such as this one is encouraging.

In the middle column, we plot the saturated field strength normalized by $B\eq$ on the left, and the saturated value of $\xi$ on the right.
For cases with dynamical $\xi$-quenching, 
it can be checked that the relationship between $\xi(z)$ and $B(z)/B\eq$ is as expected from equation~\eqref{dynamical_xi} and Figure~\ref{fig:xi},
while for cases without a dynamical $\xi$, $\xi=\xi\f$ for all $z$ (dashed lines).

Finally, in the right-most column, we present profiles of $B$, this time normalized with respect to $B(0)$, on the left,
and profiles of $\alpha\magn$ on the right, with $\alpha\kin$ shown as a dotted line for reference 
(to reiterate, the components of $\alpha$ display odd, rather than even, symmetry about the midplane).
Unlike the shape of the $p$ profile, the shape of the profile for $B$ is almost invariant under changes to the quenching formalism. 
We see that $\alpha\magn$ is smaller in magnitude when there is dynamical $\xi$-quenching and $\alpha$-quenching 
than it is when there is only $\alpha$-quenching (compare solid and dashed lines).
Likewise, a larger value of $\xi\f$ leads to a smaller $\alpha\magn$, 
even when dynamical $\xi$ is turned off so that $\xi=\xi\f$ (compare dashed lines between panels).
These results are expected because $\alpha\magn$ has to be larger to quench the field on its own 
when there is no contribution from $\xi$-quenching.

To better understand the interplay between the two types of quenching, 
we plot the saturated profiles of the relevant terms of equation~\eqref{Br} in Figure~\ref{fig:term_profile_ss_modelA}.
On the left we plot the term $-\del(\alpha\magn\mbp)/\del z$ 
and on the right the term $\del(\delta'\del\mbp/\del z)/\del z$;
both terms have even symmetry about the midplane.
Each term is normalized by $\mbp(0)\Omega\tau^2u^2/h^2$ 
to render it dimensionless and also to allow for easy comparison between the different runs.
The panels show, from top to bottom, $\xi\f=0$ and $0.2$, and the line styles are the same as for Figure~\ref{fig:profile_ss_modelA}.
We see how the role of dynamical $\alpha$-quenching becomes less important as $\xi$ increases.
This can be seen by moving from the top panel to the bottom panel, 
following the evolution of either the left-hand solid or left-hand dashed curves for the term involving $\alpha\magn$,
or by comparing dashed and solid curves in a given panel.
Even when $\xi$ is fixed at $\xi\f$ and does not evolve (dashed curves), 
the magnetic R\"{a}dler effect can still contribute significantly to the quenching of the field, if $\xi\f$ is large enough.
But in this case we would normally think of the magnetic R\"{a}dler effect as leading to a reduction in the growth rate
of the field rather than a quenching of the field, since the effect is present even in the kinematic regime \citepalias{Chamandy+Singh17}.

Finally, it is worth noting the rather sharp feature in the term involving $\delta'$ near the disc surfaces at $z=\pm h$, 
for the case of pure $\xi$-quenching and $\xi\f=0$.%
\footnote{This feature is somewhat more prominent in the profiles for Models~B and C for the same case of pure $\xi$-quenching and $\xi\f=0$.}
Specifically, this feature stems from the term $\propto(\del\xi/\del z)(\del\mbp/\del z)$, 
and is caused by a sharp maximum in $\del\xi/\del z$ near each boundary.
This, in turn, is caused by the tangling model used, and also by the boundary conditions,
which are appropriate for a disc surrounded by vacuum.
Such a feature would be expected to be less prominent in more realistic solutions that include a thick disc/halo.
However, we note that even when the term involving $\del\xi/\del z$ is (artificially) dropped, 
solutions are otherwise very similar, albeit with a somewhat stronger $\xi$-quenching.

\section{Discussion}
\label{sec:discussion}
Our results show that when combined with turbulent tangling of the large-scale magnetic field, 
the magnetic R\"{a}dler effect acts to quench the mean-field dynamo, leading to saturation of the large-scale field.
This `dynamical $\xi$-quenching'\footnote{Here $\xi$ does the quenching, but is not itself quenched!}
is generally strong enough to be competitive with dynamical $\alpha$-quenching for galaxies.
For galaxy parameter values that are similar to those expected for the Solar neighbourhood (Model~A),
the new quenching is effective even if the initial value of $\xi\f$ is very small or zero.
For parameter values more similar to those expected toward the centre of the Galaxy (Models~B and C),
dynamical $\xi$-quenching can only lead to saturation of the large-scale field at strength $\sim B\eq$
on its own, without any $\alpha$-quenching, if $\xi\f\gtrsim0.2$.
This value is less than the oft-quoted estimate of $\xi\f\approx0.3$, 
obtained from fluctuation dynamo simulations with incompressible turbulence,
which is the most relevant case here since incompressibility has been assumed in the theory on which our study has been based.
We conclude from these results that we expect dynamical $\xi$-quenching to be important in real galaxies.

\subsection{Importance vis-\`{a}-vis dynamical $\alpha$-quenching}
What does this mean for dynamical $\alpha$-quenching?
While our new mechanism provides an alternative to $\alpha$-quenching,
both mechanisms are viable and probably operate in tandem.
We have shown how the two effects can combine to lead to a stronger quenching,
resulting in smaller saturation levels for the large-scale field than what is obtained when only one of the effects operates (but still of order $B\eq$).
Our mechanism is in a sense simpler than dynamical $\alpha$-quenching in that 
while it depends on the small-scale magnetic energy density,
it is independent of the mean small-scale magnetic helicity density.

Whereas dynamical $\alpha$-quenching is a consequence of magnetic helicity conservation 
(a fundamental property of high conductivity MHD flows),
dynamical $\xi$-quenching appeals in our model to turbulent tangling of the large-scale field \citepalias{Rogachevskii+Kleeorin07}.
Such turbulent tangling is physically unavoidable and reasonably well-understood.
To a lesser extent, depending on the underlying galaxy parameter values, 
the mechanism sometimes requires the initial small-scale field to be large, that is $\xi\f\sim0.2$.
This, in turn, requires small-scale dynamo action. 
Here again we can be confident that fluctuation dynamos are present in galaxies,
but the precise value of the saturation strength of the small-scale field for such a dynamo
likely depends on galactic parameters.

However, there is another possible effect that has not been included in our models that would enhance 
the strength of dynamical $\xi$-quenching as compared to dynamical $\alpha$-quenching.
When considering the case where both $\xi$ and $\alpha\magn$ are dynamical,
we did not include in $\xi$ the small-scale magnetic energy associated with $\alpha\magn$.
To get an order of magnitude estimate, we can write
\begin{equation}
  \xi= \frac{b^2}{u^2} \sim \frac{l\mean{\bfb\cdot(\bfDel\cro\bfb)}}{u^2} \sim \frac{\alpha\magn}{u}.
\end{equation}
where we have made use of equation~\eqref{alpha_slab} and assumed $\Strouhal\sim1$, that is we have assumed $l\sim \tau u$.
Now, $\alpha\magn$ is of the same order of magnitude as $\alpha\kin$ if dynamical $\alpha$-quenching is important,
that is $\alpha\magn\sim \tau^2u^2\Omega/h$.
Thus we obtain
\begin{equation}
  \xi \sim \left(\frac{\tau u}{h}\right)\Omega\tau.
\end{equation}
The right-hand-side is equal to $0.08$, $0.12$ and $0.47$ for Models~A, B
and C, respectively.
Therefore, this effect could make a significant contribution to $\xi$, 
though the precise value of the contribution would depend
on certain unknown factors of order unity.
What this tells us though is that the saturated values of $\xi$ may
in principle be higher
than what we predicted by considering the fluctuation dynamo and tangling,
and this could result in an even stronger dynamical $\xi$-quenching.

On the other hand, one could question whether the inverse effect is important,
that is, whether $\alpha\magn$ gains a significant contribution from small-scale dynamo action or turbulent tangling.
A priori, this seems less likely to us because we would not normally expect  
magnetic noise generated through either of these effects to possess significant net current helicity,
but testing this hypothesis using numerical experiments would be useful and will be taken up elsewhere.

\subsection{Observational implications}
How do the predictions of our model (with both quenching effects) compare to those of pure dynamical $\alpha$-quenching?
Our model predicts smaller values for the saturated large-scale field strength $B$,
but this quantity is difficult to measure observationally,
and also relies on many parameters in the theory that are uncertain to within factors of order unity (such as $R_\kappa$ and $\Strouhal$).

The pitch angle $p$ of the large-scale magnetic field is more directly observable than the field strength,
and also relies on less parameters in dynamo models, making it an important quantity with which to test the theory \citep{Chamandy+16}.
To be able to compare model predictions with currently available observations, 
we must perform a weighted average of $p$ across the disc according to equation~\eqref{pitch}.
Interestingly, we find that the new quenching prescription results in average pitch angles within $1^\circ$ of those obtained
in the pure dynamical $\alpha$-quenching case, which is comparable to random uncertainties in observations.
The result that the average pitch angle is almost independent of the details of the dynamo non-linearity is 
convenient because this can provide a test of mean-field dynamo theory on the most basic ``zeroth order'' level of accuracy.

However, the profile of the pitch angle with height $p(z)$ is steeper when dynamical $\xi$-quenching is included, 
and extends to more negative values near the midplane and more positive values near the disc surfaces,
than for the case of pure dynamical $\alpha$-quenching.
This then provides a higher order test of the theory and potentially a method to distinguish between the different models for the dynamo non-linearity.
The shape of the vertical profile of the field \textit{strength}, on the other hand, 
is almost independent of the type of quenching invoked.

It is worth noting that it should be possible, even using current observations, to test the \citetalias{Rogachevskii+Kleeorin07}
turbulent tangling model (Figure~\ref{fig:xi}), which is an input to our model.
\citet{Karak+Brandenburg16} do find reasonable agreement between the predictions of \citetalias{Rogachevskii+Kleeorin07} 
and results of their DNS, used for modeling the Sun.

The solutions presented in Section~\ref{sec:dynamical_xi} 
show that the saturated large-scale field strength decreases with $\xi\f$ (the value of $\xi$ in the kinematic regime of $\meanv{B}$).
Since $\xi\f$ is expected to be determined by the small-scale (fluctuation) dynamo,
this suggests that a weaker small-scale dynamo leads to a relatively stronger large-scale field.
Statistically then, we might expect a negative correlation between large-scale magnetic field strength
and parameters that tend to enhance the saturation level of the small-scale dynamo.

This suggests an idea for non-axisymmetric galactic dynamos.
In some galaxies, the large-scale magnetic field is concentrated in magnetic spiral arms akin to the gaseous spiral arms,
but sometimes phase-shifted from them (\citealt{Beck+Hoernes96}; see \citealt{Beck+Wielebinski13} for a review).
Several effects have been proposed to explain these features 
\citep{Moss98,Shukurov98,Rohde+99,Chamandy+13a,Chamandy+13b,Moss+13,Chamandy+15,Moss+15}.
Now consider the possibility that small-scale dynamo action is more intense within the gaseous spiral arms
than within the interarm regions, causing $\xi\f$ to be larger within the arms.
Then (all other parameters being equal) we would expect the saturation strength of the large-scale field
to be larger in the interarm regions, where $\xi\f$ is smaller.
Take, for example, the panels in the middle column of Figure~\ref{fig:profile_ss_modelA},
imagine that the top row with $\xi\f=0$ loosely corresponds to interarm regions and the bottom row with $\xi\f=0.2$ to arm regions,
and consider the black solid curves for the most realistic case that includes both types of quenching.
Not only is the large-scale field larger for $\xi\f=0$, 
but the average saturated value of $\xi$ across the disc is somewhat larger in the $\xi\f=0.2$ case.
These features are consistent with the results of Models~B and C which are not shown for the sake of brevity.
These features would also be in general agreement with observations of arm/interarm large-scale and small-scale magnetic field strengths in NGC~6946, 
which is the prototypical example of a galaxy showing (inter-arm) magnetic arms \citep{Beck07,Basu+Roy13}. 
But this effect would rely on small-scale dynamo action being stronger in the gaseous arms than in between them, 
and whether that is the case in reality has not yet been explored.

Galactic outflows associated with fountain flow, winds, or magnetic buoyancy
are capable of producing several effects in large-scale galactic dynamos
\citep{Brandenburg+93,Brandenburg+95,Moss+99,Shukurov+06,Moss+10,Gressel+13a,Bendre+15,Chamandy+15}. 
Principally, outflows tend to reduce the growth rate and saturation strength by enhancing the critical dynamo number.
On the other hand, they help to make the field strength larger in the saturated state by producing an advective flux of $\alpha\magn$
(see \citealt{Sur+07b}; \citetalias{Chamandy+14b} for models that include both of these effects).
In this work, we have prescribed the vertical component of the mean velocity to vanish in order to keep the models as simple as possible,
but it would be interesting to exploring the combined effects of outflows and dynamical $\xi$ in future work.

\subsection{Testing the effect through direct numerical simulations}
It is important to investigate the new effects explored in this
paper using direct numerical simulations (DNS).
An oft-used approach to study the standard $\alpha^2\Omega$ dynamos,
where a stochastic helical forcing in the momentum equation is
used in a three-dimensional Cartesian shearing box
\citep[see, e.g.,][]{Kapyla+Brandenburg09}, would be
sufficient to explore this if we also include the uniform rotation.
Treating both, the rotation and the strength of small-scale magnetic
fluctuations, as parameters in DNS, we could study systematically their
effects on the kinematic growth rate as well as the non-linear
saturation phase of the large-scale dynamos that are
expected to be excited in such a setup.
However, we must be cautious in the interpretation of DNS results
as possible effects arising due to interaction between the background
shear and fluctuating magnetic field are not fully known and must be
accounted for before making any useful comparison.
In a series of work, \citet{Squire+Bhattacharjee15c,Squire+Bhattacharjee15b}
proposed a new mechanism, called the magnetic shear
current effect, whereby an off-diagonal component of magnetic diffusivity
tensor leads to generation of large-scale magnetic fields in presence
of shear and strong magnetic fluctuations.
Their results were based on quasilinear calculations as well as
low Reynolds number simulations which required somewhat unphysically
strong magnetic fluctuations for the effect to exist, and therefore
it is not clear if the same can be expected at astrophysically relevant
large Reynolds number in presence of naturally produced
magnetic fluctuations.

It would also be useful to simulate galactic dynamo models involving
both large and small scale dynamos in a more realistic setup
\citep{Kapyla+18} where the turbulence in the interstellar medium
is driven mainly by supernova explosions, leading to self-consistent
generation of vorticity and helicity in the compressible medium.
Isolating the kinematic phase of the large scale dynamo and understanding
its growth rate characteristics in the light of new effects discussed
in the present manuscript will be valuable and could significantly improve
our understanding of the non-linear saturation of mean magnetic fields
in such systems.

\subsection{Extensions to the theory}
As noted above, the role of shear in presence of magnetic fluctuations
is necessary to advance our understanding of the large scale dynamos.
Therefore, similar to the models presented in \citetalias{Radler+03}
and Section 10.3 of \citetalias{Brandenburg+Subramanian05a}, there is a
need to have a theoretical formulation which includes shear in the
calculations. This would be more difficult compared to the case
of uniform rotation as shear renders the evolution equation of magnetic
fields explicitly inhomogeneous in the fixed laboratory frame.
But we envisage that the extension can still be made by employing the
shearing coordinate transformation
\citep{Sridhar+Subramanian09a,Sridhar+Subramanian09b,Sridhar+Singh10},
which was used to determine the Galilean invariant expressions
for the turbulent transport coefficients, albeit in the absence of any
$\alpha$ effect \citep{Singh+Sridhar11}. Thus, by exploiting the ideas and
techniques presented in these works, we could explore properly the
role of shear in a nonperturbative manner and assess its role in a
larger context.

\subsection{Placing this work in a broader context}

\begin{figure}
  \centering
  \includegraphics[width=75mm, clip=true, trim=0 0 0 0]{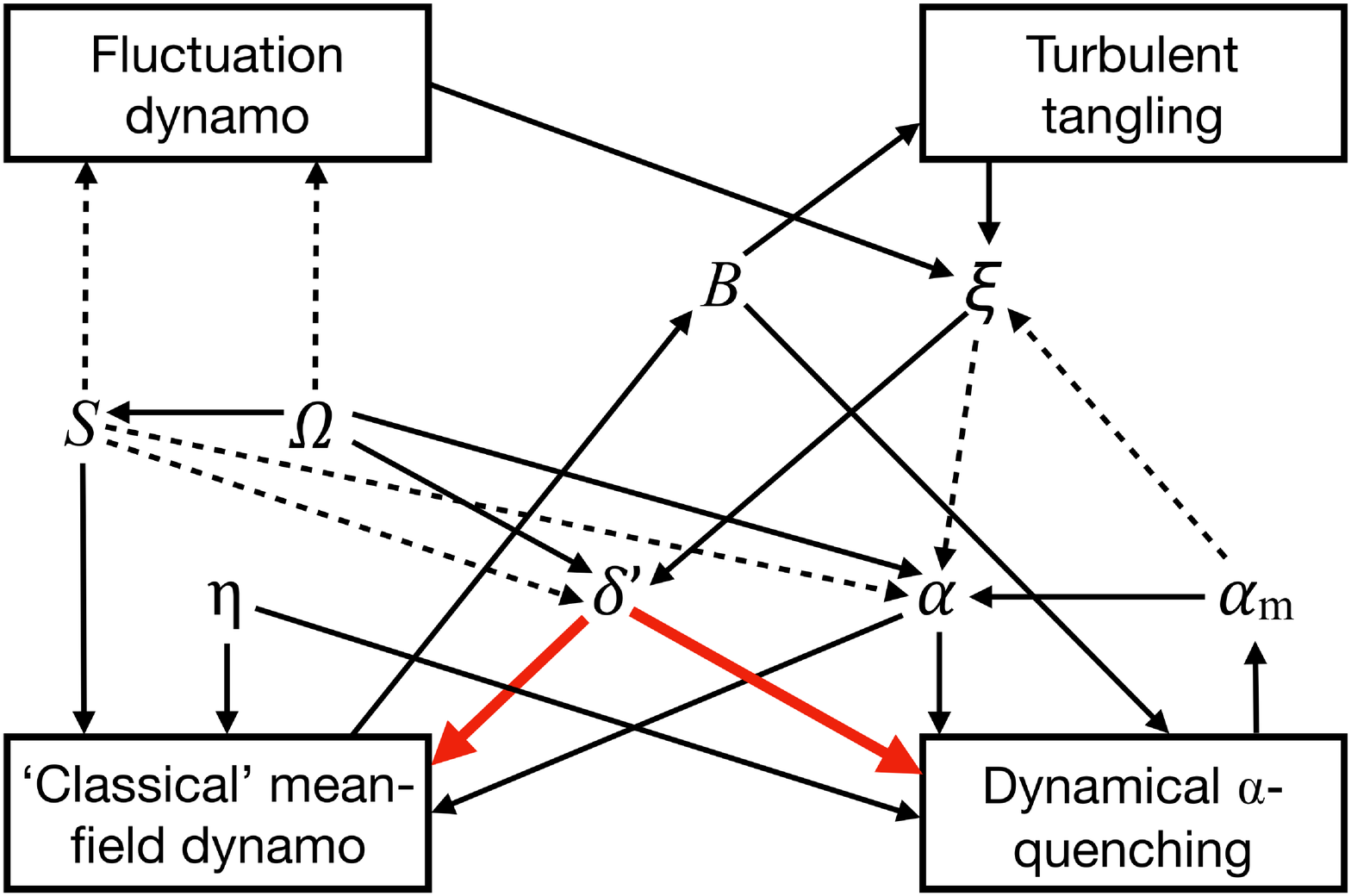}
  \caption{Schematic showing the relations between various processes and quantities discussed in the text.
           The four major mechanisms invoked in our model are shown in rectangles,
           while quantities are represented by their symbols (grouped in rows according to the type of quantity).
           Arrows show the direction of influence between the various quantities and processes.
           Thick red arrows designate those effects on which this work focuses,
           solid arrows designate known effects included in the model,
           while dashed arrows represent (likely) effects that are not included in our model
           because they require further study.
           \label{fig:flow}
          }
\end{figure}

This work and the theory on which it is based \citepalias{Radler+03,Brandenburg+Subramanian05a,Rogachevskii+Kleeorin07} 
establishes and characterizes multiple connections between physical entities, and it can be difficult to keep track of all of these effects.
Thus, in Figure~\ref{fig:flow} we present a schematic diagram that summarizes the relations between physical processes and quantities in this work.
This flow chart acts as a conceptual aid and also makes one aware of the possibilities and needs for future studies,
though we emphasize that it is limited to the ideas discussed in this work and is not meant to be comprehensive in its scope.

The rectangles with text represent the main processes, the symbols the key quantities, and the arrows the direction of influence.
Symbols are organized in rows by type. 
In the top row we have mean magnetic field strength $B$ and normalized small-scale magnetic energy density $\xi=b^2/B\eq^2$.
The middle row contains kinematic quantities, namely the angular rotation speed $\Omega$ and shear rate $S$.
The bottom row contains the turbulent transport coefficients included in our model.
Thick red arrows show the main influences that are explored in this study.
Dotted arrows show influences that are likely important, but are not included in this study.

Turbulent tangling (equation~\eqref{dynamical_xi}) takes as input $B$ and affects $\xi$,
but $\xi$ also depends on the fluctuation dynamo (which sets $\xi\f$ in our model), 
as well as on the dynamical quenching non-linearity (equation~\eqref{dynamical_quenching})
through the small-scale currently helicity density term $\alpha\magn$. 
At the same time, $\xi$ affects $\alpha$ through the second term on the right-hand-side of equation~\eqref{alpha_slab}, 
and $\delta'$ through equation~\eqref{deltakappa_slab}.
Meanwhile $B$ is an output of the classical mean-field dynamo (equation~\eqref{dynamo}), 
and an input into dynamical quenching as well as turbulent tangling.

The shear rate $S$ is determined by the differential rotation, and is thus dependent on $\Omega$.
Further, $\Omega$, but probably also $S$, is important in setting $\alpha$ and $\delta'$, 
$S$ is responsible for the $\Omega$ effect in mean-field dynamos, 
and both $\Omega$ and $S$ also likely play important roles in the fluctuation dynamo.
For ease of presentation, we have omitted the flux density $\Flux$ of $\alpha\magn$,
which would feed into the dynamical quenching, and would be affected by $\alpha\magn$, $\Omega$, $S$, $B$, 
and probably by $\xi$ as well.

Note that the structure of this diagram is chosen to emphasize the mechanisms and connections that are important for this study,
and in doing so glosses over other interesting related phenomena.
For example, the chart would be more symmetrical and comprehensive
if it included a theory of the saturation mechanism for the fluctuation dynamo.

\subsection{Could the magnetic R\"{a}dler effect drive a dynamo?}
In \citetalias{Chamandy+Singh17} we found that the magnetic R\"{a}dler effect can be interpreted as partially suppressing the $\alpha$ effect
in the kinematic regime of mean-field galactic dynamo action.
One reason is that like the $\alpha$ terms, the key terms involving $\delta'$ couple equations for magnetic field components, 
e.g., leading to $\del^2\mbp/\del z^2$ in the $r$-component \eqref{Br} of the induction equation.
So in at least one way the $\delta'$ effect behaves like an $\alpha$ effect with opposite sign,
though it may behave differently in other ways.
It is known that an $\alpha$ effect with opposite sign 
($\alpha_{\phi\phi}<0$ in the northern hemisphere), can drive a dynamo \citepalias[e.g.][]{Brandenburg+Subramanian05a}.
Then one might expect that for large enough negative values of $\delta'$ a dynamo could possibly be obtained, irrespective of the value of $\alpha$.
One could hypothetically imagine such a case arising in nature if very large values of $\xi$ are present, for instance.

To answer this question we obtain numerical solutions for this large $\xi$ case,
and compare them with analytical solutions of a simplified model to aid interpretation,
in Appendix~\ref{sec:driving}.
Our numerical results show that a dynamo is not excited by the magnetic R\"{a}dler effect operating along
with differential rotation and turbulent diffusion for the specific galactic disc dynamo model explored.
If the sign of the effect could somehow be reversed, then a dynamo could be excited
since the new term would look like the traditional R\"{a}dler effect term.
But with the sign that has been derived for $\delta'$, we only get decaying solutions, even for large $\xi$.

\section{Summary and conclusions}
\label{sec:conclusions}
In this work and in \citetalias{Chamandy+Singh17} we have described an effect stemming 
from the theoretical work of \citetalias{Radler+03} and \citetalias{Brandenburg+Subramanian05a}, which we call the `magnetic R\"{a}dler effect,'
since it is the magnetic analogue of the well-known R\"{a}dler or $\bfOmega\cro\bfJ$ effect.
This effect relies on the presence of rotation-induced anisotropy in the turbulence 
and a ratio of small-scale magnetic to kinetic energy densities $\xi$ of order a few$\times0.1$.
The effect potentially has importance in many astrophysical contexts, 
but we have chosen to focus on its application to the galactic dynamo problem.
Below we summarize the main findings.
\begin{itemize}

\item On its own, this effect results in a significant decrease in the kinematic dynamo growth rate,
and slightly smaller saturated large-scale field strengths (top row of Figure~\ref{fig:Btime_log_ss}).

\item When combined with a realistic model for the turbulent tangling of the mean magnetic field,
the magnetic R\"{a}dler effect can lead to efficient quenching and saturation of the mean field as it nears equipartition.
This ``dynamical $\xi$-quenching'' is stronger the larger the initial 
(in the kinematic regime of $\meanv{B}$) value of $\xi$, that is $\xi\f$,
but in some cases can be significant even when $\xi\f=0$.

\item This new saturation mechanism is comparable in strength to the well-studied dynamical $\alpha$-quenching mechanism,
and the two effects complement one another, resulting in a combined quenching that is stronger than each individual contribution
and a saturated field strength that is lower than when only one of the contributions is invoked.

\item We derive a new term in the dynamical $\alpha$-quenching equation 
that has the same form as a generalized Vishniac-Cho flux \citep{Vishniac+Cho01,Subramanian+Brandenburg06},
but is proportional to the small-scale magnetic, rather than kinetic, energy density. 
As both the VC flux and new $\xi$-dependent term hail ultimately from rotation-induced (or shear-induced) turbulence anisotropy,
this formal correspondence hints at a deeper connection still to be unravelled, 
and also suggests that including the shear in the calculation of $\Emf$ is likely to be important.

\item The magnetic pitch angle $p=\arctan(\mbr/\mbp)$ averaged across the disc turns out to be insensitive to the dynamo non-linearity 
(dynamical $\xi$-quenching or dynamical $\alpha$-quenching), 
as does the shape of the magnetic field strength profile plotted against height $z$.
These observational quantities can therefore be used to test the theory to ``zeroth order.''
The profile $p(z)$ \textit{is} sensitive to differences in the non-linearity, 
and therefore potentially provides a ``higher order'' test.

\item The magnetic R\"{a}dler effect cannot be used to drive a galactic dynamo in our model, 
since the turbulent transport coefficient $\delta'$ has the wrong sign.

\end{itemize}

By necessity, our model is rather simple in some respects. 
It treats the turbulent transport coefficients somewhat ``asymmetrically'' in that a heuristic prescription is adopted 
for $\alpha\kin$ and the turbulent pumping term involving $\gamma$ is neglected, 
while for the $\eta$, $\delta$ and $\kappa$ coefficients (which are independent of the stratification of the turbulence),
the expressions from \citetalias{Radler+03} and \citetalias{Brandenburg+Subramanian05a} are directly adopted.
This choice was made to make the problem tractable and to isolate the effect under study,
but it would be interesting to explore the magnetic R\"{a}dler effect in the context of a more realistic dynamo
model that makes parameters like the turbulent speed $u$ depend on height $z$.
At the same time, one could relax the assumption of vacuum boundary conditions and allow for the thin disc to merge into a thick disc or halo,
but then global 2D axisymmetric solutions would be more appropriate.
Large-scale outflows could be included to make the model more general.

Our model is based on theory that assumes the turbulence to be incompressible, 
does not include shear in the derivation of the mean electromotive force $\Emf$, 
assumes the rotation to be slow $\Omega\tau\ll1$,
and does not include the feedback of the small-scale magnetic field onto the small-scale velocity field.
Here we are limited by the underlying theory available, but future generalizations of such calculations would be valuable.

We have not attempted to explain the magnetic R\"{a}dler effect using a physical picture,
for example involving the motions of field lines;
though doing so would be desirable.
Progress has already been made by \citet{Pipin07,Pipin+Seehafer09} in this regard.

Although we have included dynamical $\alpha$-quenching in our model,
we have not attempted to include a contribution to the small-scale magnetic energy $\xi$ 
associated with the growth of small-scale magnetic helicity.
Doing so would likely make the dynamical $\xi$-quenching effect stronger.
Equally interesting would be to test the effect at a more fundamental level using direct numerical simulations,
for instance in a (shearing) periodic box with varying levels of rotation, shear and magnetic fluctuations.

More generally, one cannot rule out the possibility that the effect studied in this work is only the tip of the iceberg, so to speak,
and that there exist other effects involving the influence of small-scale magnetic field on large-scale dynamos
that are yet to be discovered or else discovered \citep{Subramanian+Brandenburg06,Vishniac12a} but not yet explored in detail.

\section*{Acknowledgements}
We are indebted to Eric Blackman for providing detailed comments on the manuscript which led to significant improvements.
We thank the referee for a helpful report which led to improvements in the presentation.
We also thank Kandaswamy~Subramanian, Igor~Rogachevskii, J\"{o}rn~Warnecke and Pallavi~Bhat for stimulating and helpful discussions.
LC is grateful to MPS for hosting him in July 2017, during which a portion of this work was completed.
NKS gratefully acknowledges the hospitality provided by University of Rochester during his visit from $29^{\rm th}$ of April to $4^{\rm th}$ of May, 2018.

\appendix
\section{Saturated state for case without tangling: analytic treatment}
\label{sec:analytic}
Here we obtain an approximate analytic solution for the mean magnetic field in the saturated state,
for the case that includes finite $\xi$.
The solution is only approximate in that it uses the `no-$z$' treatment, along with the $\alphatilde$$\Omega$ approximation \citepalias{Chamandy+Singh17}.
Under the no-$z$ approximation, equation~\eqref{dynamical_quenching_modified} becomes
\begin{equation}
  \label{dynamical_quenching_noz}
  \begin{split}
    \frac{\del\alpha\magn}{\del t} 
      \sim &-\frac{\Strouhal^2\eta}{B\eq^2}\Bigg\{\frac{2}{3\tau}\left(\frac{\alpha}{\eta}(\mbr^2+\mbp^2) 
        +\frac{3\sqrt{-\pi \Dtilde}}{8h}\mbr\mbp\right)\\
        &-\frac{\Omega}{h}
        \left[ \left(C\VC(q -1) +\frac{4}{5}\xi\right)\mbp^2-\left(C\VC(q -1) -\frac{4}{5}\xi\right)\mbr^2\right]\Bigg\}\\
        &-\frac{\pi^2R_\kappa\eta}{h^2}\alpha\magn,
  \end{split}
\end{equation}
where $\Dtilde_\mathrm{gen}=\alpha q\Omega h^3/\eta^2$ is a generalized effective dynamo number for which $\alphatilde\kin$ 
has been replaced by $\alpha= \alphatilde\kin +\alpha\magn$,
and where we have substituted $\del\mbr/\del z \sim -\mbr/h$ and $\del\mbp/\del z \sim -\mbp/h$, 
in addition to the other `no-$z$' relations used in \citetalias{Chamandy+14b,Chamandy+Singh17}.

Let us assume $C\VC=0$ for simplicity.  
For the steady state, $\del\alpha\magn/\del t=0$ and $\Dtilde=D\crit$, so we have
\begin{equation}
  \label{dynamical_quenching_noz_steady_state}
  \begin{split}
    0= &-\frac{\Strouhal^2\eta}{B\eq^2}\Bigg[\frac{2}{3\tau}\left(\frac{\alpha}{\eta}(\mbr^2+\mbp^2) 
       +\frac{3\sqrt{-\pi D\crit}}{8h}\mbr\mbp\right)\\
       &-\frac{4}{5}\frac{\Omega}{h}\xi(\mbr^2+\mbp^2)\Bigg] -\frac{\pi^2R_\kappa\eta}{h^2}\alpha\magn.
  \end{split}
\end{equation}
The critical dynamo number $D\crit$ can be obtained from equations~\eqref{Br} and \eqref{Bp} by requiring time derivatives to vanish,
and works out to $D\crit= -(\pi/2)^5$.

For the first term in equation~\eqref{dynamical_quenching_noz_steady_state}, we can write $\alpha= \alpha\kin +\alpha\magn$.
Now the critical value $\alpha\crit= \alphatilde\kin +\alpha\magn$,
where $\alphatilde\kin= [1 -(\pi^3/20)\xi]$ \citepalias{Chamandy+Singh17}.
It follows that 
\begin{equation}
  \alpha= \alpha\crit +\frac{\pi^3}{20}\xi\alpha\kin= \frac{\eta}{hR_\Omega}\left(D\crit +\frac{\pi^3}{20}\xi D\right).
\end{equation}
So we then have for the first term in equation~\eqref{dynamical_quenching_noz_steady_state},
\begin{equation}
  \frac{\alpha}{\eta}B^2= \frac{B^2}{hR_\Omega}\left(D\crit +\frac{\pi^3}{20}\xi D\right).
\end{equation}
Note that $D$ and $R_\Omega$ are both negative so their ratio is positive.

For the second term we first write, using the definition of the pitch angle $\tan p= \mbr/\mbp$ with $-\pi/2< p\leq \pi/2$,
\begin{equation}
  \mbr\mbp= B^2\tan p \cos^2p.
\end{equation}
Now, $\tan p$ is obtained by solving equations \eqref{Br} and \eqref{Bp} in the no-$z$ approximation, 
to give \citepalias{Chamandy+14b}
\begin{equation}
  \tan p= \frac{1}{R_\Omega}\sqrt{-\frac{2}{\pi}D\crit}.
\end{equation}
Note that $p$ is predicted to be independent of $\xi$ in the saturated state because $\alphatilde$ gets set equal to the critical value
$\alpha\crit$, which is independent of $\xi$, and it is the ratio $R_{\alphatilde}/R_\Omega$ which sets $p$.
Thus, we find for the second term of equation~\eqref{dynamical_quenching_noz_steady_state},
\begin{equation}
  \frac{3\sqrt{-\pi D\crit}}{8h}\mbr\mbp= -\frac{3 B^2}{4\sqrt{2}}\frac{D\crit}{h R_\Omega}\cos^2p.
\end{equation}

The third term is the new VC flux-like term.
We can use the relation $\alpha\kin= \tau^2u^2\Omega/h$ to write
\begin{equation}
  \Omega= \frac{R_\alpha}{3\tau}= \frac{D}{3\tau R_\Omega}.
\end{equation}
Substituting this into equation~\eqref{dynamical_quenching_noz_steady_state},
and factoring out $2/(3\tau)$, we obtain
\begin{equation}
  -\frac{4}{5}\frac{\Omega}{h}\xi B^2= -\frac{2}{3\tau} \left(\frac{2}{5}\frac{B^2}{hR_\Omega}\xi D\right). 
\end{equation}

Finally, for the fourth term we first write
\begin{equation}
  \alpha\magn= \alpha\crit -\alpha\kin\left(1-\frac{\pi^3}{20}\xi\right)= 
    \alpha\crit -\alphatilde\kin= -\frac{\eta(\Dtilde -D\crit)}{h R_\Omega}.
\end{equation}
Substituting the latter expression into the fourth term of equation~\eqref{dynamical_quenching_noz_steady_state} we find
\begin{equation}
  \begin{split}
    -\frac{\pi^2R_\kappa \eta}{h^2}\alpha\magn=& \frac{\pi^2R_\kappa\eta^2}{h^3R_\Omega}\left(\Dtilde -D\crit\right)\\
       =& \frac{2\eta}{3\tau}\frac{\pi^2}{2} \left(\frac{\tau u}{h}\right)^2 \frac{R_\kappa}{hR_\Omega}\left(\Dtilde -D\crit\right),
  \end{split}
\end{equation}
where for the final equality we have used the relation $\eta= \tau u^2/3$ to write the expression in a convenient form.

Multiplying the equation by $3\tau hR_\Omega/(2\Strouhal^2\eta D\crit)$, and rearranging, we obtain
\begin{equation}
  \left(\frac{B}{B\eq}\right)^2= \frac{\pi^2 R_\kappa}{C\chi(p,\xi)}\left(\frac{\Dtilde}{D\crit} -1\right),
\end{equation}
where
\begin{equation}
  \label{chi}
  \chi(p,\xi)\equiv 1 -\frac{3}{4\sqrt{2}}\cos^2 p +\left(\frac{\pi^3}{20} -\frac{2}{5}\right)\xi\frac{D}{D\crit},
\end{equation}
and
\begin{equation}
  C\equiv 2\left(\frac{h}{l}\right)^2.
\end{equation}
This expression agrees with expression~(18) of \citetalias{Chamandy+14b} for the limiting case $\xi\rightarrow0$ 
and negligible advective helicity flux (here the latter vanishes because there is no outflow in the model).
The parameter $\xi$ enters in two places: through $\Dtilde= D[1-(\pi^3/20)\xi]$, and through $\chi(p,\xi)$. 
Both effects suppress the saturated field strength.
Note that the $\delta'$-dependent contribution to $\Emf\cdot\meanv{B}$ leads to the term $-(2/5)\xi(D/D\crit)$,
which helps to offset the effect of the term $(\pi^3/20)\xi(D/D\crit)$.
Therefore, were the new term proportional to $\xi$ and involving $\mbp$
in equation~\eqref{dynamical_quenching_noz} or \eqref{dynamical_quenching_modified}
to be omitted, $B$ would be suppressed more strongly.
This prediction is borne out in the numerical solutions of the full equations,
as can be seen by comparing the top (including this term)
and middle (not including this term) rows of Figure~\ref{fig:Btime_log_ss}.

\section{A new non-linear effect within the dynamical $\alpha$-quenching framework: the general case}
\label{sec:xiVC_general}
If radial and azimuthal derivatives are not neglected, 
then the VC flux term becomes, in cylindrical coordinates,
\begin{equation}
  \label{VC}
  \begin{split}
    \frac{\del\alpha\magn}{\del t}=&
      \ldots +\frac{C\VC\Strouhal^2\eta}{B\eq^2}(q-1)\Omega
       \Bigg[
           \mbr\frac{\del\mbr}{\del z} -\mbp\frac{\del\mbp}{\del z} -\frac{1}{2}\mbz\frac{\del\mbr}{\del r}\\ 
         &-\frac{1}{2}\mbr\frac{\del\mbz}{\del r} +\frac{1}{2}(q-1)\frac{\mbr\mbz}{r}
          -\frac{1}{2}\frac{r}{q-1}\frac{\del q}{\del r}\frac{\mbr\mbz}{r}\\
         &+\frac{1}{2}\frac{\mbz}{r}\frac{\del\mbp}{\del\phi} +\frac{1}{2}\frac{\mbp}{r}\frac{\del\mbz}{\del\phi}
        \Bigg].
  \end{split}
\end{equation}
Likewise,
equations~(10.56--10.61) of \citetalias{Brandenburg+Subramanian05a} can be written as
\begin{align}
  \Emfr=&\:\alpha\f\mbr 
         +\alpha_{rz}\mbz
         -\eta\left(\frac{1}{r}\frac{\del\mbz}{\del\phi} 
         -\frac{\del\mbp}{\del z}\right) \nonumber\\
        &+\gamma_\phi\mbz 
         -\gamma_z\mbp 
         +(-\delta +\kappa)\frac{\del\mbr}{\del z} 
         +( \delta +\kappa)\frac{\del\mbz}{\del r},\\
  \Emfp=&\:\alpha\f\mbp 
         +\alpha_{\phi z}\mbz
         -\eta\left( \frac{\del\mbr}{\del z}              
         -\frac{\del\mbz}{\del r}\right) \nonumber\\
        &-\gamma_r\mbz    
         +\gamma_z\mbr
         +(-\delta +\kappa)\frac{\del\mbp}{\del z} 
         +( \delta +\kappa)\frac{1}{r}\frac{\del\mbz}{\del\phi},\\
  \Emfz=&\:\alpha_{zz}\mbz 
         +\alpha_{zr}\mbr +\alpha_{z\phi}\mbp
         -\eta\left[ \frac{1}{r}\frac{\del(r\mbp)}{\del r}
         -\frac{1}{r}\frac{\del\mbr}{\del\phi}\right] \nonumber\\ 
        &+\gamma_r\mbp -\gamma_\phi\mbr
         +2\kappa\frac{\del\mbz}{\del z},
\end{align}
where 
\begin{align}
  \alpha\f&=    \tfrac{1}{3}\tau\mean{\bm{j}\cdot\bm{b}} 
               -\tfrac{4}{5}\Omega\tau^2\frac{\del}{\del z}\left(u^2 
               -\tfrac{1}{3}b^2\right),\\
  \alpha_{rz}    &=\alpha_{zr}       = \tfrac{11}{30}\Omega\tau^2\frac{\del}{\del r}\left(u^2+\tfrac{3}{11}b^2\right),\\
  \alpha_{\phi z}&=\alpha_{z\phi}= \tfrac{11}{30}\Omega\tau^2\frac{1}{r}\frac{\del}{\del\phi}\left(u^2+\tfrac{3}{11}b^2\right),\\
  \alpha_{zz}&= \tfrac{1}{3}\tau\mean{\bm{j}\cdot\bm{b}} 
               -\tfrac{1}{15}\Omega\tau^2\frac{\del}{\del z}\left( u^2 -7b^2\right),\\
  \eta&=        \tfrac{1}{3}\tau u^2,\\
  \gamma_r&=   -\tfrac{1}{6}\tau\frac{\del}{\del r}\left( u^2 -b^2\right) 
               +\tfrac{1}{6}\Omega\tau^2\frac{1}{r}\frac{\del}{\del\phi}\left( u^2 +b^2\right),\\
  \gamma_\phi&=-\tfrac{1}{6}\tau\frac{1}{r}\frac{\del}{\del\phi}
                  \left( u^2 -b^2\right) -\tfrac{1}{6}\Omega\tau^2\frac{\del}{\del r}\left( u^2 +b^2\right),\\
  \gamma_z&=   -\tfrac{1}{6}\tau\frac{\del}{\del z}\left( u^2 -b^2\right),\\
  \delta&=      \tfrac{1}{6}\Omega\tau^2\left( u^2 -b^2\right),\\
  \kappa&=      \tfrac{1}{6}\Omega\tau^2\left( u^2 +\tfrac{7}{5}b^2\right).
\end{align}
Considering only the terms containing $\delta$ or $\kappa$ and using dots to denote other terms,
and also making use of equations \eqref{xi}, \eqref{Beq} and \eqref{dynamical_quenching}, we obtain
\begin{equation}
  \begin{split}
    \label{deltakappaterm_general}
    \frac{\del\alpha\magn}{\del t}=&
      \ldots -\frac{\Strouhal^2\eta\Omega}{B\eq^2}
        \Bigg\{
           \frac{4}{5}\xi\left(\mbr\frac{\del\mbr}{\del z} +\mbp\frac{\del\mbp}{\del z}\right)\\
         &+\frac{2}{3}\left(1+\frac{1}{5}\xi\right)
             \left(\mbr\frac{\del\mbz}{\del r} +\frac{\mbp}{r}\frac{\del\mbz}{\del\phi}\right)\\
         &-\frac{2}{3}\left(1+\frac{7}{5}\xi\right)
             \frac{\mbz}{r}\left[\frac{\del}{\del r}(r\mbr) +\frac{\del\mbp}{\del\phi}\right]
        \Bigg\},
  \end{split}
\end{equation}
where we have used
\begin{equation}
  \bfDel\cdot\meanv{B}=
    \frac{1}{r}\frac{\del}{\del r}(r\mbr) +\frac{1}{r}\frac{\del\mbp}{\del\phi} +\frac{\del\mbz}{\del z}
      =0.
\end{equation}
Finally, combining equations~\eqref{VC} and \eqref{deltakappaterm_general} we obtain
\begin{equation}
  \begin{split}
  \frac{\del\alpha\magn}{\del t}=
    &\ldots +\frac{\Strouhal^2\eta\Omega}{B\eq^2}
      \Bigg\{
          \left[C\VC(q-1) -\frac{4}{5}\xi\right]\mbr\frac{\del\mbr}{\del z}\\
        &-\left[C\VC(q-1) +\frac{4}{5}\xi\right]\mbp\frac{\del\mbp}{\del z}\\
        &-\left[ \frac{1}{2}C\VC(q-1)
                -\frac{2}{3}\left(1+\frac{7}{5}\xi\right)\right]\mbz\frac{\del\mbr}{\del r}\\
        &+\left[ \frac{1}{2}C\VC\left((q-1)^2 -r\frac{\del q}{\del r}\right)
                +\frac{2}{3}\left(1+\frac{7}{5}\xi\right)\right]\frac{\mbr\mbz}{r}\\
        &+\left[ \frac{1}{2}C\VC(q-1)
                +\frac{2}{3}\left(1+\frac{7}{5}\xi\right)\right]\frac{\mbz}{r}\frac{\del\mbp}{\del\phi}\\
        &-\left[ \frac{1}{2}C\VC(q-1)
                +\frac{2}{3}\left(1+\frac{1}{5}\xi\right)\right]\mbr\frac{\del\mbz}{\del r}\\
        &+\left[ \frac{1}{2}C\VC(q-1)
                -\frac{2}{3}\left(1+\frac{1}{5}\xi\right)\right]\frac{\mbp}{r}\frac{\del\mbz}{\del\phi}
      \Bigg\}.
  \end{split}
\end{equation}
Therefore, the divergence of the Vishniac-Cho flux and the $\Emf\cdot\meanv{B}$ component 
stemming from the generalized R\"{a}dler effect give contributions to the $\alpha\magn$ equation that have the same form.

\section{Solutions for large $\xi$}
\label{sec:driving}

\subsection{Analytic solution for a simpler illustrative case}
\label{sec:large_xi_analytic}
To explore the possibility of a dynamo \textit{driven} by the magnetic R\"{a}dler effect, 
we first derive the dispersion relation for equations \eqref{Br} and \eqref{Bp}.
For simplicity and tractability, we explore analytically the simpler case 
where the domain is infinite and $\alpha=\const$.
This generalizes the treatment of \citetalias{Brandenburg+Subramanian05a}, 
who considered separately an $\alpha\Omega$-type dynamo and a $\delta\Omega$-type dynamo.
In Section~\ref{sec:large_xi_numerical} we compare qualitatively the analytical results of this simple model 
with numerical results of our more realistic galactic dynamo model which has imposed vacuum boundary conditions and $\alpha$ which depends on $z$.
For now, we look for solutions of the form
\begin{equation}
  \label{Bcomplex}
  \meanv{B}= \mathrm{Re} \left(\meanv{B}\f\Exp{\lambda t +ikz}\right)
\end{equation}
in an infinite domain, with all transport coefficients taken as constants.
Substituting equation~\eqref{Bcomplex} into equations~\eqref{Br} and \eqref{Bp} we obtain
\begin{align}
  \label{lambda_B0r}
  \lambda\mbfr&=               -ik\alpha\mbfp -k^2\delta'\mbfp -k^2\eta\mbfr,\\
  \label{lambda_B0p}
  \lambda\mbfp&= -q\Omega\mbfr +ik\alpha\mbfr +k^2\delta'\mbfr -k^2\eta\mbfp.
\end{align}
Rearranging, multiplying the left sides and right sides together, and dividing by $\mbfr\mbfp$, 
we arrive at the dispersion relation
\begin{equation}
  \left(\lambda+\eta k^2\right)^2= \left(-ik\alpha -k^2\delta'\right)\left(ik\alpha +k^2\delta' -q\Omega\right).
\end{equation}
Below we neglect terms related to $\alpha$ and $\delta'$ in equation~\eqref{Bp} or \eqref{lambda_B0p} 
because they are generally subdominant compared with the term involving $q\Omega$.
Then we obtain
\begin{equation}
  \left(\lambda +\eta k^2\right)^2= q\Omega\left[ik\alpha +k^2\delta'\right],
\end{equation}
which gives
\begin{equation}
  \label{lambda}
  \lambda= -\eta k^2 \pm (q\Omega)^{1/2}\left[ik\alpha +k^2\delta'\right]^{1/2}.
\end{equation}
Now for convenience we define
\begin{equation}
  v\equiv k^2\delta', \qquad w\equiv k\alpha,
\end{equation}
and solve $x+iy=(v+iw)^{1/2}$.
We obtain
\begin{equation}
  x^2-y^2=v, \qquad 2xy= w, \qquad x^2+y^2= (v^2+w^2)^{1/2},
\end{equation}
where the first (second) relation comes from equating real (imaginary) parts 
and the last relation comes from equating the moduli of the left-hand and right-hand sides.
The first and third relations can be used to solve for $x$ and $y$ modulo a sign.
By straightforward algebra we then find
\begin{align}
  x&= \pm\frac{1}{\sqrt{2}}\left[v+(v^2+w^2)^{1/2}\right]^{1/2},\\
  y&= \pm\frac{1}{\sqrt{2}}\left[-v+(v^2+w^2)^{1/2}\right]^{1/2}.
\end{align}
From the relation $2xy=w$ we find that $x$ and $y$ must have the same sign.
Then by comparison with equation~\eqref{lambda} we obtain
\begin{align}
  \label{real}
  \mathrm{Re}(\lambda)&= -\eta k^2 \pm\left|\frac{q\Omega k\alpha}{2}\right|^{1/2}(X+Y)^{1/2}\\
  \label{imag}
  \mathrm{Im}(\lambda)&= -\omega\cyc= \pm\left|\frac{q\Omega k\alpha}{2}\right|^{1/2}(X-Y)^{1/2},
\end{align}
where
\begin{equation}
  X\equiv \left(1+\frac{v^2}{w^2}\right)^{1/2}= \left[1+\left(\frac{k\delta'}{\alpha}\right)^2\right]^{1/2},
\end{equation}
and
\begin{equation}
  Y\equiv \frac{v}{w} = \frac{k\delta'}{\alpha}.
\end{equation}
Here $\mathrm{Re}(\lambda)$ is the growth rate and $\omega\cyc$ is the cycle frequency of dynamo waves.

In the limit $\delta'\rightarrow0$, we have $X\rightarrow1$ and $Y\rightarrow0$, 
and we recover the standard $\alpha\Omega$ case \citepalias[equations 6.39-6.40 of][]{Brandenburg+Subramanian05a}.
In the limit $\delta'\rightarrow\infty$, we obtain $(X+Y)\rightarrow2Y$ and $(X-Y)\rightarrow0$,
which is like a standard non-oscillatory $\alpha\Omega$ dynamo but with $\alpha$ in expression~\eqref{real} replaced by $2k\delta'$.
This case was also studied by \citetalias{Brandenburg+Subramanian05a} (their equation (6.54)).
If $\delta'<0$, a dynamo is not obtained, as pointed out by \citetalias{Brandenburg+Subramanian05a}.
We see that when $\delta'\rightarrow-\infty$, $(X+Y)\rightarrow0$ while $(X-Y)\rightarrow-2Y$.
Therefore, as $-\delta'$ is increased, 
for large enough values we should expect to see the decay rate approach an asymptotic limit and the frequency of oscillations continue to increase.
Do these features also obtain in the case of our more realistic galactic dynamo with boundaries and $z$-dependent $\alpha$?
Below we turn to answering this question using numerical solutions.

\subsection{Simulations in the kinematic regime for large values of $\xi$}
\label{sec:large_xi_numerical}
\begin{figure}
  \includegraphics[width=1.0\columnwidth,clip=true,trim=30 55 0  0]{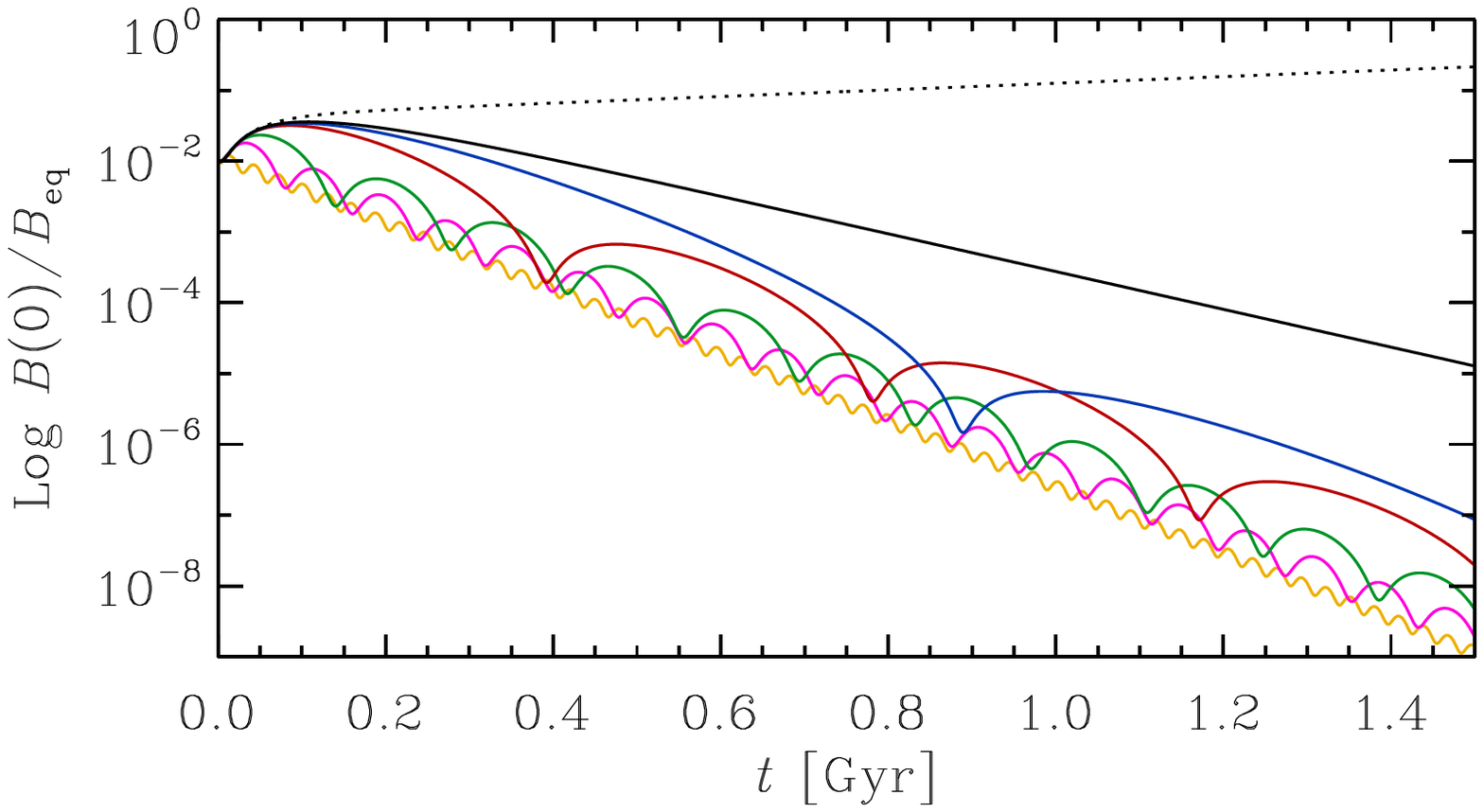}\\
  \includegraphics[width=1.0\columnwidth,clip=true,trim=30 0  0 10]{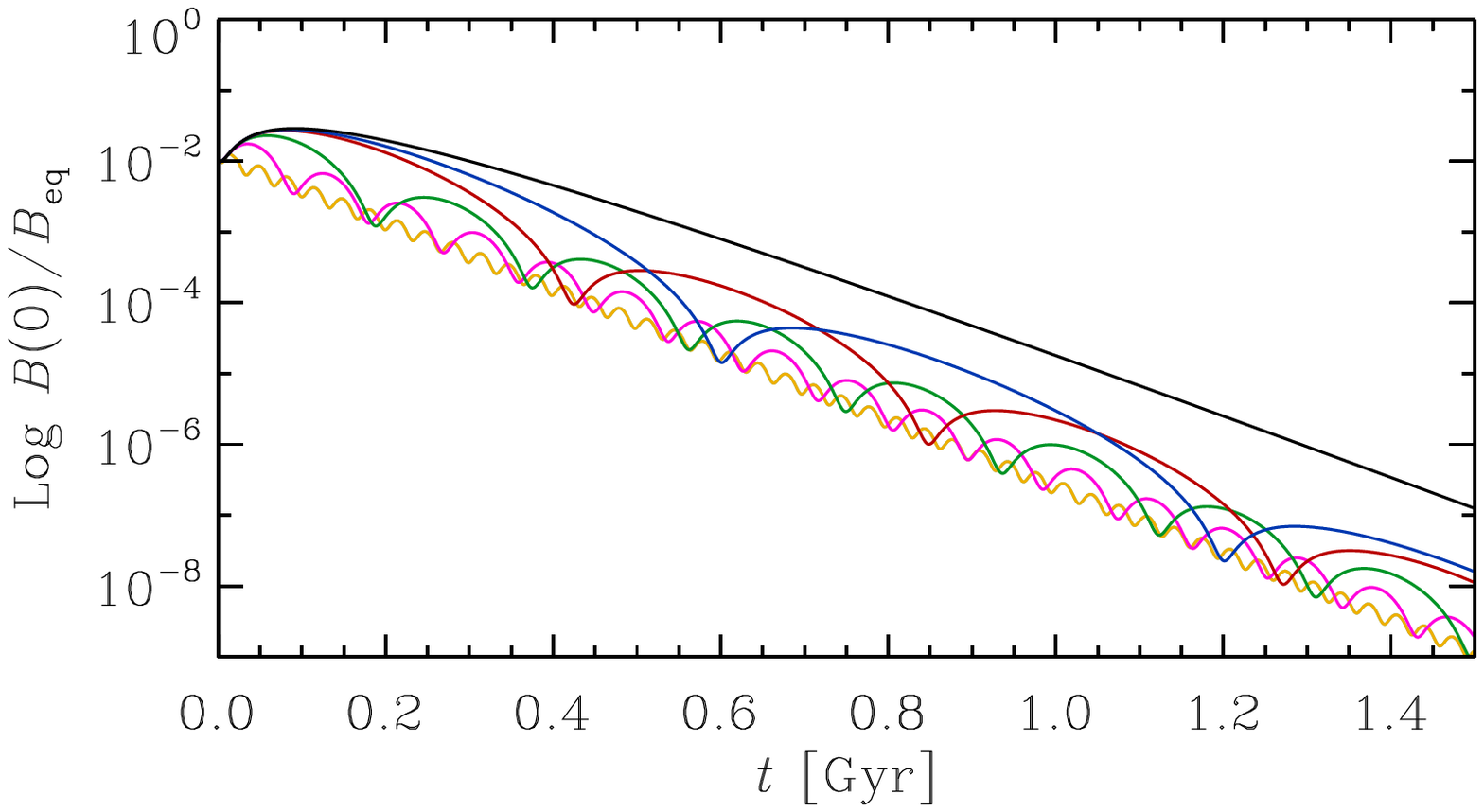}
  \caption{Top: Evolution of magnetic field strength at the midplane, relative to the equipartition value, 
           for Model~B with $\alpha\f=\tau^2u^2\Omega/h$ and $\xi=0.45$ (dotted), 
           $0.65$ (black solid), $0.7$ (blue), $0.8$ (red), $1.6$ (green), $3.2$ (magenta) and $12.8$ (orange).
           Bottom: Here $\alpha\f=0$ and $\xi=0$ (black), $0.05$ (blue), $0.1$ (red), $0.5$ (green), $2$ (magenta) and $10$ (orange).
           \label{fig:large_xi}
          }
\end{figure}

Using the parameter values of Model~B, we explore what happens when $\delta'$ 
is made to be large and negative. 
We do this by ramping up $\xi$ and plotting the solution for each value.
The large-scale field strength at the midplane is plotted against time for 
simulations with different values of $\xi$ in Figure~\ref{fig:large_xi}.
The results for $\alpha\f=\tau^2u^2\Omega/h$ (\citealt{Krause+Radler80}; ch.~VI of \citealt{Ruzmaikin+88})
are shown in the top panel, while those for $\alpha\f=0$ (no $\alpha$ effect) are shown in the bottom panel.
For all solutions presented, $\mbr$ and $\mbp$ have even symmetry about the midplane.
In the first case, as we increase $\xi$, 
what had been a growing non-oscillating solution ($\xi=0.45$, dotted line in the top panel)
changes to a decaying non-oscillating solution ($\xi=0.65$, solid black in the top panel) 
and then to a decaying oscillating solution.
The average decay rate increases slightly as $\xi$ is increased, just as suggested by equation~\eqref{real},
while the cycle frequency increases markedly, also in qualitative agreement with equation~\eqref{imag}.
The amplitude of oscillations decreases with $\xi$.

In the second case, when $\alpha=0$, we initially have a non-oscillating decaying solution when $\xi=0$,
since $\alpha$ and $\delta'$ terms vanish and we are left with a diffusion equation for $\meanv{B}$.
As $\xi$ is increased, the solution becomes oscillatory, as expected,
and solutions resemble those of the finite $\alpha$ case.
However, in the vanishing $\alpha$ case, the average decay rate does not change when $\xi$ is increased, 
which is just what happens for the analytic solution of Section~\ref{sec:large_xi_analytic}, 
as seen from equation~\eqref{real} for the case $\alpha=0$.
This feature can be recognized in the numerical solutions 
by noting that the minima of all the curves in the bottom panel of Figure~\ref{fig:large_xi} fall along the same line.
We also find that when the simulation is run to very long times $\sim50\Gyr$, the decay rate for the $\xi=0$ non-oscillating solution
approaches the average decay rate of the oscillating solutions, as would be expected from equation~\eqref{real}.
Therefore, the numerical solutions presented display some of the same qualitative features 
as the analytic solutions obtained for an infinite domain and constant $\alpha$ in Section~\eqref{sec:large_xi_analytic}.

\footnotesize{
\noindent
\bibliographystyle{mnras}
\bibliography{refs}
}

\label{lastpage}

\end{document}